\documentclass[twocolumn]{28onr_art}
\usepackage{ulem}
\usepackage{color}
\usepackage{natbib}
\usepackage[pdftex]{graphicx}
\usepackage{amsmath}
\usepackage[colorlinks=true,
            urlcolor=blue,
            linkcolor=docgreen,
            citecolor=docgreen,
            bookmarks=true,
            pdftitle={Three-Dimensional Simulations of Deep-Water Breaking Waves},
            pdfauthor={Kyle A. Brucker},
            pdfsubject={Proceedings of the 28th Naval Hydrodynamics
                         Symposium},
            pdfkeywords={Breaking Waves, Air Entrainment, Numerical
                         Simulations, High Performance Computing},
	    bookmarksopen=false,
            pdfpagemode=None]{hyperref}

\definecolor{docgreen}{rgb}{0,.5,0}

\def\be{\begin{eqnarray}}
\def\ee{\end{eqnarray}}
\def\benl{\begin{eqnarray*}}
\def\eenl{\end{eqnarray*}}

\newcommand{\nwc}{\newcommand}
\nwc{\bm}{\boldmath}
\nwc{\m}{\mbox}
\nwc{\ubm}{\unboldmath}
\nwc{\bmU}{\m{\bm$U$\ubm}}
\nwc{\bmX}{\m{\bm$X$\ubm}}
\nwc{\bmu}{\m{\bm$u$\ubm}}
\nwc{\bmx}{\m{\bm$x$\ubm}}
\nwc{\bmz}{\m{\bm$z$\ubm}}
\nwc{\bmv}{\m{\bm$v$\ubm}}
\nwc{\bmw}{\m{\bm$w$\ubm}}
\nwc{\bmW}{\m{\bm$W$\ubm}}
\nwc{\bmn}{\m{\bm$n$\ubm}}
\nwc{\bmG}{\m{\bm$G$\ubm}}
\nwc{\bmF}{\m{\bm$F$\ubm}}
\nwc{\bmI}{\m{\bm$I$\ubm}}
\nwc{\bmN}{\m{\bm$N$\ubm}}
\nwc{\bmP}{\m{\bm$P$\ubm}}
\nwc{\bmcalP}{\m{\bm $\cal P$\ubm}}
\nwc{\bmV}{\m{\bm$V$\ubm}}
\nwc{\bmS}{\m{\bm$S$\ubm}}

\begin{document} 
\pagenumbering{arabic} 

\title{Three-Dimensional Simulations of Deep-Water Breaking Waves}
\author{Kyle A. Brucker$^1$, Thomas T. O'Shea$^1$, Douglas G.\ Dommermuth$^1$,
and Paul Adams$^2$} 
\affiliation{\small $^1$Naval Hydrodynamics Division, Science Applications
International Corporation, \\ 10260 Campus Point Drive, MS 34, San Diego, CA
92121 \\ $^2$Unclassified Data Analysis and Assessment Center, \\U.S. Army
Engineering Research and Development Center, MS 39180}
\maketitle

\begin{abstract} 
The formulation of a canonical deep-water breaking
wave problem is introduced, and the results of a set of
three-dimensional numerical simulations for deep-water
breaking waves are presented. In this paper fully nonlinear
progressive waves are generated by applying a normal
stress to the free surface. Precise control of the
forcing allows for a systematic study of four types of
deep-water breaking waves, characterized herein as weak
plunging, plunging, strong plunging, and very strong
plunging.

The three-dimensional iso contours of vorticity exhibit
intense streamwise vorticity shortly after the initial
ovular cavity of air is entrained during the primary
plunging event. An array of high resolution images are
presented as a means to visually compare and contrast
the major events in the breaking cycles of each case.
The volume-integrated energy shows 50\% or more of
the peak energy is dissipated in strong and very strong
plunging events. The volume of air entrained beneath
the free surface is quantified. For breaking events characterized
by plunging, strong plunging and very strong
plunging, significant quantities of air remain beneath the
free surface long after the initial breaking event. The
rate at which the air beneath the free surface degasses
is linear and the same in all cases. The use of volume-weighted
(Reynolds) and mass-weighted (Favre) averages
are compared, and it is found that statistics obtained
by Favre averaging show better agreement with respect
to the position of free surface than do those obtained by
Reynolds averaging. The average volume fraction plotted
on a log scale is used to visually elucidate small volumes
of air entrained below the free surface. For the
strong plunging and very strong plunging cases significant
air is also entrained after the initial plunging event
at the toe of spilling breaking region. Improvements to
the Numerical Flow Analysis code, which expand the
types of problems it can accurately simulate are discussed,
along with the results of a feasibility study which
shows that simulations with 5-10 billion unknowns are
now tractable.
\end{abstract}

\section{Introduction} 

Surface wave breaking in deep water plays an important
role in many engineering and geophysical problems. The
interested reader is referred to review articles by \cite{melville96}
and \cite{duncan01} and the detailed discussion of
deep-water breaking waves in \cite{rapp90}.
The focus here remains on aspects relevant to the numerical
simulation of breaking waves.

Historically, the two main types of simulations to
simulate free-surface flows are the boundary integral
equations method (BIEM) and high-order spectral (HOS)
methods. These calculations fail at the point at which the
surface impacts upon itself, if not sooner. They also employ
a single-phase approximation in which the effects
of the air on the water are neglected.

\begin{figure}
\centering
\includegraphics[width=\linewidth]{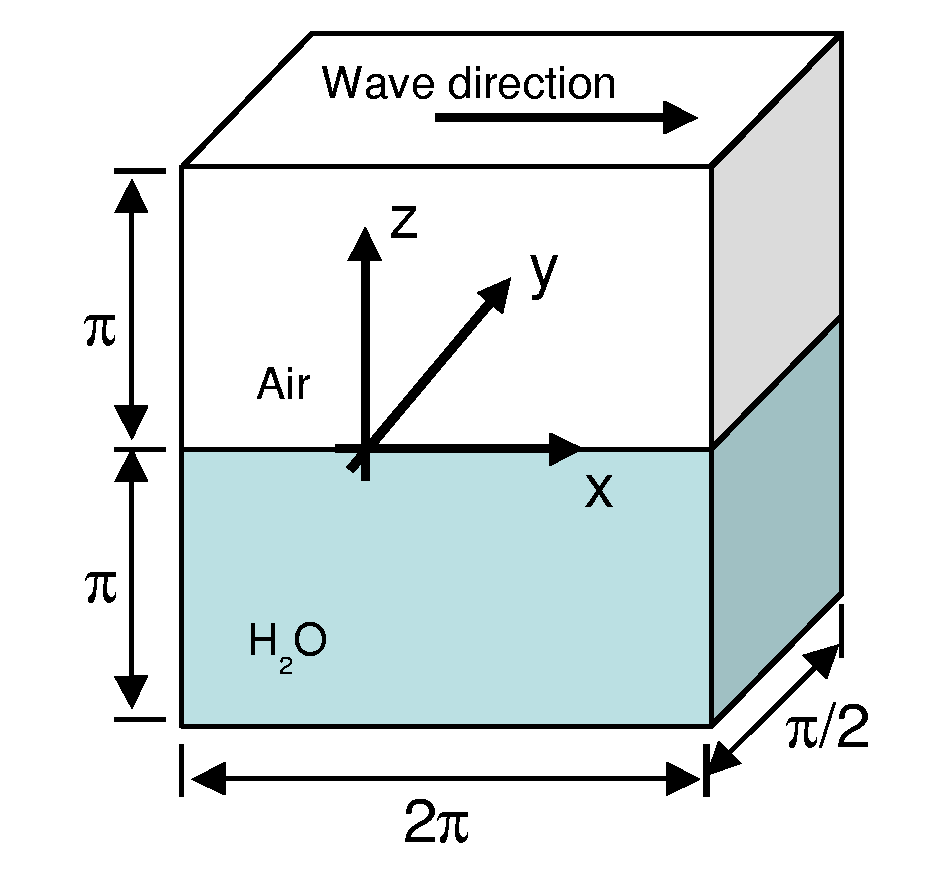}
\caption{\label{fig:setup} Schematic of computational domain.}
\end{figure}

Here, we advance the numerical study of breaking
waves by introducing a numerical method to simulate
three-dimensional, non-linear deep-water breaking
waves, and a consistent statistical framework in which
to analyze the results. The following framework is introduced
for studying a canonical deep-water breaking
wave. Consider a single wave, with wavenumber $k$,
which is propagating as part of a train of waves in deepwater.
The computational domain, shown in Figure \ref{fig:setup},
moves with the linear crest speed of the wave, and is
periodic in the stream-wise, $x$, direction. The temporally
evolving frame-work is conceptually similar to the
frame-work employed to simulate fully-developed turbulent
channel flow by \cite{moin82}, plain mixing
layers by \cite{rogers94}, and plane wakes by
\cite{moser98}.

The organization of the paper is as follows: First, the
theoretical frame work for studying a canonical breaking
wave is developed. Second, a method to generate
non-linear progressive waves and its validation are discussed,
along with the other computational details specific
to simulating breaking waves. Third, the results of a
parametric study of deep-water plunging breaking events
are discussed. This discussion is split into three sections,
(1) Three-dimensional flow visualization, (2) Integrated
energetics and the quantification of air entrainment, (3)
Statistical decomposition of the energy into kinetic, turbulent
kinetic and potential energy modes. Prior to the
conclusions, the results of a feasibility study on simulations
with 5-10 billion unknowns are discussed.

\section{Numerical Method} 

Four types of wave breaking are studied. For the sake of discussion herein they are defined as weak plunging (WP), plunging (P), strong plunging (SP), and very strong plunging (VSP).

Physical quantities are normalized by characteristic velocity ($U_o$), length ($L_o$), time ($L_o/U_o$), water density ($\rho_w$), and pressure ($\rho_w U_o^2$) scales. Consider the turbulent flow at the interface between air and water with $\rho_a$ and $\rho_w$ respectively denoting the densities of air and water

Let $\alpha$ denote the volume fraction of fluid that is inside a cell. By definition, $\alpha = 0$ for a cell that is totally filled with air, and $\alpha = 1$ for a cell that is totally filled with water. In terms of $\alpha$, the normalized density is express as
\begin{equation} 
\rho(\alpha)  =  \gamma + (1 - \gamma) \alpha \; ,
\label{eq:alpha} 
\end{equation}
where $\gamma = \rho_a/\rho_w$ is the density ratio between air and water.  
Let $u_i$ denote the normalized three-dimensional velocity field as a function
of normalized space ($x_i$) and normalized time ($t$).  The conservation of mass
is
\begin{equation} 
\frac{\partial \rho}{\partial t} +\frac{\partial u_j
\rho}{\partial x_j} = 0 \;\; . 
\label{eq:mass} 
\end{equation}
For incompressible flow,
\begin{equation} 
\frac{\partial \rho}{\partial t} +u_j \frac{\partial
\rho}{\partial x_j} = 0 \;\; .  
\label{eq:density} 
\end{equation}
Subtracting Equation \eqref{eq:density} from \eqref{eq:mass} gives a solenoidal
condition for the velocity:
\begin{equation} 
\frac{\partial u_i}{\partial x_i} = 0 \;\; .
\label{eq:solenoidal} 
\end{equation}
Substituting Equation \eqref{eq:alpha} into \eqref{eq:mass} and making use of
\eqref{eq:solenoidal}, provides an advection equation for  the volume fraction:
\begin{equation} 
\frac{\partial \alpha}{\partial t}+ \frac{\partial}{\partial
x_j} \left(u_j \alpha \right)= 0 \;\; .  
\label{eq:vof} 
\end{equation}
For an infinite Reynolds number, viscous stresses are negligible, and the
conservation of momentum is
\begin{equation} 
\frac{\partial u_i}{\partial t}+\frac{\partial}{\partial x_j}
\left(u_j u_i \right)  =  -\frac{1}{\rho} \frac{\partial p}{\partial x_i}
-\frac{p_s}{\rho} \frac{\partial H(\alpha)}{\partial x_i} -
\frac{\delta_{i3}}{F_r^2}  \;\; , 
\label{eq:momentum} 
\end{equation}
\noindent where $F_r^2 = U_o^2/(g L_o)$ is the Froude number, and $g$ is the
acceleration of gravity.  $p$ is the pressure and $p_s$ is a stress that acts
normal to the interface.  $H(\alpha)$ is a Heaviside function, and $\delta_{ij}$
is the Kronecker delta function.   

The divergence of the momentum equations \eqref{eq:momentum} in combination with
the solenoidal condition \eqref{eq:solenoidal} provides a Poisson equation for
the dynamic pressure:
\begin{equation} 
\frac{\partial}{\partial x_i} \frac{1}{\rho} \frac{\partial
p}{\partial x_i} = \Sigma \;\; , 
\label{eq:poisson} 
\end{equation}
\noindent where $\Sigma$ is a source term.  The pressure is used to project the
velocity onto a solenoidal field.  Details of the volume fraction advection, the
pressure projection, and the numerical time integration are provided in
\citet{dommermuth07} and \citet{dommermuth08}. Sub-grid scale stresses
are modeled using an implicit model that is built into the treatment of
convective terms.  The performance of the implicit SGS model is provided in
\citet{nfa3}. Specific issues relevant to the current
study include free-surface smoothing and interfacial
forcing, and are discussed in the next sections.

\subsection{Smoothing}
\label{sec:smth}
The free-surface boundary layer is not resolved in volume of fluid (VOF)
simulations at high Reynolds numbers with large density jumps such as air and
water.  Under these circumstances, the tangential velocity is discontinuous
across the free-surface interface and the normal component is continuous.   As a
result, unphysical tearing of the free surface tends to occur.   Favre-like
filtering can be used to alleviate this problem by forcing the air velocity
slightly above the interface to be driven by the water velocity slightly below
the interface in a physical manner.    Consider the following filter,
\begin{equation}
\label{eq:smooth_velo} 
u^*_i=\frac{\overline{ \rho u_i
}}{\overline{\rho}} \;\; {\rm for} \; \alpha \ge 0.5 \;\; , 
\end{equation}
where $u^*_i$ is the smoothed velocity field, $u_i$ is the unfiltered velocity
field, $\rho$ is the density, and $\alpha$ is the volume fraction.  Overline
symbols  denote averaging over a small control volume.
\begin{equation} 
\overline{F(x)} = \int_{\rm v} W(\xi) F(x-\xi) {\rm dv}  \;\; .
\label{eq:smooth} 
\end{equation}
Here, $F(x)$ is a general function, ${\rm v}$ is a control volume that surrounds
a cell, and $W(x)$ is a weighting function that neither overshoots or
undershoots the maximum or minimum allowable density.  Due to the  high density
ratio between water and air, Equation \eqref{eq:smooth_velo} tends to push the
water-particle velocity into the air.   Once the velocity is filtered, we need
to project it back onto a solenoidal field in the fluid volume (V):
\begin{equation} 
\label{eq:project1} 
u_i = u^*_i - \frac{1}{\rho} \frac{\partial
\phi}{\partial x_i}  \;\; {\rm in \; V} \; , 
\end{equation}
where $\phi$ is a potential function.   For an incompressible flow, we require
that $u_i$ is solenoidal care of Equation \eqref{eq:solenoidal}.  Substituting
\eqref{eq:project1} into \eqref{eq:solenoidal} gives a Poisson equation for
$\phi$:
\begin{equation} 
\frac{\partial}{\partial x_i} \frac{1}{\rho} \frac{\partial
\phi}{\partial x_i} = \frac{\partial u^*_i}{\partial x_i}   \;\; {\rm in \; V}
\; .  
\label{eq:project2} 
\end{equation}
Details of the implementation and performance of the preceding filter
are provided in \citet{nfa6}.

\subsection{Atmospheric Forcing}
\label{sec:atm_frc}
The surface stress ($p_s$) in Equation \eqref{eq:momentum} can be used to apply a
pressure to the interface to generate a known disturbance.  The formulation in
terms of the gradient of Heaviside function ensures that the stress is applied
only at the the free surface.   The stress is applied for a finite amount of
time with an amplitude that is slowly ramped up and down to minimize transients.
The surface stress can be used to generate a linear superposition of waves in
the following manner: 
\begin{multline} 
p_s=G(t) \left\{ A_r+ \sum_{n=1}^N A_n \cos \biggl[ k_n ( x+U_c
t)- \biggr. \right.\\ \omega_n (t-T_u) \biggl. \biggr] \Biggl.\Biggr\}  \, ,
\label{eq:pa1}
\end{multline}
where $A_n$, $k_n$, and $\omega_n$ are respectively the Fourier amplitude,
wavenumber, and frequency.  Typically, the wavenumber and wave frequency satisfy
a linear dispersion relationship, $\omega_n^2= k_n/F^2_r$.   $U_c$ is the
current velocity.   In a frame of reference that is fixed with the crest of the
wave, $U_c$ equals the phase speed ($\omega_n/k_n$).  $T_u$ is an unwinding time
that can be used to generate steep events at $t=T_u$.   $G(t)$ ramps up and down
the stress for $0 \leq t \leq T_f$:
\begin{equation}
G(t) =\frac{1}{2}\left[1-cos\left(\frac{2 \pi
t}{T_f}\right)\right]  \; .
\label{eq:pa2}
\end{equation}
$G(t)=0$ for $t > T_f$.  A three-dimensional uniform
random disturbance is passed through a low-pass filter
and added to the flow. The r.m.s. amplitude of the $A_r$
integrated over the fluid volume is $a_o$:
\begin{equation}
\frac{1}{V}\int_V dV A^2_r \left(x_i \right) =a^2_0  \; .
\label{eq:pa3} 
\end{equation}

\subsection{Simulation Parameters}

For all the simulations in this study $L_o = 1/k = 1$,
$U_o = (g/k)^{1/2} = 1$, and $Fr = 1$. The length of the
domain is $2 \pi$. The depth of the water and the height of
the air are set equal to $\pi$, and the width of the domain is $\pi/2$, as illustrated in Figure 1. Coarse, medium, and fine
NFA simulations are performed with respectively $256 \times 64 \times 256$, $512 \times 128 \times 256$, and $1024 \times 256 \times 512$ grid
points.

For the finest resolution cases, the time step $\Delta t =
0.002$ for $t \leq 9$ when the wave is developing, and $\Delta t = 0.001$ for $t > 9$ when the wave is breaking.
Density-weighted velocity smoothing is applied every 20
time steps using simple averaging over $3\times3\times3$ grid cells.
The finest resolution simulations are run up to $t = 36.5$,
corresponding to 32,000 time steps.

The pressure forcing that is used in these simulations
consists of four values of $A_o/(2 \pi)=
0.017, 0.018, 0.019, 0.020$ corresponding to initial overturning
events characterized by weak plunging (WP),
plunging (P), strong plunging (SP) and very strong
plunging (VSP). Each breaking-wave amplitude is simulated
using coarse (C), medium (M), and fine (F) resolution.
Cases discussed in the results section will be referred
to by the type of breaking and grid resolution. For
example, $VSP^F$ would refer to a very strong plunging
simulation at the fine grid resolution. The other parameters
in the atmospheric forcing are $U_c = -1$, $k_o = 1$,
$\omega_c = 1$, $T_u = 0$, $T_f = 4\pi$, and $a_o = 0.1 A_0$.

\subsection{Validation}
\label{sec:val}

\begin{figure}
(a) \includegraphics[width=\linewidth]{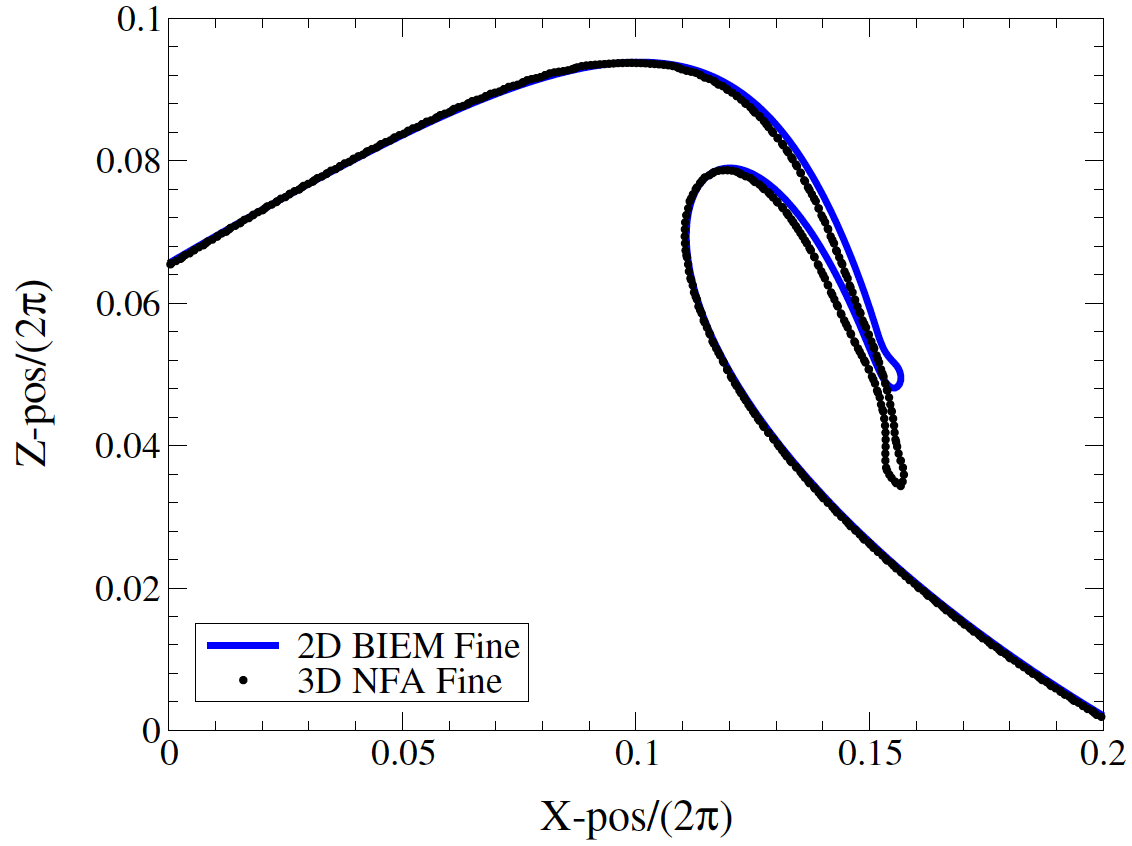} \\
(b) \includegraphics[width=\linewidth]{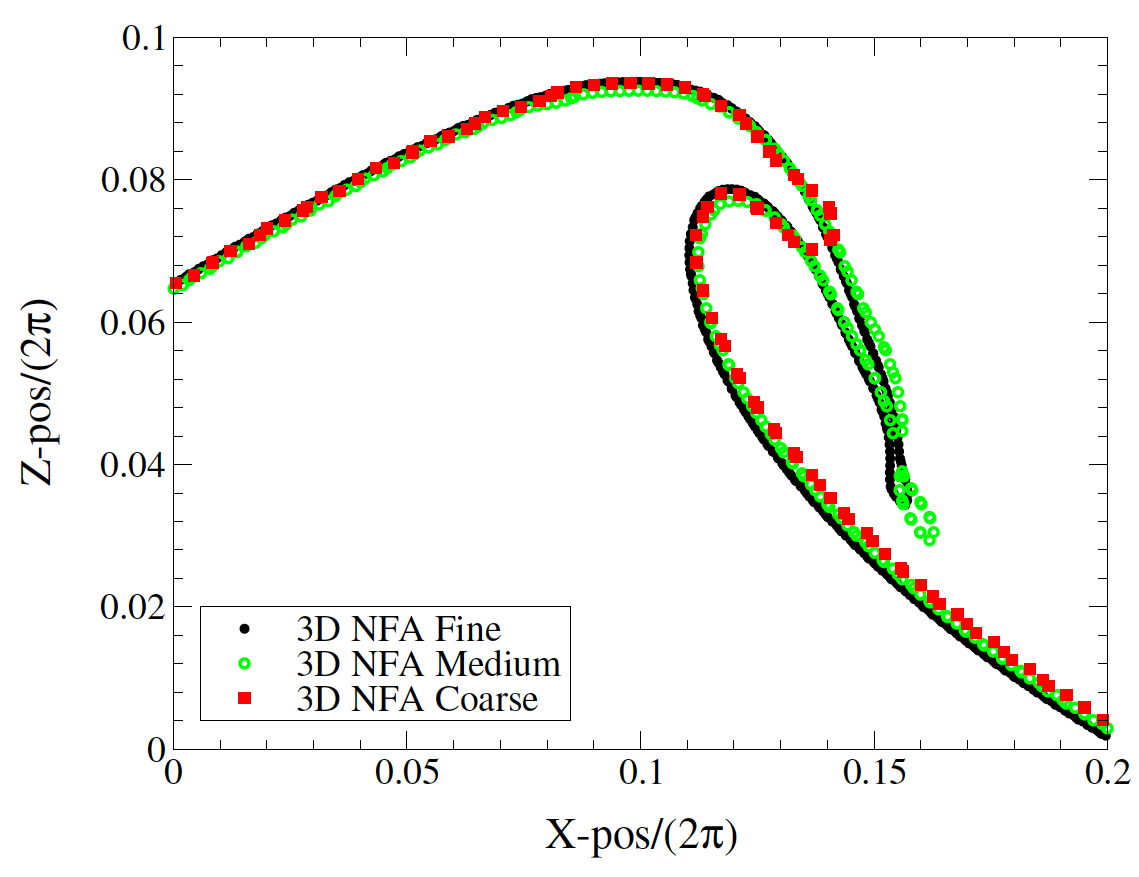} \\
(c) \includegraphics[width=\linewidth]{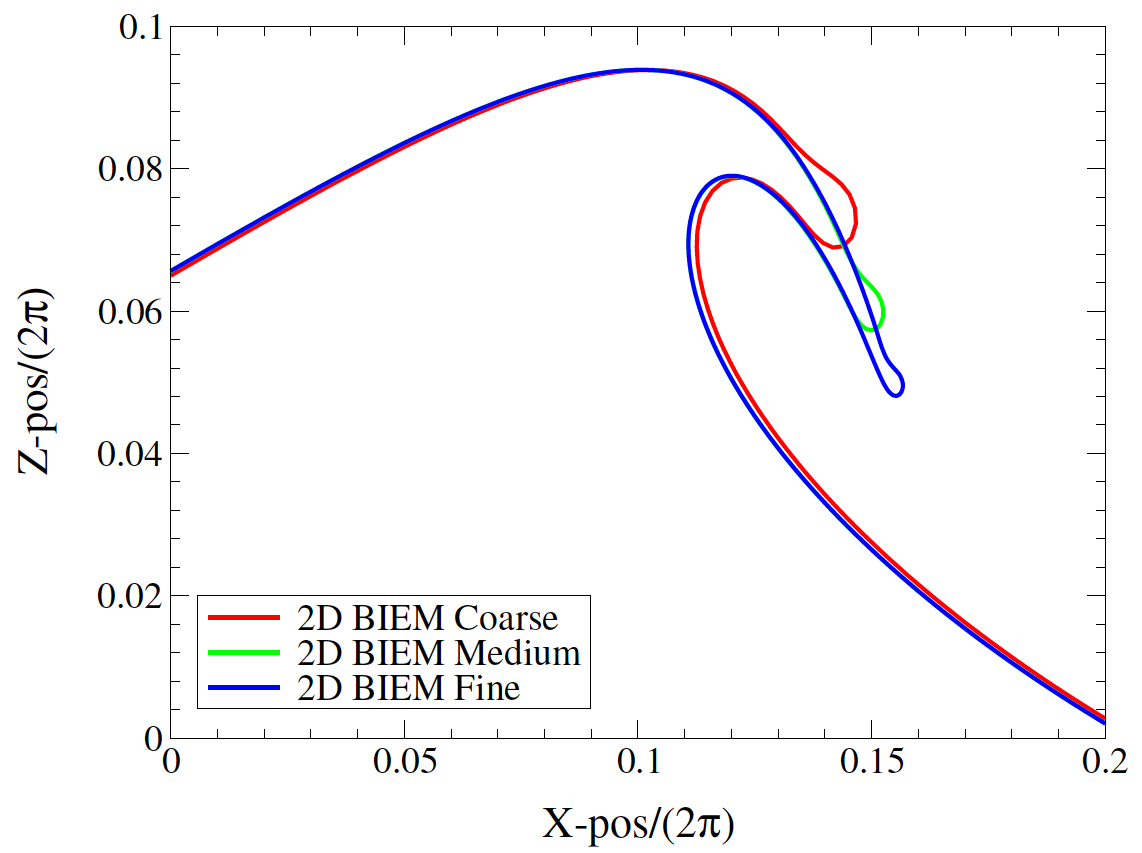}
\caption{\label{fig:biem_comp} (a) Comparison of fine resolution 3-D NFA simulations
to the fine resolution 2-D BIEM calculations just before
reentry. (b) Comparison of three-dimensional NFA simulations
at different grid resolutions just before reentry. (c) Comparison
of two-dimensional BIEM simulations at different grid resolutions
just before reentry.} 
\end{figure}

A comparison to a Boundary Integral Equation Method (BIEM) \cite{dommermuth88}
is presented to show that the aforementioned atmospheric pressure forcing
technique generates fully non-linear progressive
waves.   Cuts of the free-surface at $y = 0$ from cases
$VSP^C$, $VSP^M$, and $VSP^F$, which are the coarse,
medium and fine resolution cases corresponding to a very strong plunging event ($A_o/2\pi=0.02$), are
compared. In addition coarse, medium and fine BIEM
simulations are performed with respectively 512,
1024, and 2048 nodes along the free surface. Figure~\ref{fig:biem_comp}(a) compares the finest resolution NFA and BIEM
simulations. The slight difference between NFA and
BIEM in the tip of the plunging breaking wave may
be due to the effects of air in the NFA simulations.
Figure~\ref{fig:biem_comp}(b) and Figure~\ref{fig:biem_comp}(c) show grid studies of three-dimensional
NFA simulations and two-dimensional
BIEM simulations, respectively. The three-dimensional
NFA simulations, which have less grid resolution than
the two-dimensional BIEM simulations, resolve the
plunging jet better than the BIEM simulations. Table~\ref{table:mass}
shows that mass is conserved over the duration of the
simulation. The method proposed here including the
smoothing and projection operation is mass conservative.

Having shown agreement with BIEM methods up to
reentry, attention is now focused on the flow after reentry.

\begin{table}
\begin{tabular}{c|cc}
\hline
\hline 
Case                 &Resolution & $\frac{|\int dV  (\phi-\phi_0) |}{\int dV \phi_0}$   \\ 
\hline 
Very Strong Plunging & Coarse & $2.10E^{-4}$    \\ 
$\cdot$              & Medium & $3.14E^{-4}$    \\
$\cdot$              & Fine   & $3.31E^{-4}$    \\
Strong Plunging      & Coarse & $1.59E^{-4}$    \\
$\cdot$              & Medium & $2.61E^{-4}$    \\
$\cdot$              & Fine   & $2.22E^{-4}$    \\ 
Plunging             & Coarse & $1.37E^{-4}$    \\ 
$\cdot$              & Medium & $1.91E^{-4}$    \\
$\cdot$              & Fine   & $5.16E^{-5}$    \\
Weak Plunging        & Coarse & $1.32E^{-4}$    \\ 
$\cdot$              & Medium & $1.08E^{-4}$    \\
$\cdot$              & Fine   & $6.35E^{-4}$    \\  
\hline 
\hline 
\end{tabular}
\caption{\label{table:mass} Mass conservation over total duration of simulation.}
\end{table}

\section{Flow Visualizations}

\begin{figure*}
\centering
{\bf A \hspace{102pt} B  \hspace{100pt} C \hspace{102pt} D}\\
\vspace{5pt}
\pdfsavepos
\hyperlink{viz11}{\includegraphics[trim = 0mm .00cm 0mm .00cm,clip=true,angle=0,scale=0.06]{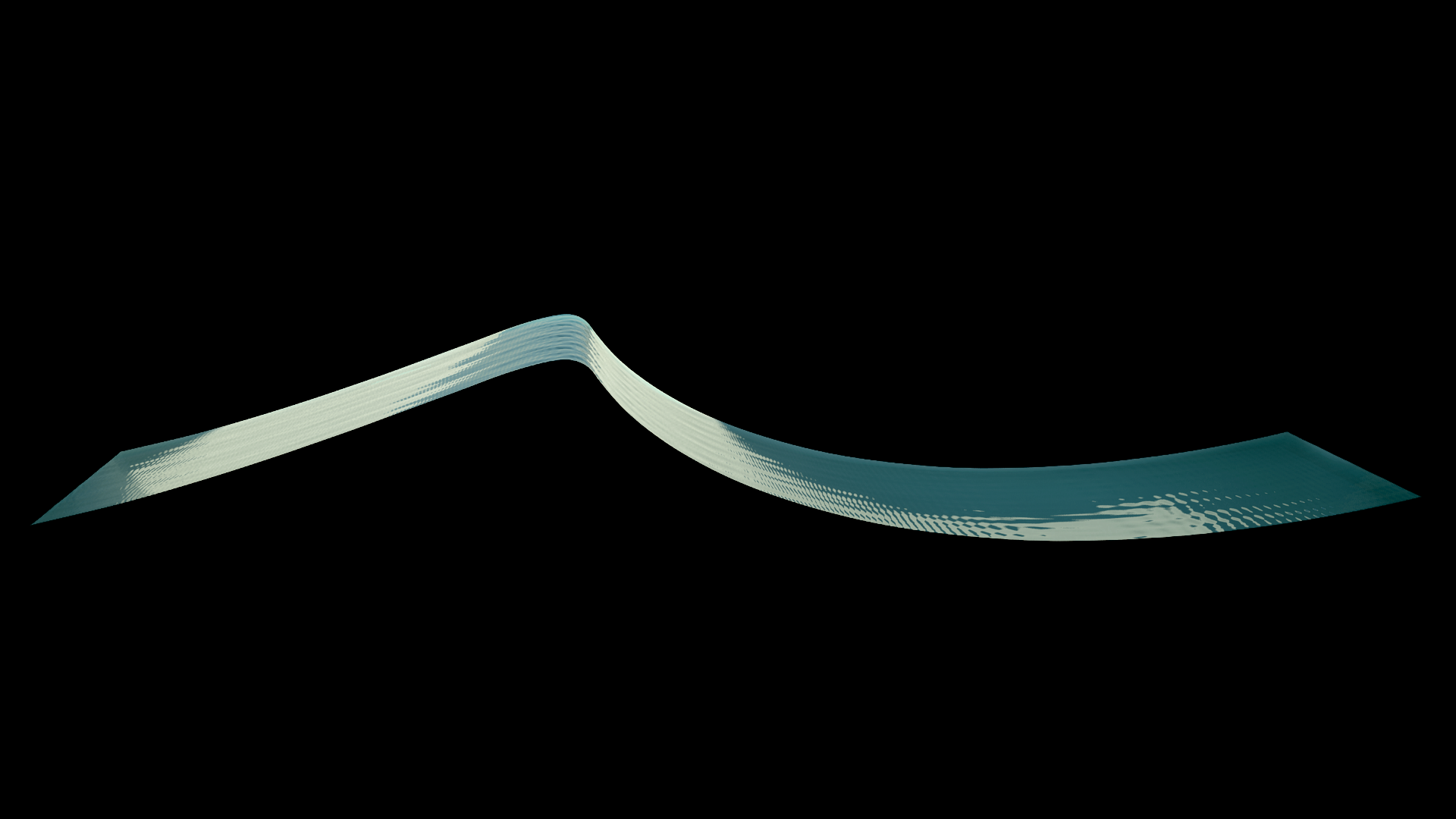}}
\pdfdest name{viz11} FitR width 2.3cm 
\hyperlink{viz12}{\includegraphics[trim = 0mm .00cm 0mm
 .00cm,clip=true,angle=0,scale=0.06]{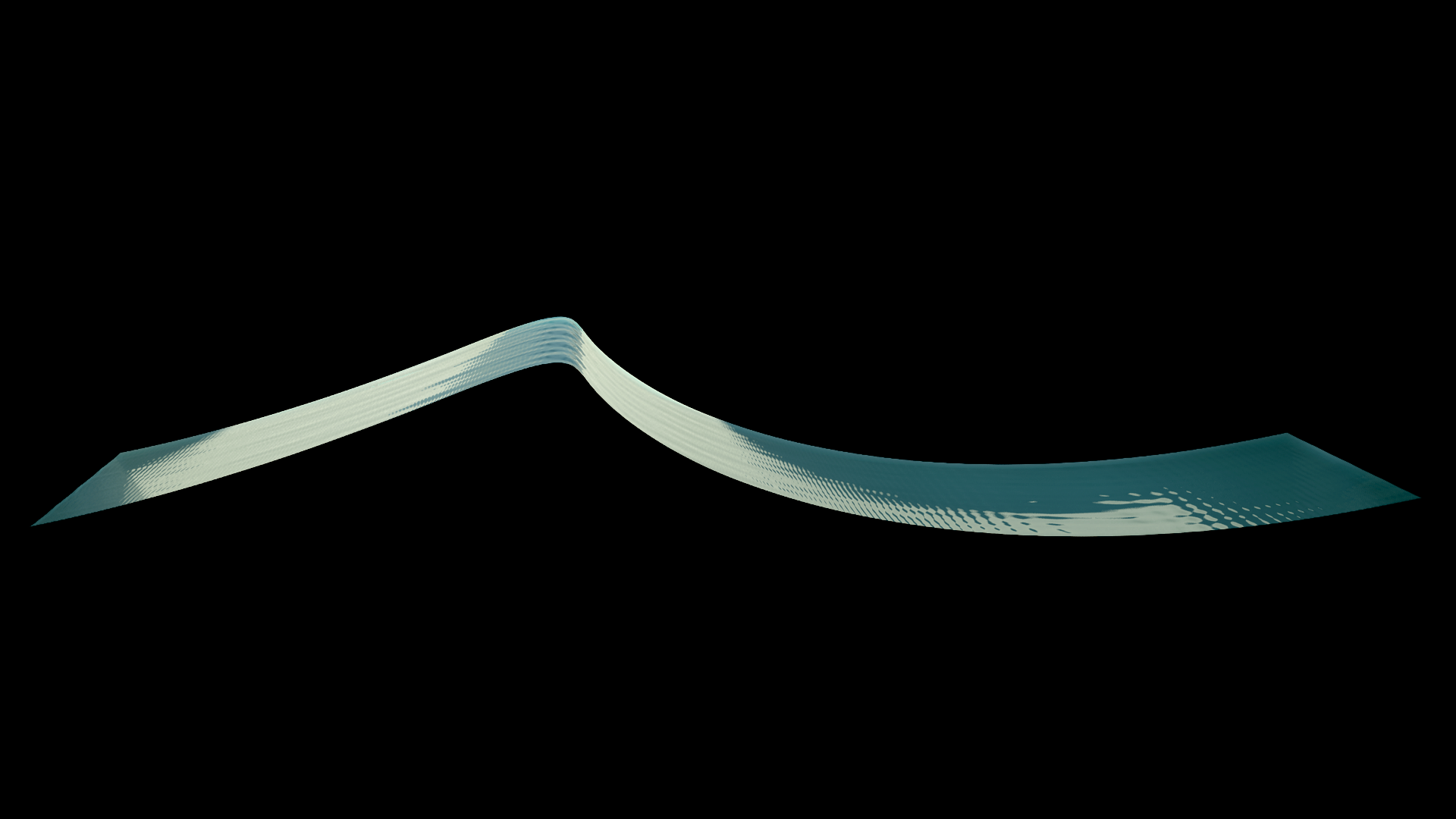}}
\pdfdest name{viz12} FitR width 2.3cm 
\hyperlink{viz13}{\includegraphics[trim = 0mm .00cm 0mm
 .00cm,clip=true,angle=0,scale=0.06]{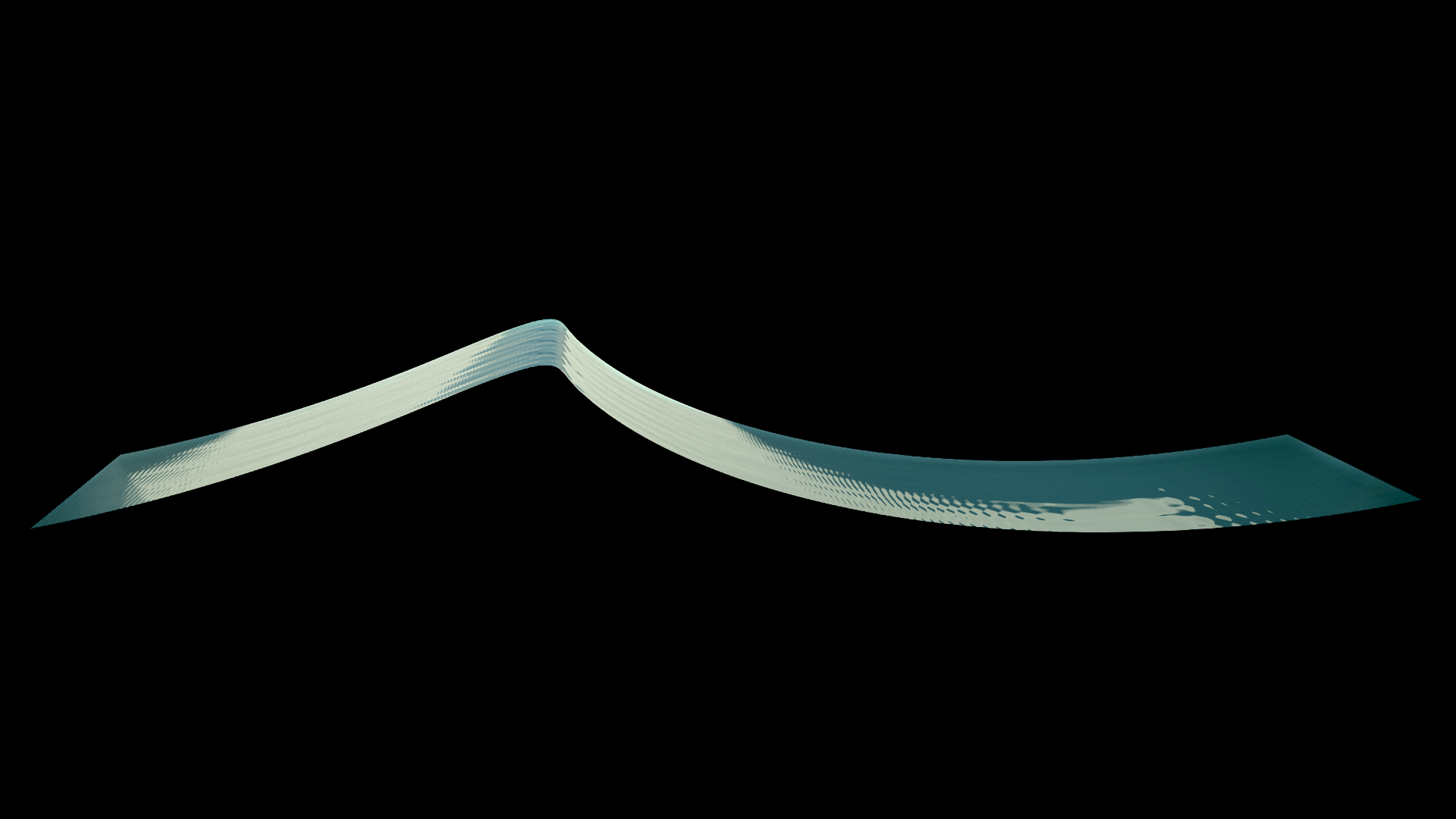}}
\pdfdest name{viz13} FitR width 2.3cm 
\hyperlink{viz14}{\includegraphics[trim = 0mm .00cm 0mm
 .00cm,clip=true,angle=0,scale=0.06]{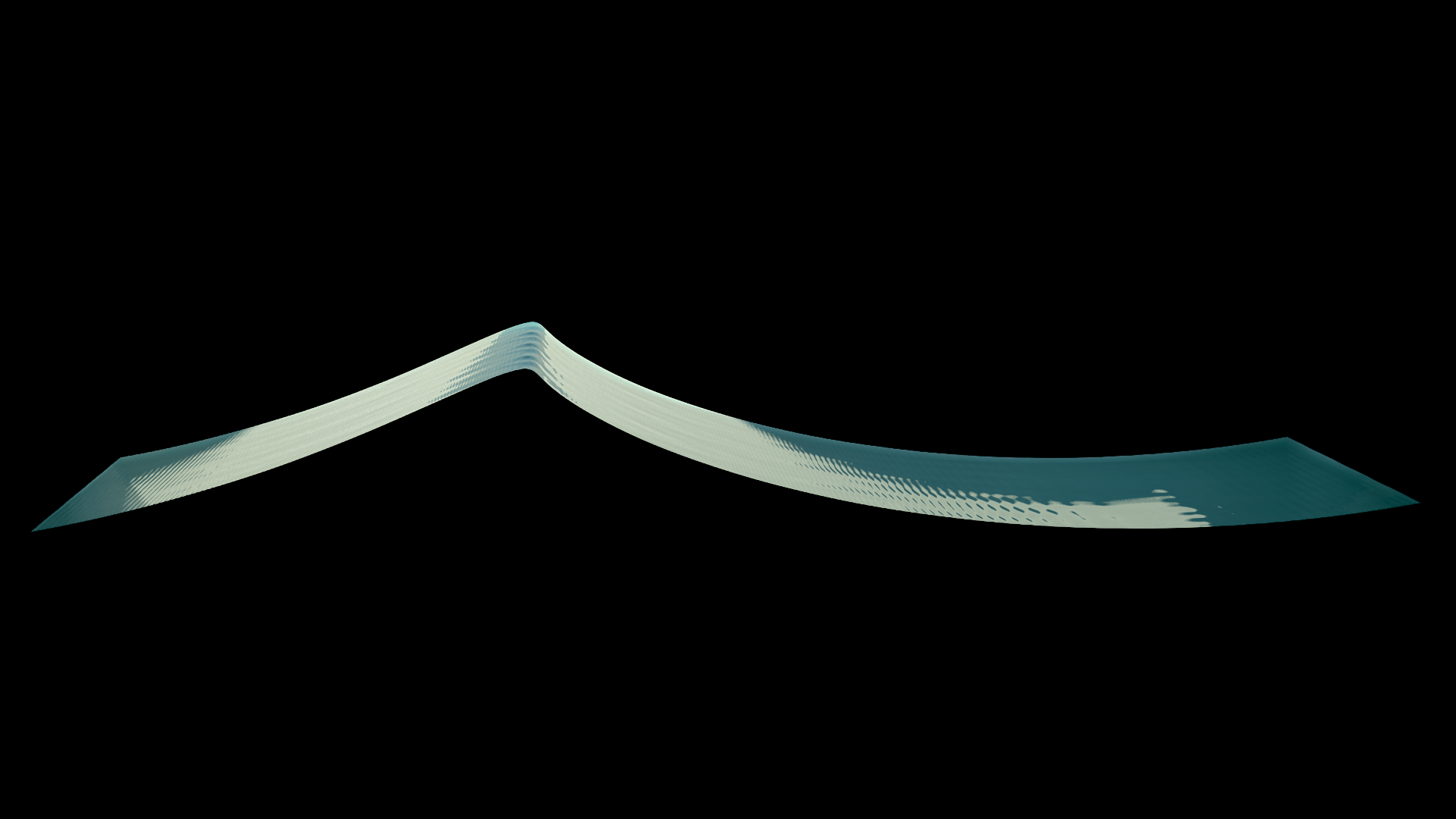}}\\
\pdfdest name{viz14} FitR width 2.3cm 
\vspace{1pt}
\hyperlink{viz21}{\includegraphics[trim = 0mm .00cm 0mm .00cm,clip=true,angle=0,scale=0.06]{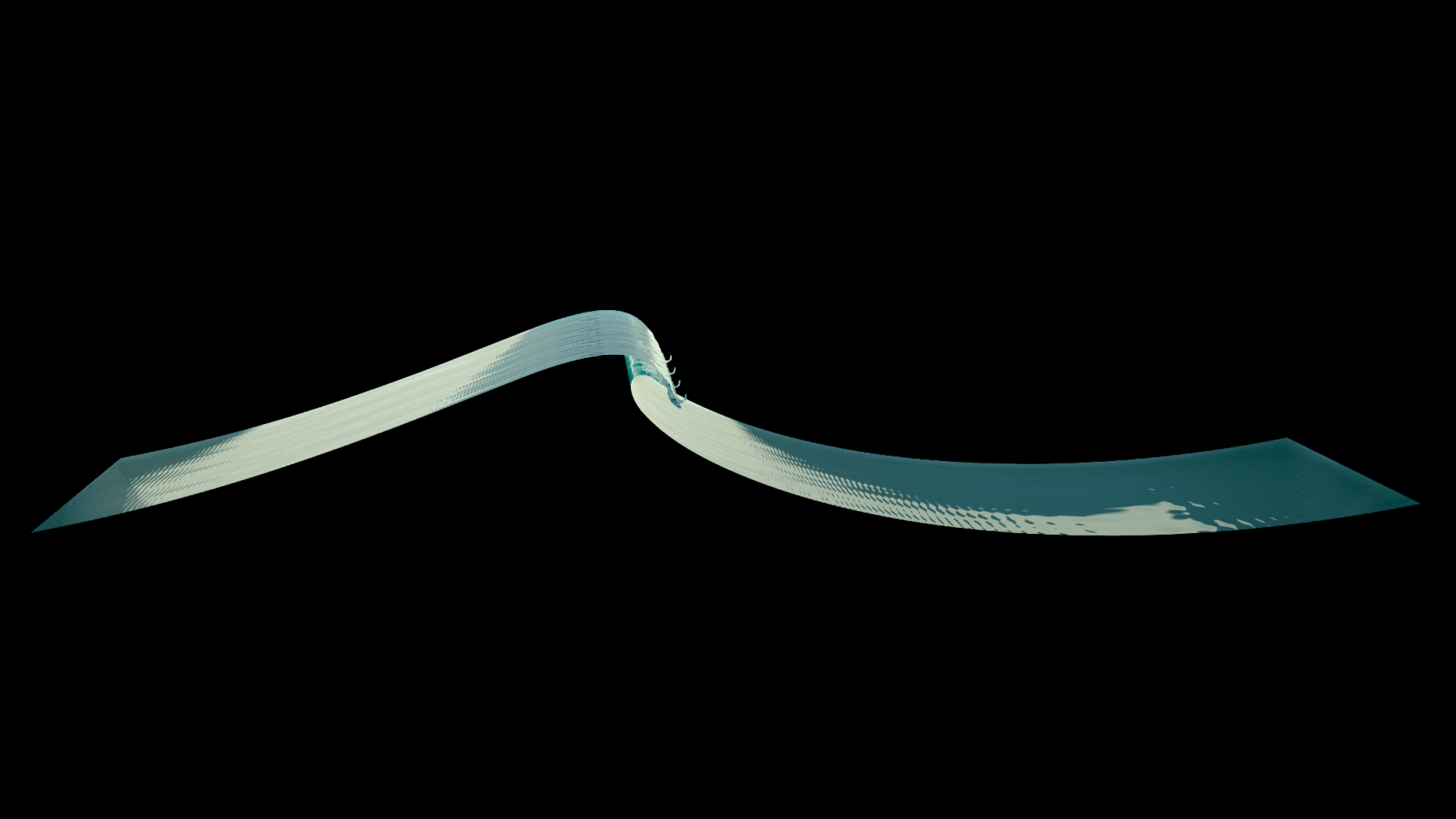}}
\pdfdest name{viz21} FitR height 2.3cm 
\hyperlink{viz22}{\includegraphics[trim = 0mm .00cm 0mm .00cm,clip=true,angle=0,scale=0.06]{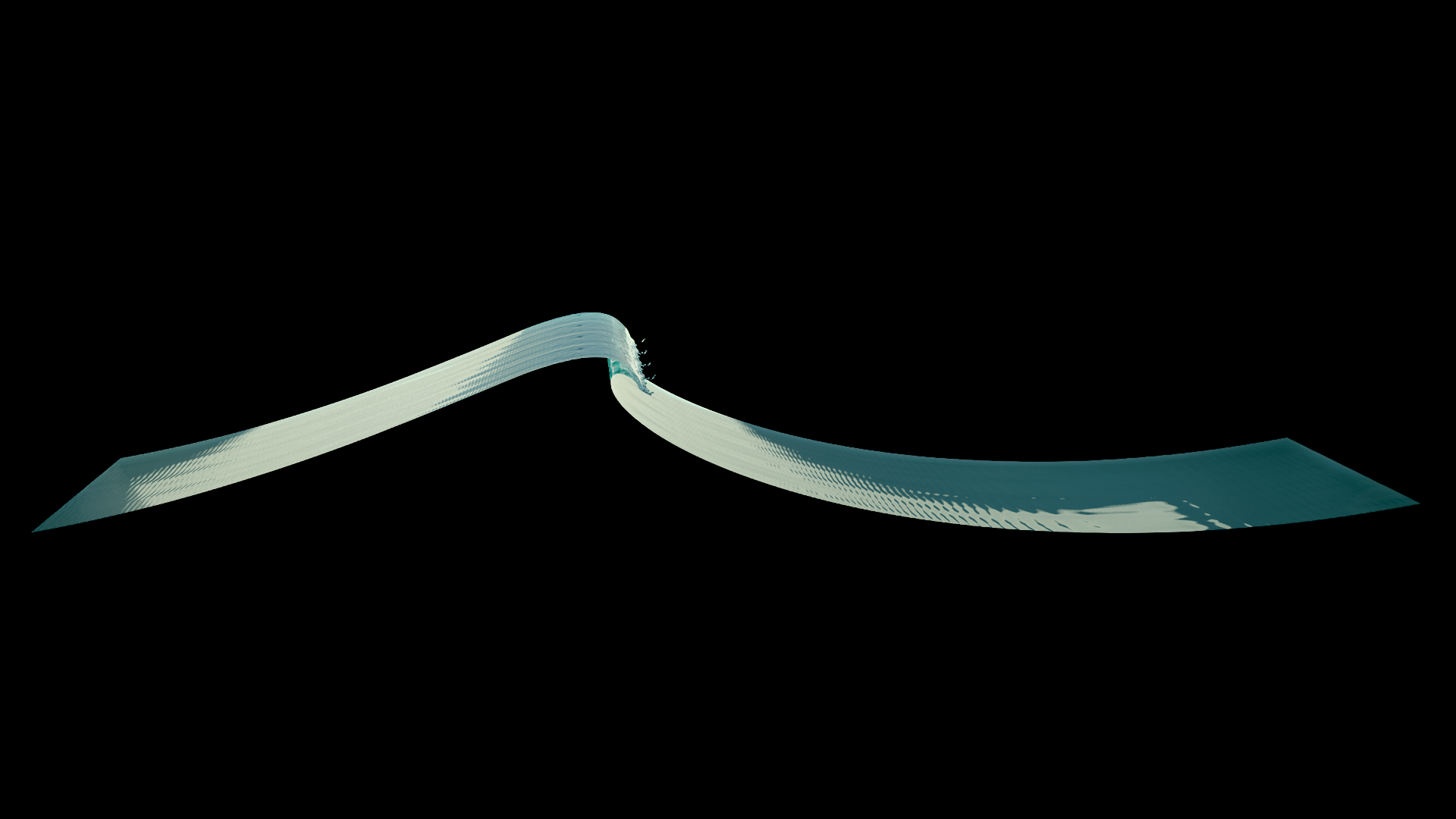}}
\pdfdest name{viz22} FitR width 2.3cm 
\hyperlink{viz23}{\includegraphics[trim = 0mm .00cm 0mm  .00cm,clip=true,angle=0,scale=0.06]{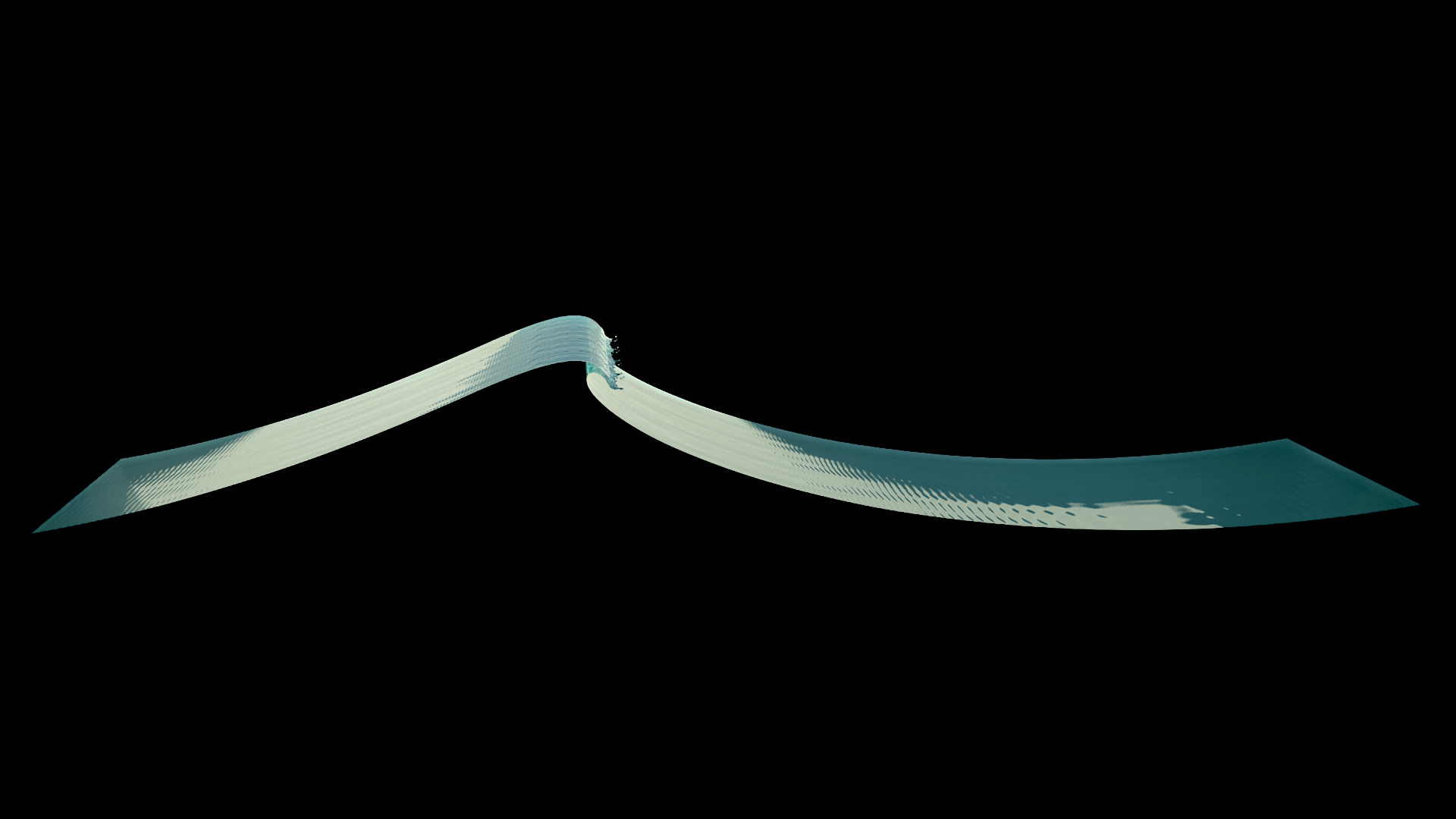}}
\pdfdest name{viz23} FitR width 2.3cm 
\hyperlink{viz24}{\includegraphics[trim = 0mm .00cm 0mm  .00cm,clip=true,angle=0,scale=0.06]{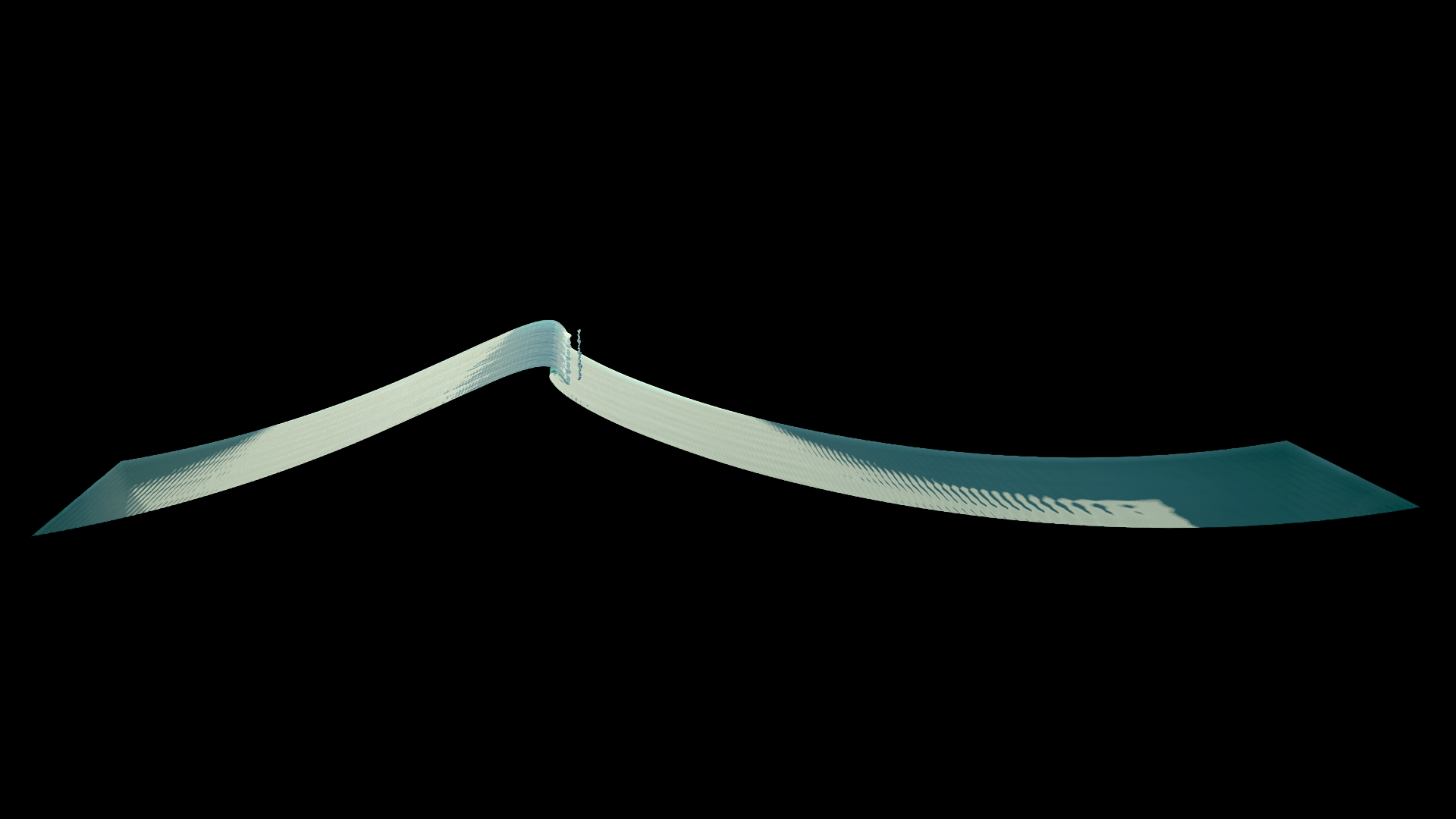}}\\
\pdfdest name{viz24} FitR width 2.3cm 
\vspace{1pt}
\hyperlink{viz31}{\includegraphics[trim = 0mm .00cm 0mm
.00cm,clip=true,angle=0,scale=0.06]{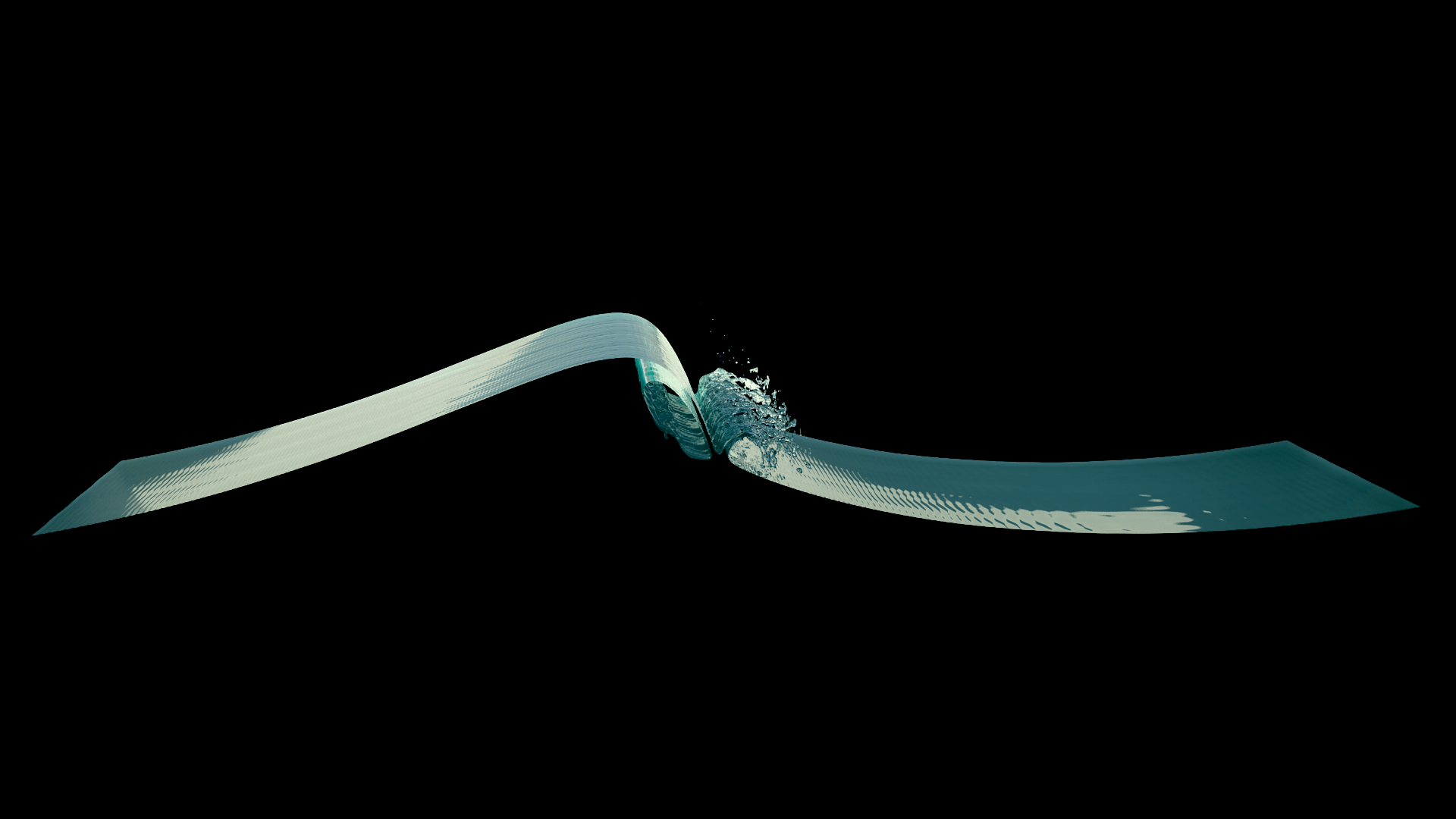}}
\pdfdest name{viz31} FitR height 2.3cm 
\hyperlink{viz32}{\includegraphics[trim = 0mm .00cm 0mm
.00cm,clip=true,angle=0,scale=0.06]{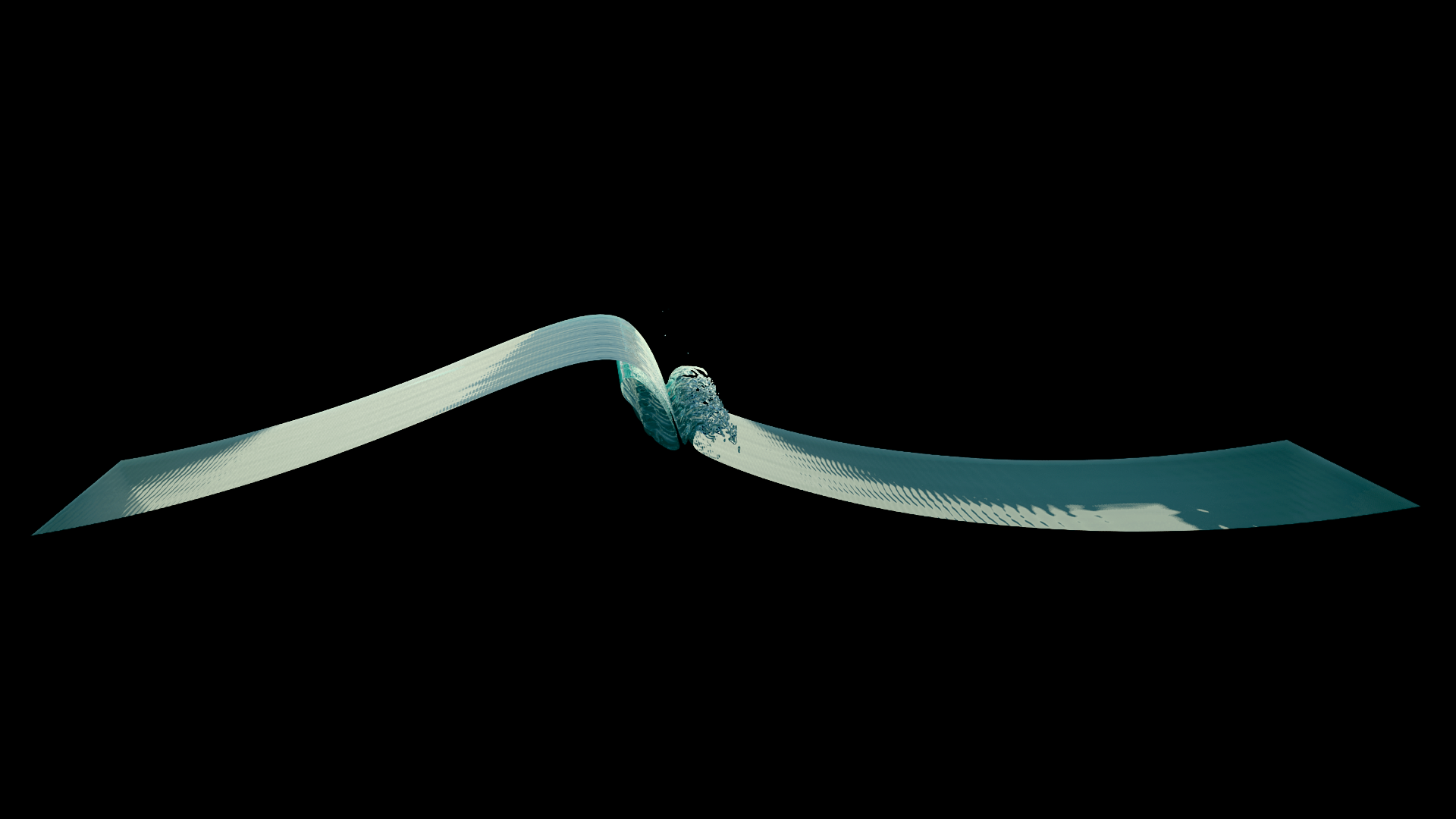}}
\pdfdest name{viz32} FitR height 2.3cm 
\hyperlink{viz33}{\includegraphics[trim = 0mm .00cm 0mm
.00cm,clip=true,angle=0,scale=0.06]{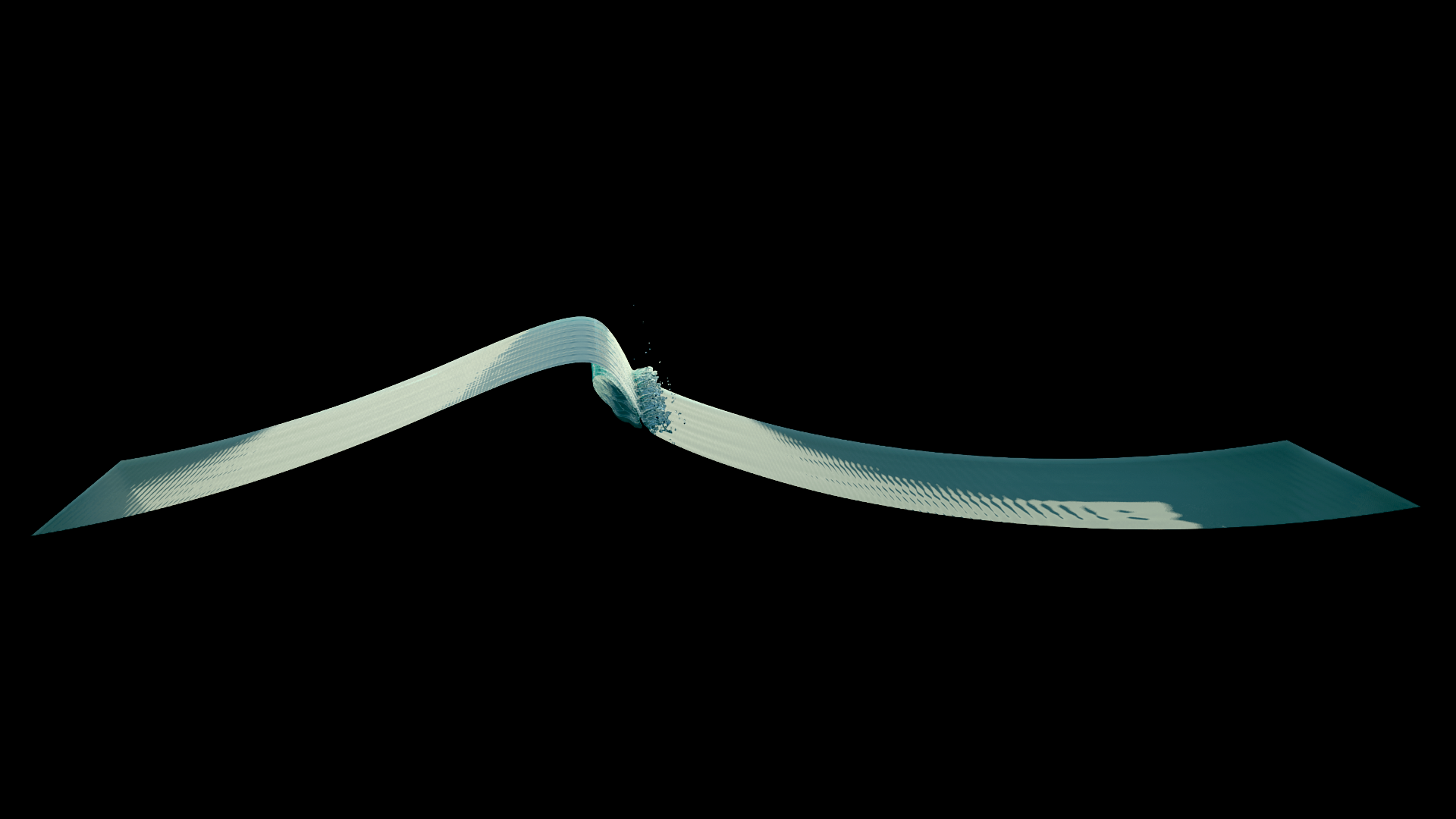}}
\pdfdest name{viz33} FitR height 2.3cm 
\hyperlink{viz34}{\includegraphics[trim = 0mm .00cm 0mm
.00cm,clip=true,angle=0,scale=0.06]{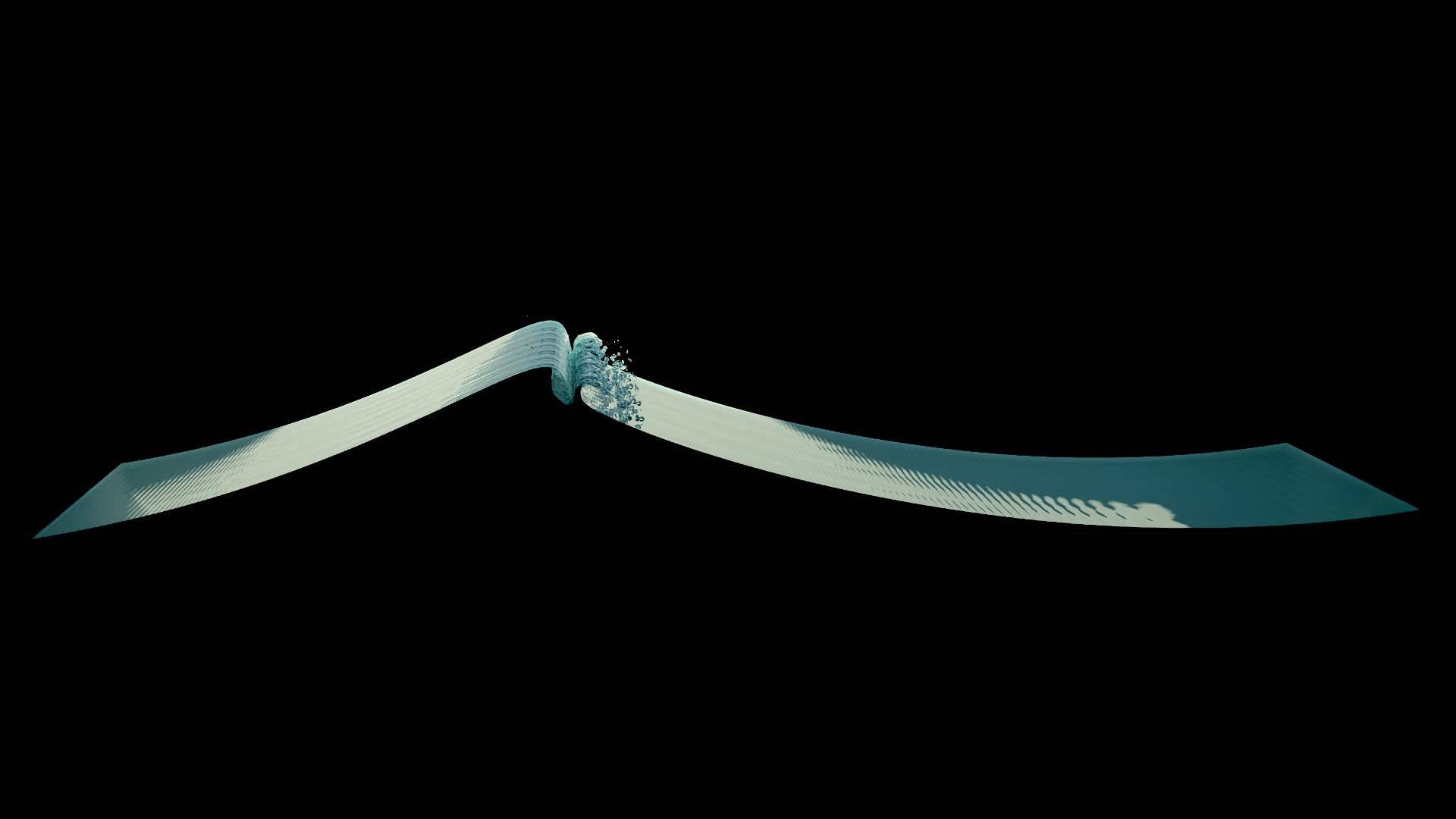}}\\
\pdfdest name{viz34} FitR height 2.3cm 
\vspace{1pt}
\hyperlink{viz41}{\includegraphics[trim = 0mm .00cm 0mm
.00cm,clip=true,angle=0,scale=0.06]{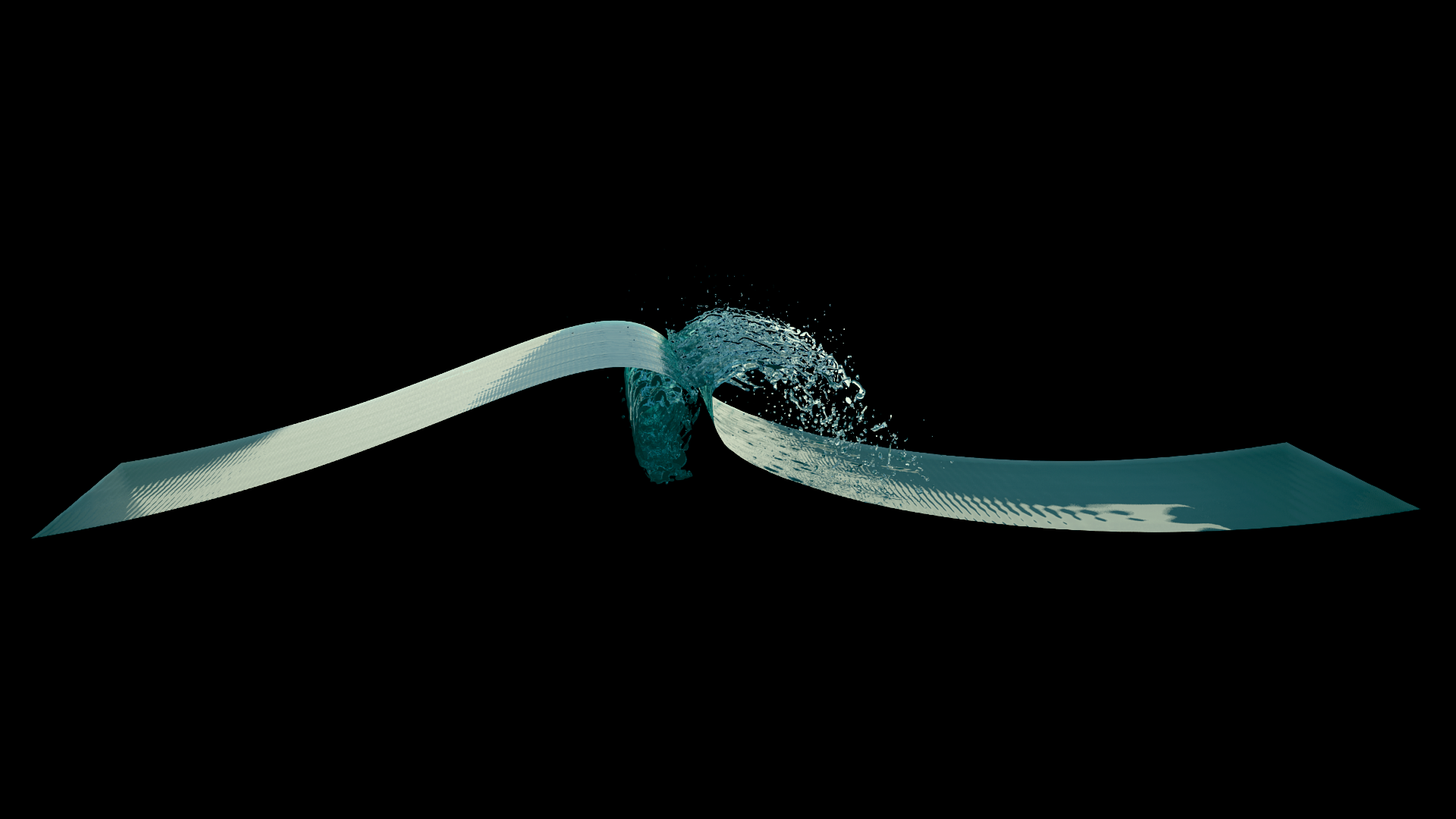}}
\pdfdest name{viz41} FitR height 2.3cm 
\hyperlink{viz42}{\includegraphics[trim = 0mm .00cm 0mm
.00cm,clip=true,angle=0,scale=0.06]{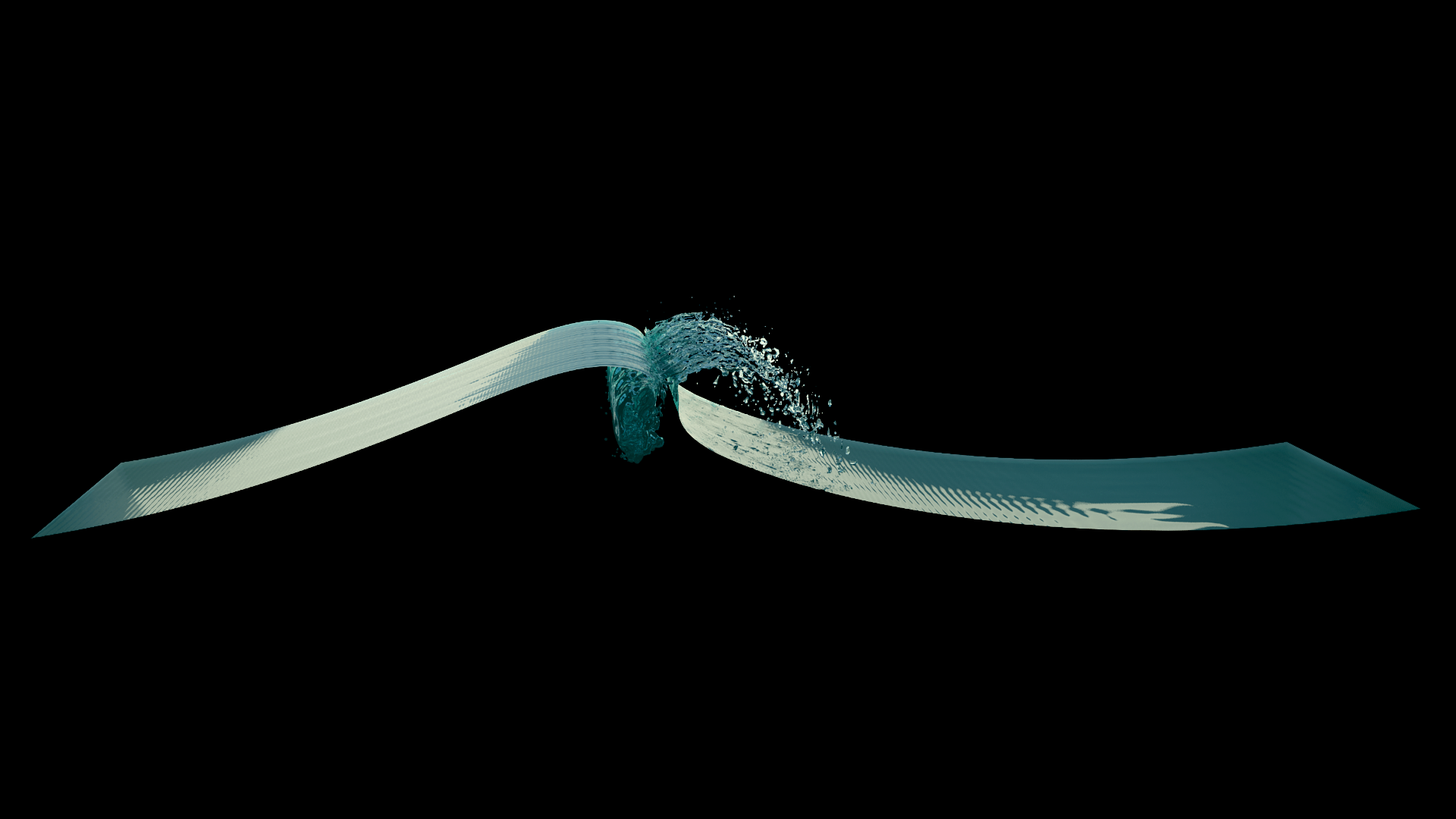}}
\pdfdest name{viz42} FitR height 2.3cm 
\hyperlink{viz43}{\includegraphics[trim = 0mm .00cm 0mm
.00cm,clip=true,angle=0,scale=0.06]{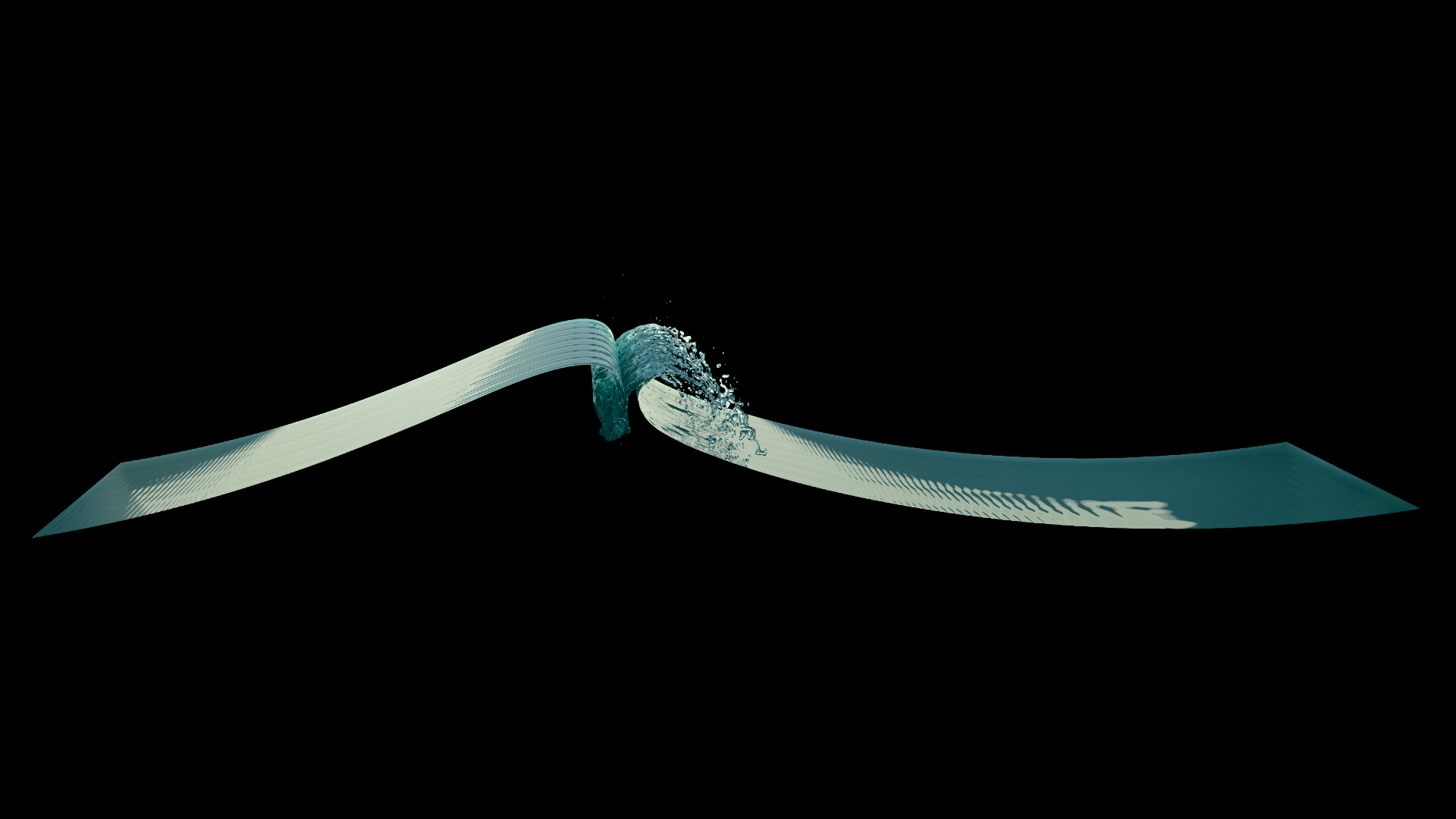}}
\pdfdest name{viz43} FitR height 2.3cm 
\hyperlink{viz44}{\includegraphics[trim = 0mm .00cm 0mm
.00cm,clip=true,angle=0,scale=0.06]{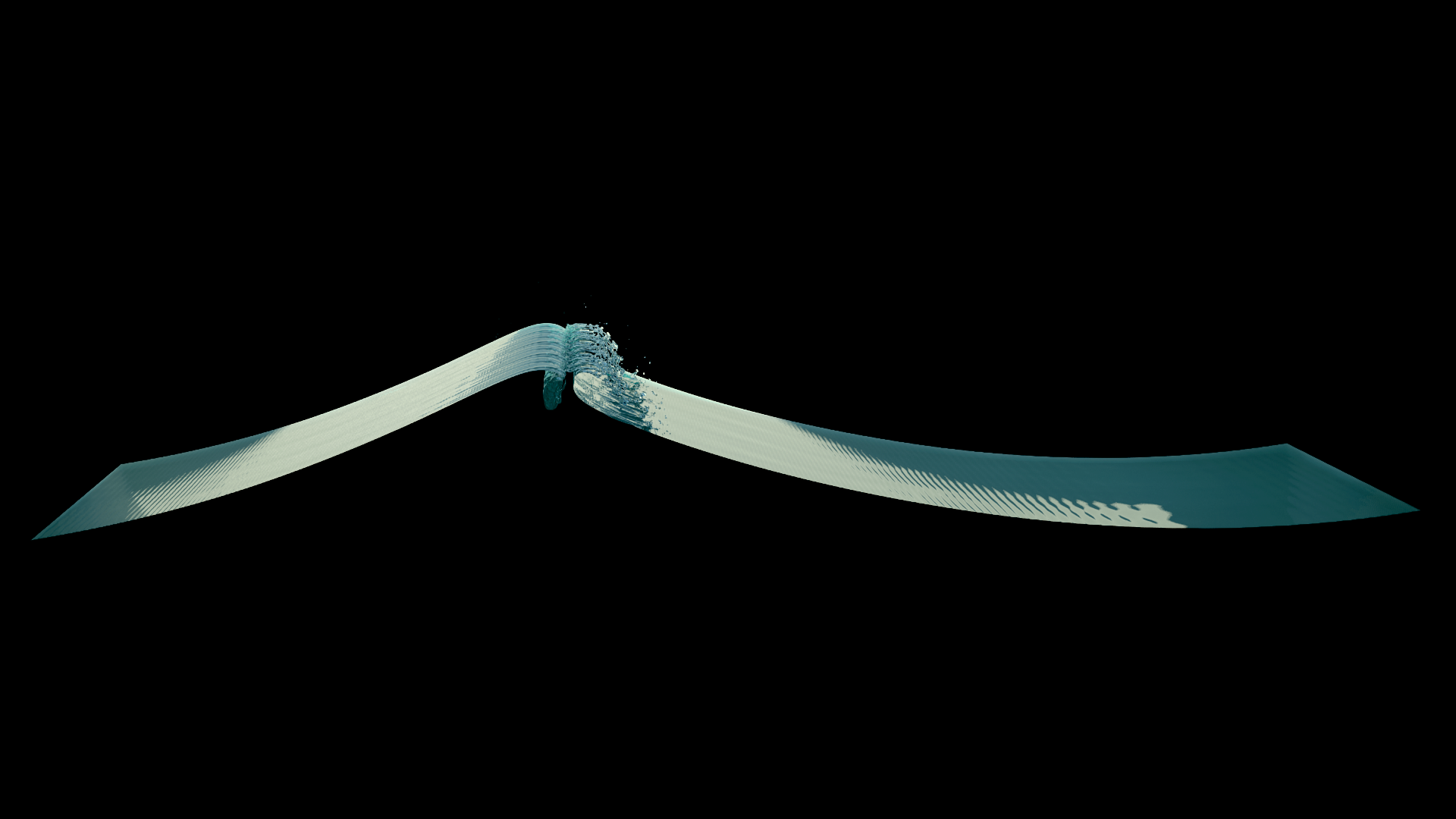}}\\
\pdfdest name{viz44} FitR height 2.3cm 
\vspace{1pt}
\hyperlink{viz51}{\includegraphics[trim = 0mm .00cm 0mm
.00cm,clip=true,angle=0,scale=0.06]{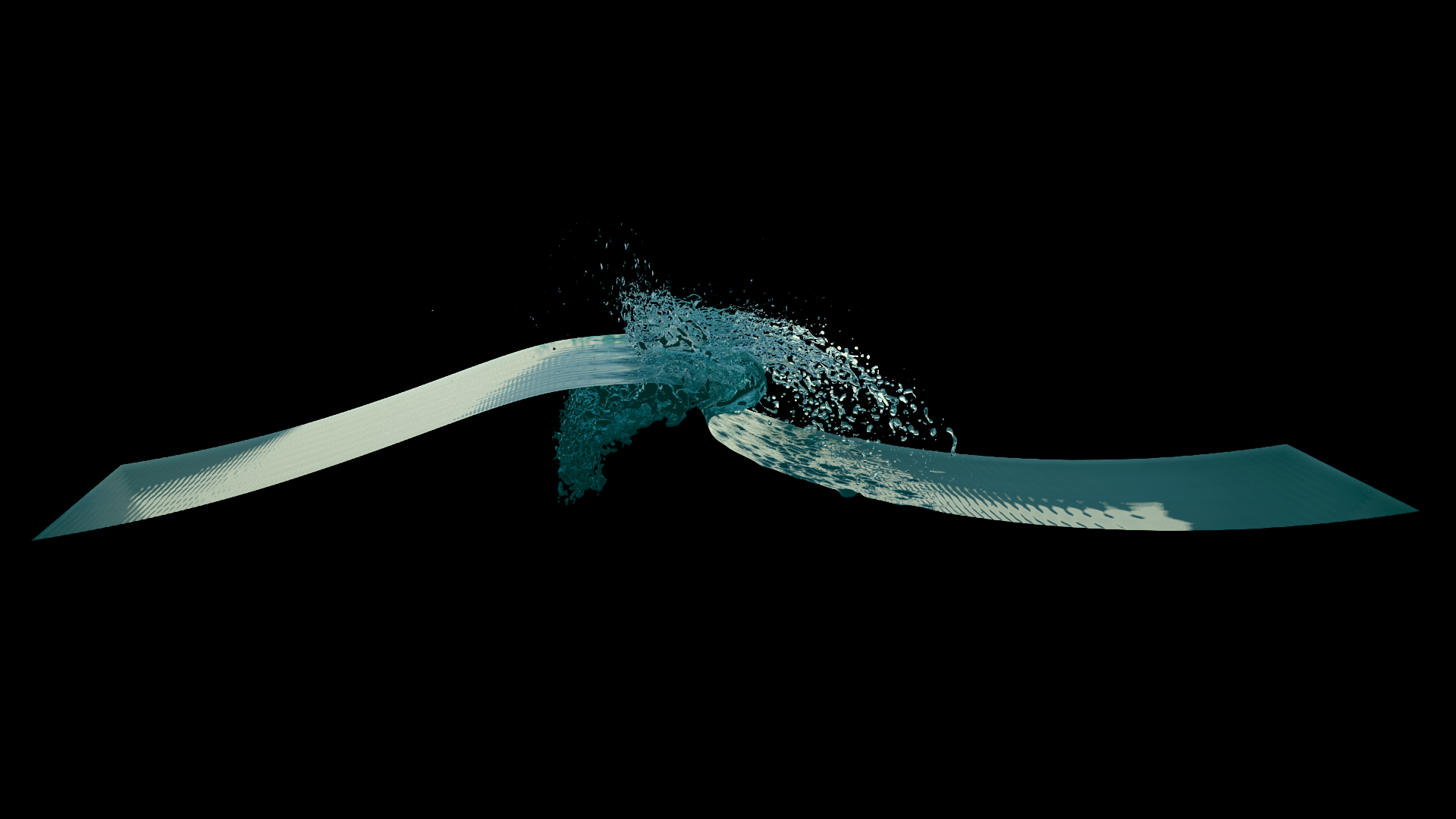}}
\pdfdest name{viz51} FitR height 2.3cm 
\hyperlink{viz52}{\includegraphics[trim = 0mm .00cm 0mm
.00cm,clip=true,angle=0,scale=0.06]{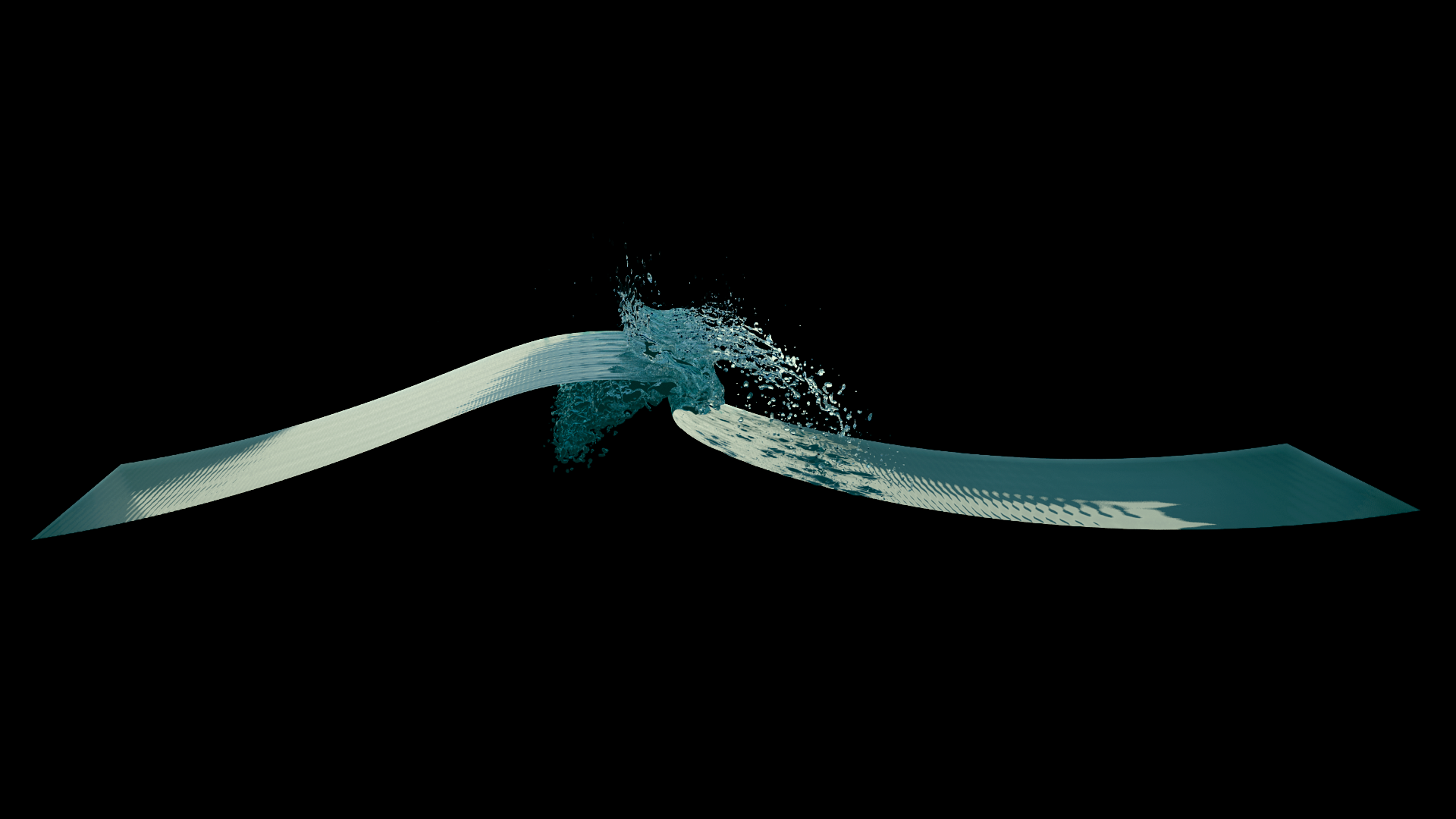}}
\pdfdest name{viz52} FitR height 2.3cm 
\hyperlink{viz53}{\includegraphics[trim = 0mm .00cm 0mm
.00cm,clip=true,angle=0,scale=0.06]{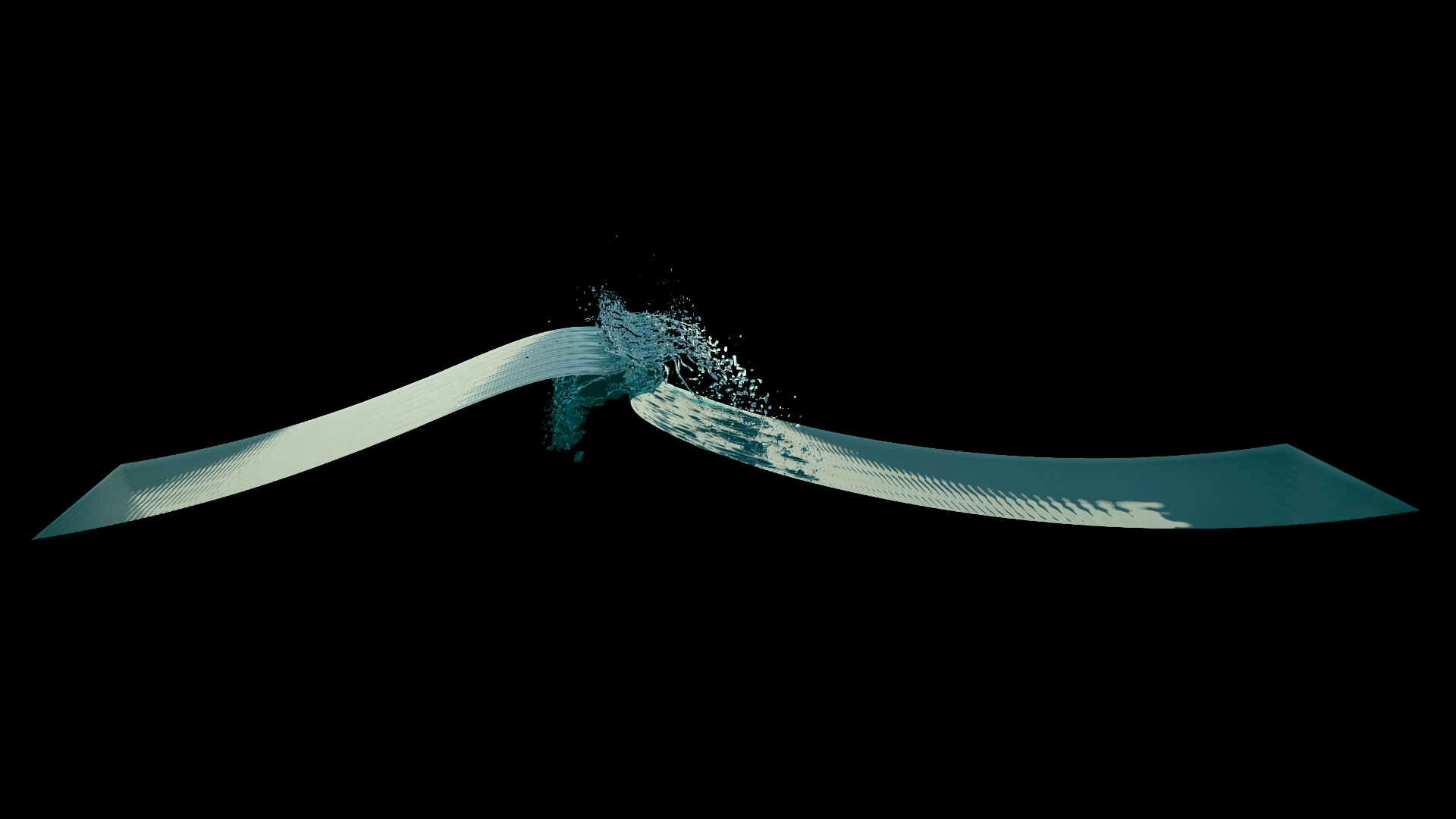}}
\pdfdest name{viz53} FitR height 2.3cm 
\hyperlink{viz54}{\includegraphics[trim = 0mm .00cm 0mm
.00cm,clip=true,angle=0,scale=0.06]{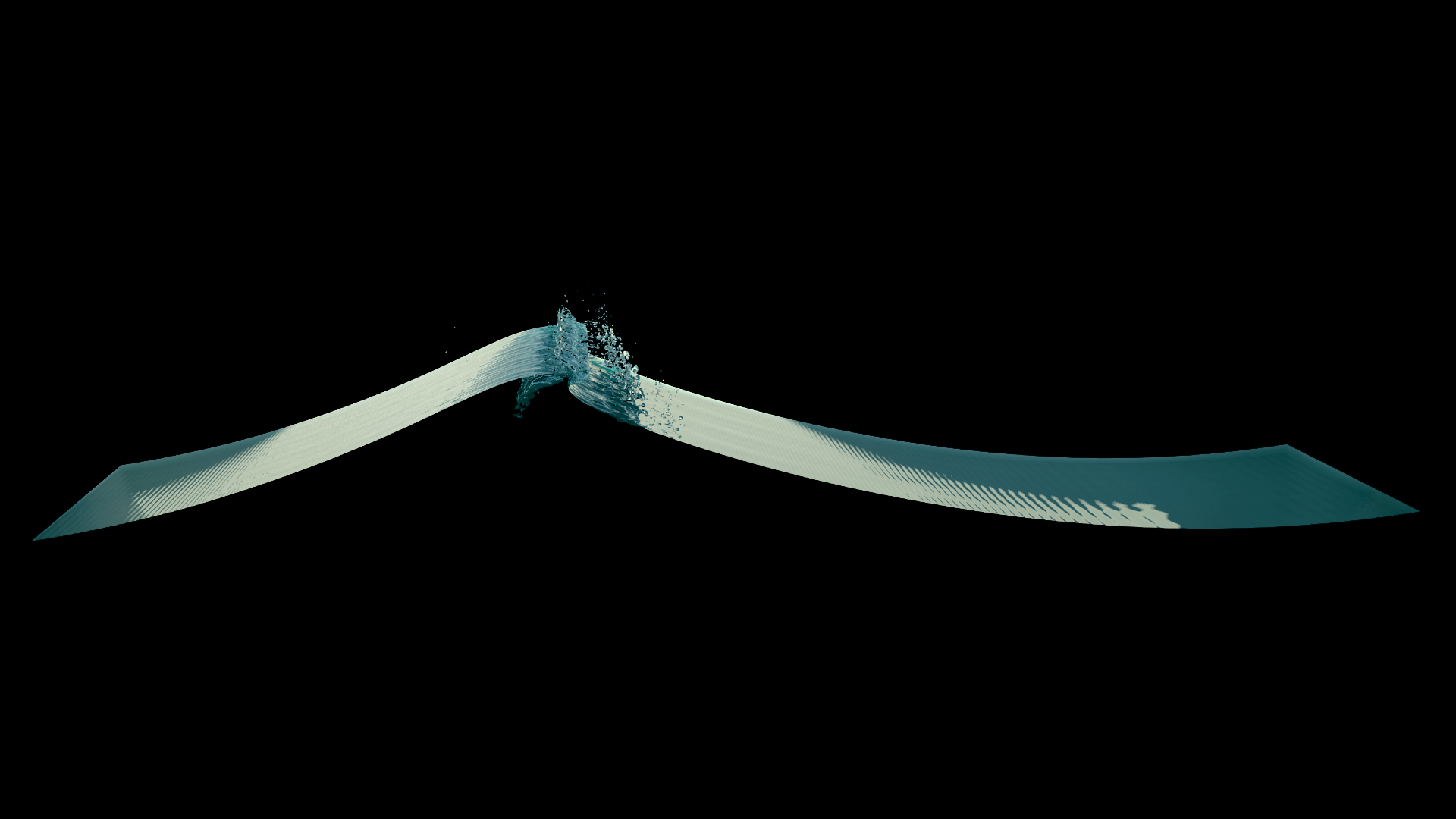}}\\
\pdfdest name{viz54} FitR height 2.3cm 
\vspace{1pt}
\hyperlink{viz61}{\includegraphics[trim = 0mm .00cm 0mm
.00cm,clip=true,angle=0,scale=0.06]{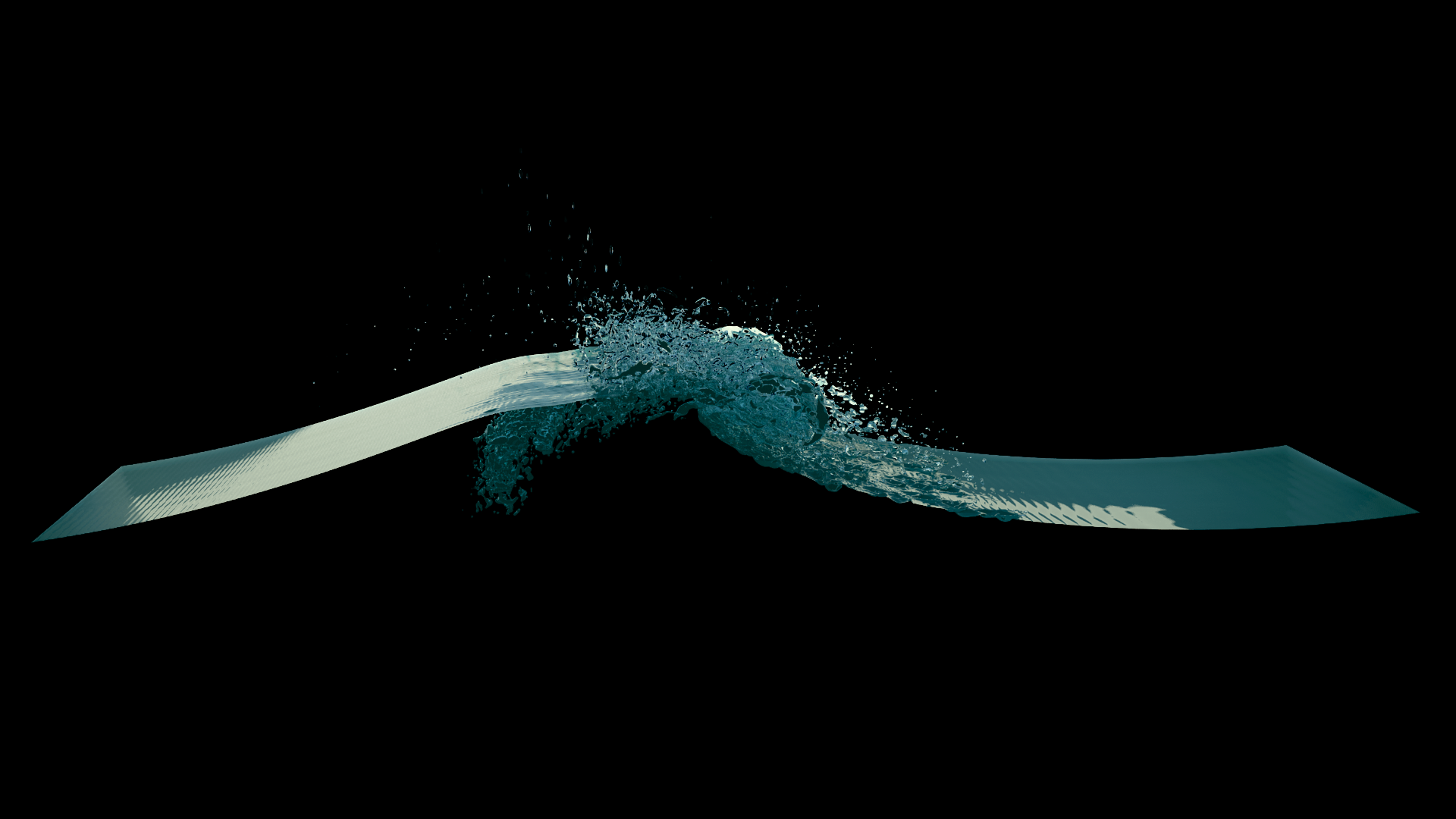}}
\pdfdest name{viz61} FitR height 2.3cm 
\hyperlink{viz62}{\includegraphics[trim = 0mm .00cm 0mm
.00cm,clip=true,angle=0,scale=0.06]{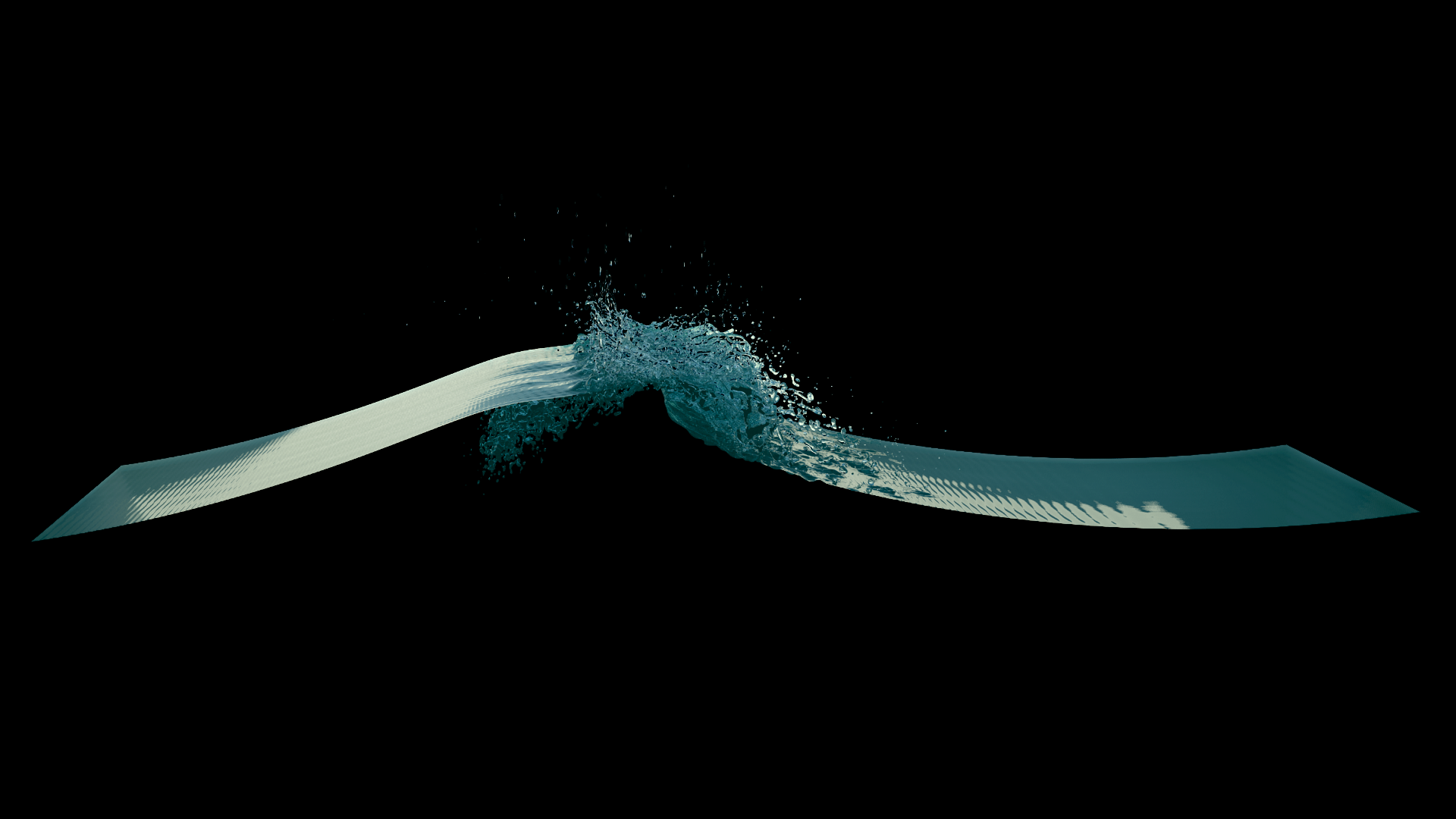}}
\pdfdest name{viz62} FitR height 2.3cm 
\hyperlink{viz63}{\includegraphics[trim = 0mm .00cm 0mm
.00cm,clip=true,angle=0,scale=0.06]{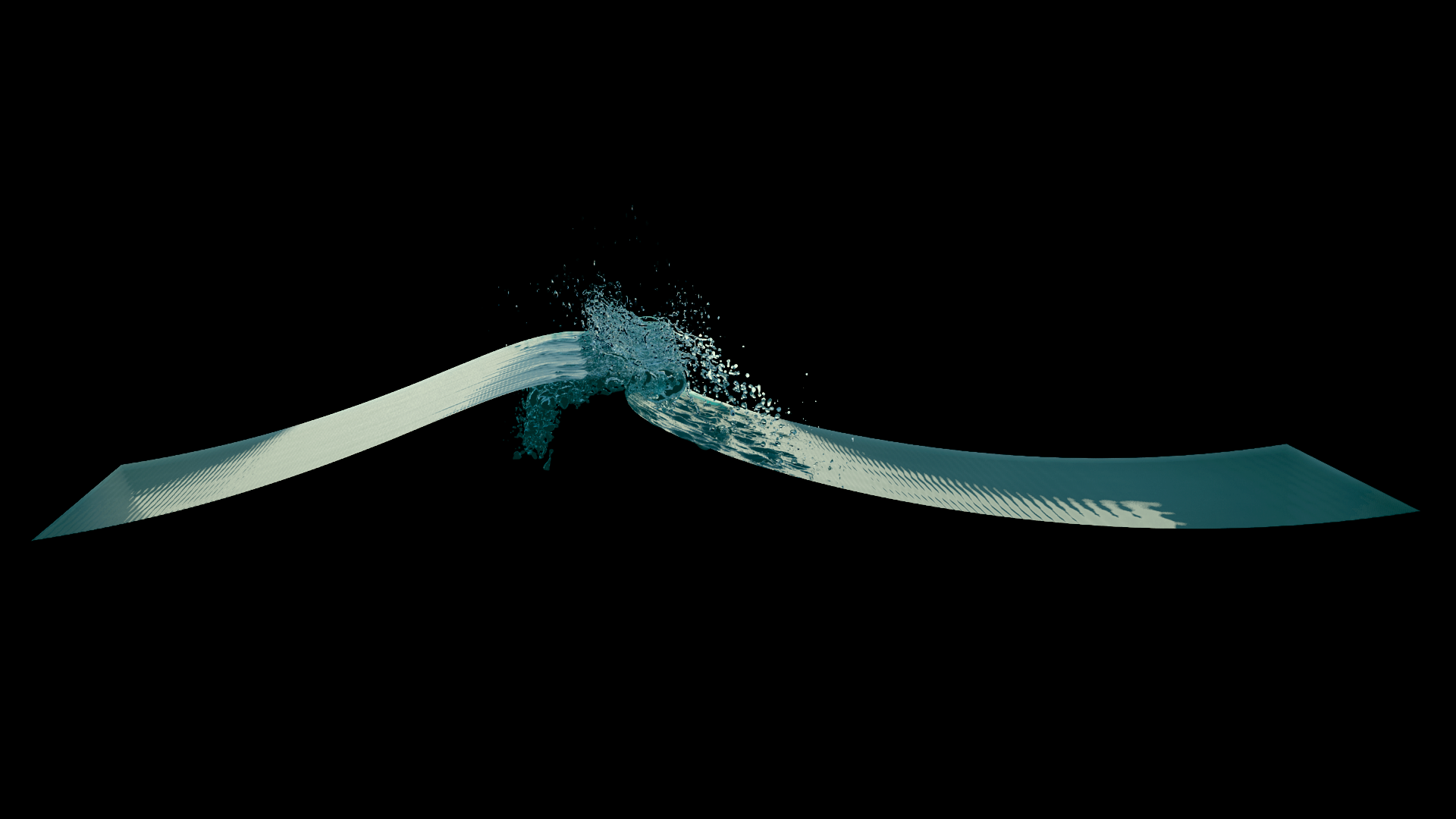}}
\pdfdest name{viz63} FitR height 2.3cm 
\hyperlink{viz64}{\includegraphics[trim = 0mm .00cm 0mm
.00cm,clip=true,angle=0,scale=0.06]{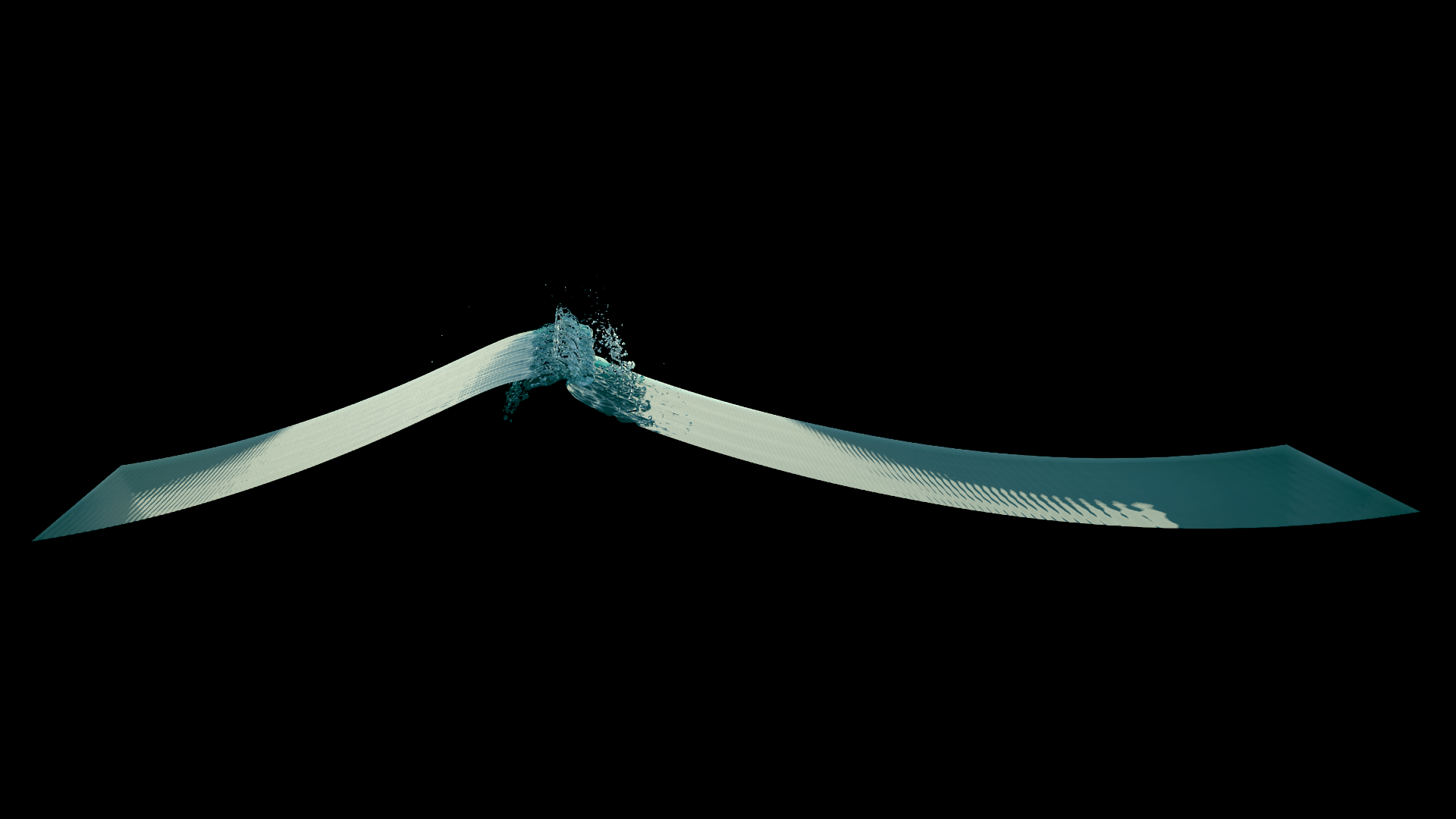}}\\
\pdfdest name{viz64} FitR height 2.3cm 
\vspace{1pt}
\hyperlink{viz71}{\includegraphics[trim = 0mm .00cm 0mm
.00cm,clip=true,angle=0,scale=0.06]{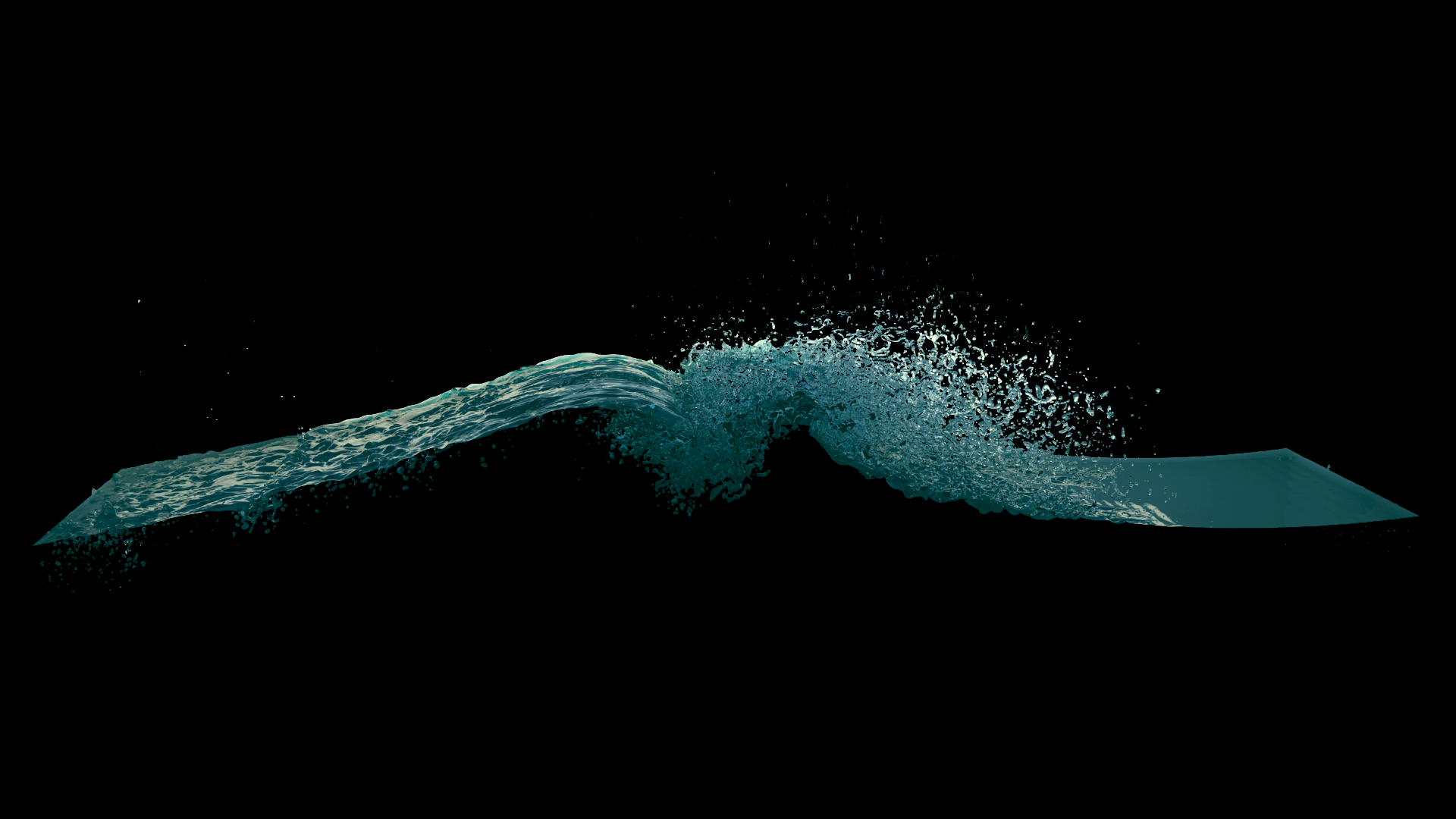}}
\pdfdest name{viz71} FitR height 2.3cm 
\hyperlink{viz72}{\includegraphics[trim = 0mm .00cm 0mm
.00cm,clip=true,angle=0,scale=0.06]{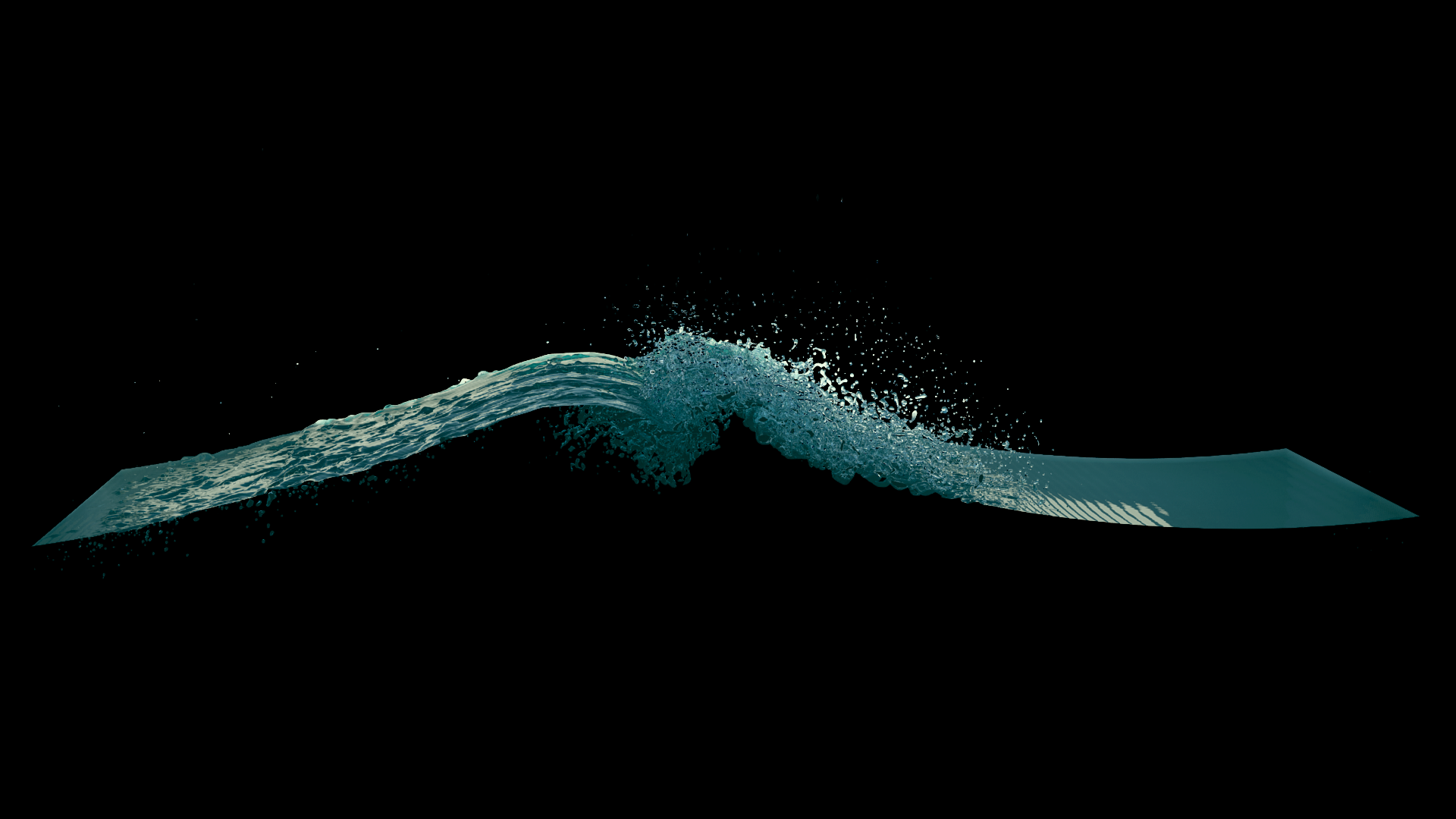}}
\pdfdest name{viz72} FitR height 2.3cm 
\hyperlink{viz73}{\includegraphics[trim = 0mm .00cm 0mm
.00cm,clip=true,angle=0,scale=0.06]{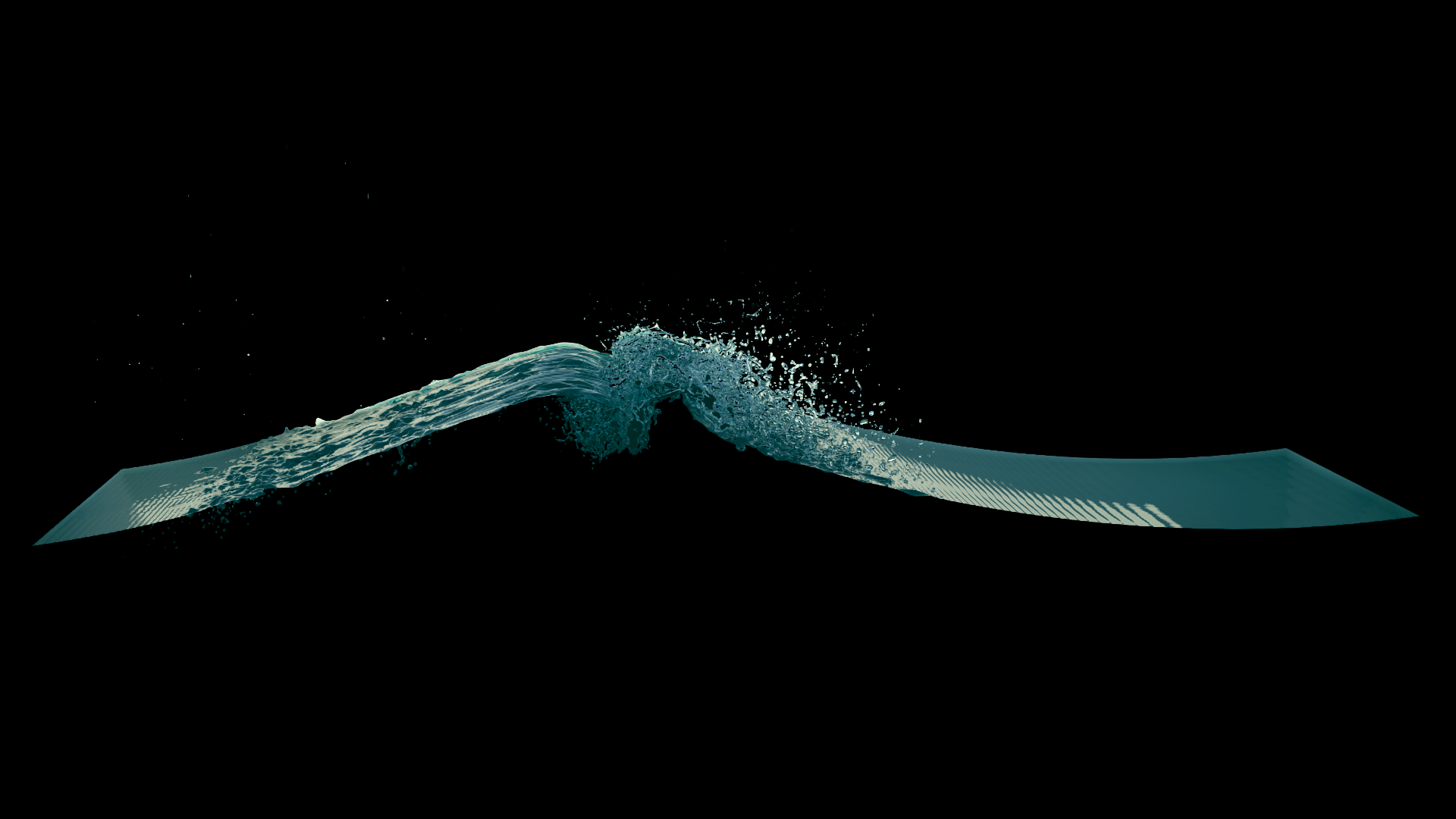}}
\pdfdest name{viz73} FitR height 2.3cm 
\hyperlink{viz74}{\includegraphics[trim = 0mm .00cm 0mm
.00cm,clip=true,angle=0,scale=0.06]{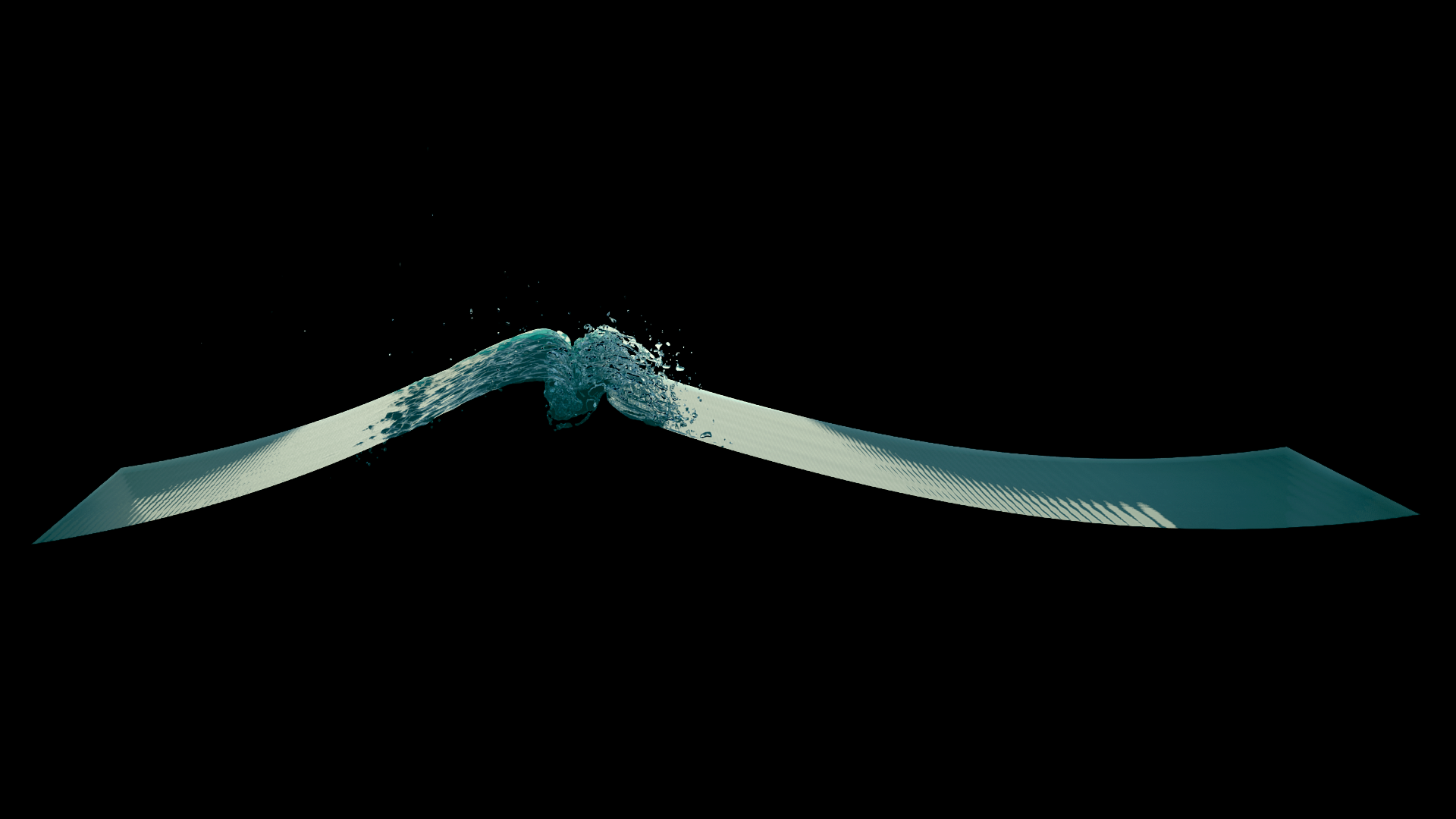}}\\
\pdfdest name{viz74} FitR height 2.3cm 
\vspace{1pt}
\hyperlink{viz81}{\includegraphics[trim = 0mm .00cm 0mm
.00cm,clip=true,angle=0,scale=0.06]{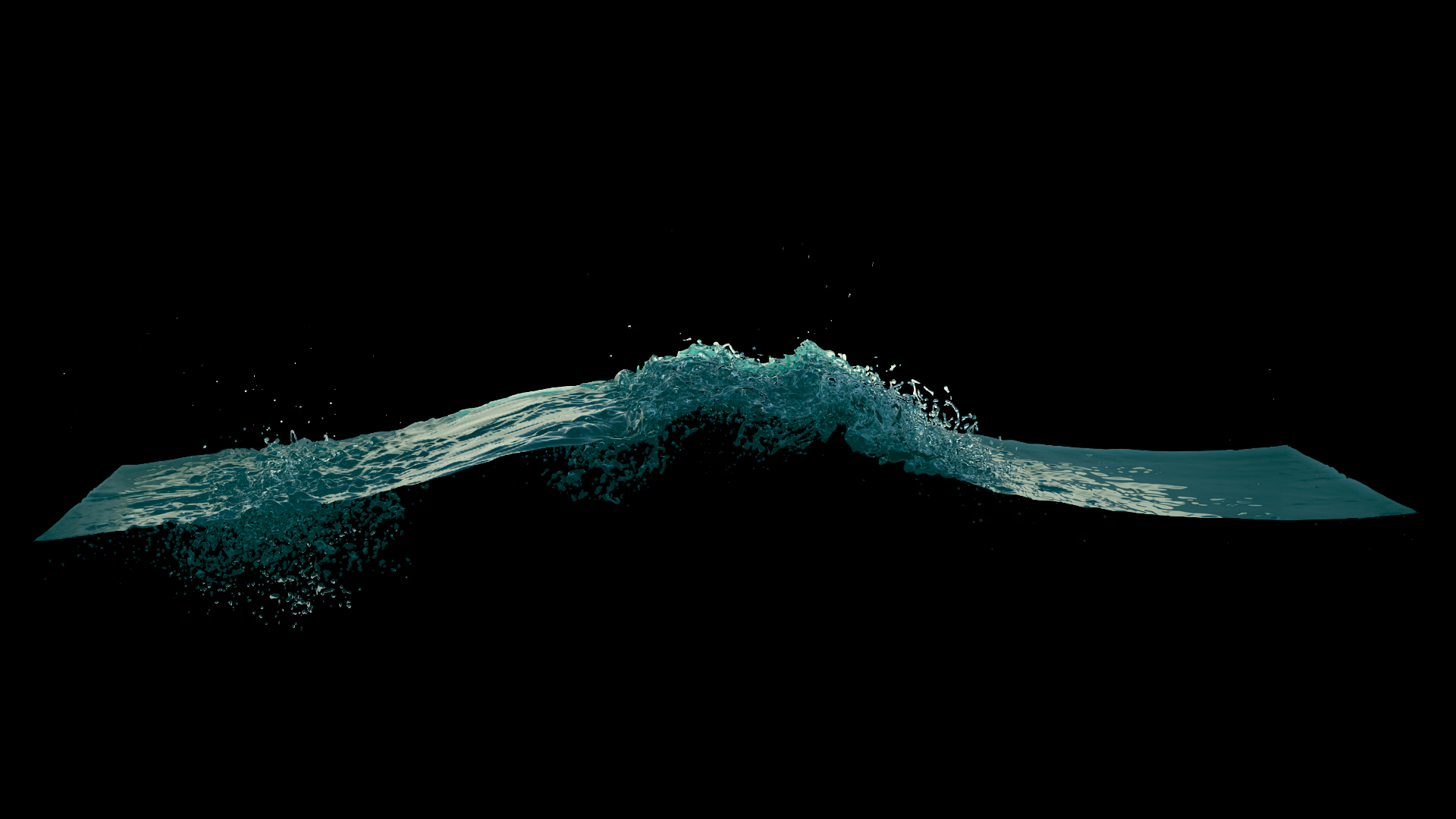}}
\pdfdest name{viz81} FitR height 2.3cm 
\hyperlink{viz82}{\includegraphics[trim = 0mm .00cm 0mm
.00cm,clip=true,angle=0,scale=0.06]{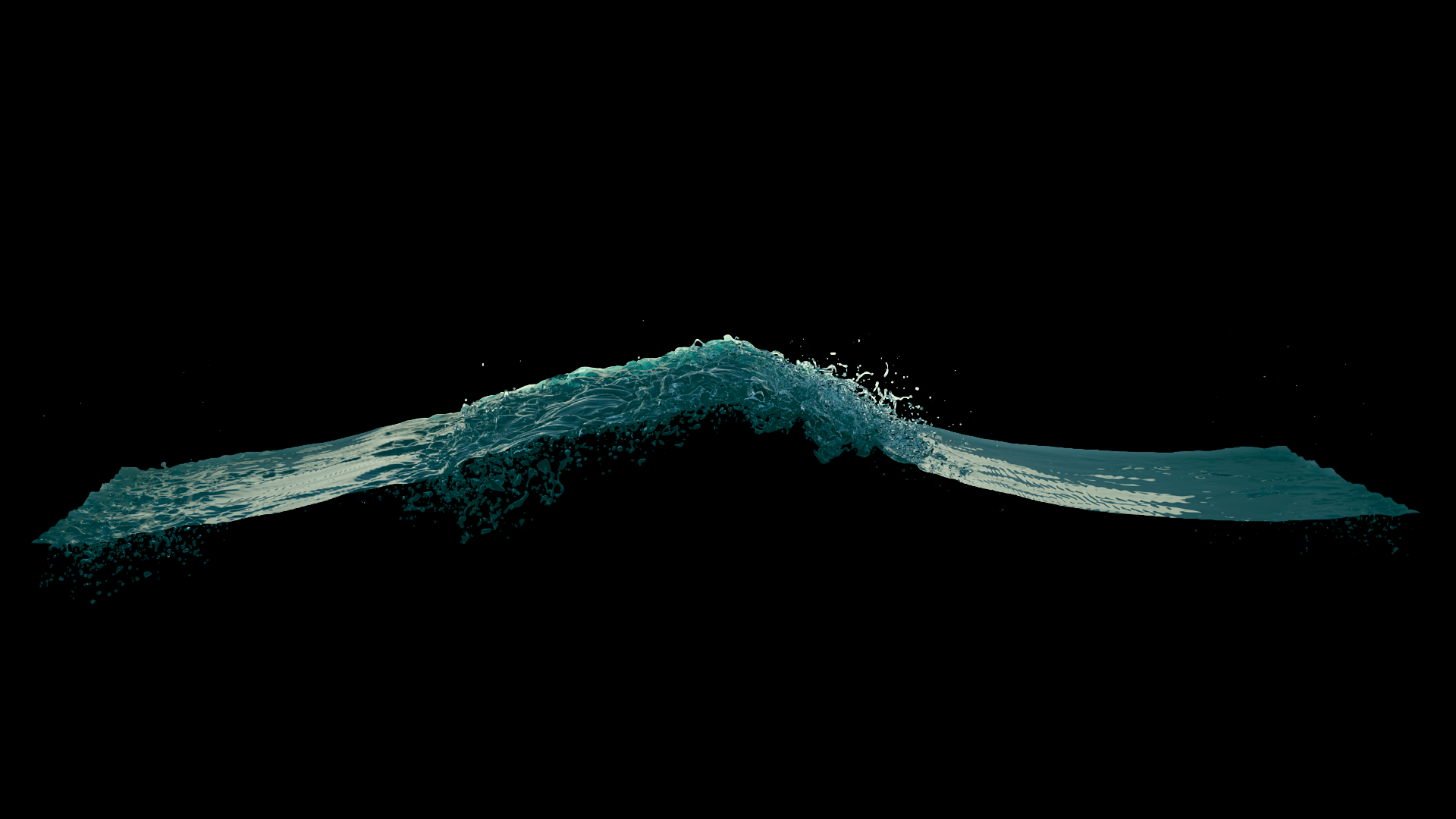}}
\pdfdest name{viz82} FitR height 2.3cm 
\hyperlink{viz83}{\includegraphics[trim = 0mm .00cm 0mm
.00cm,clip=true,angle=0,scale=0.06]{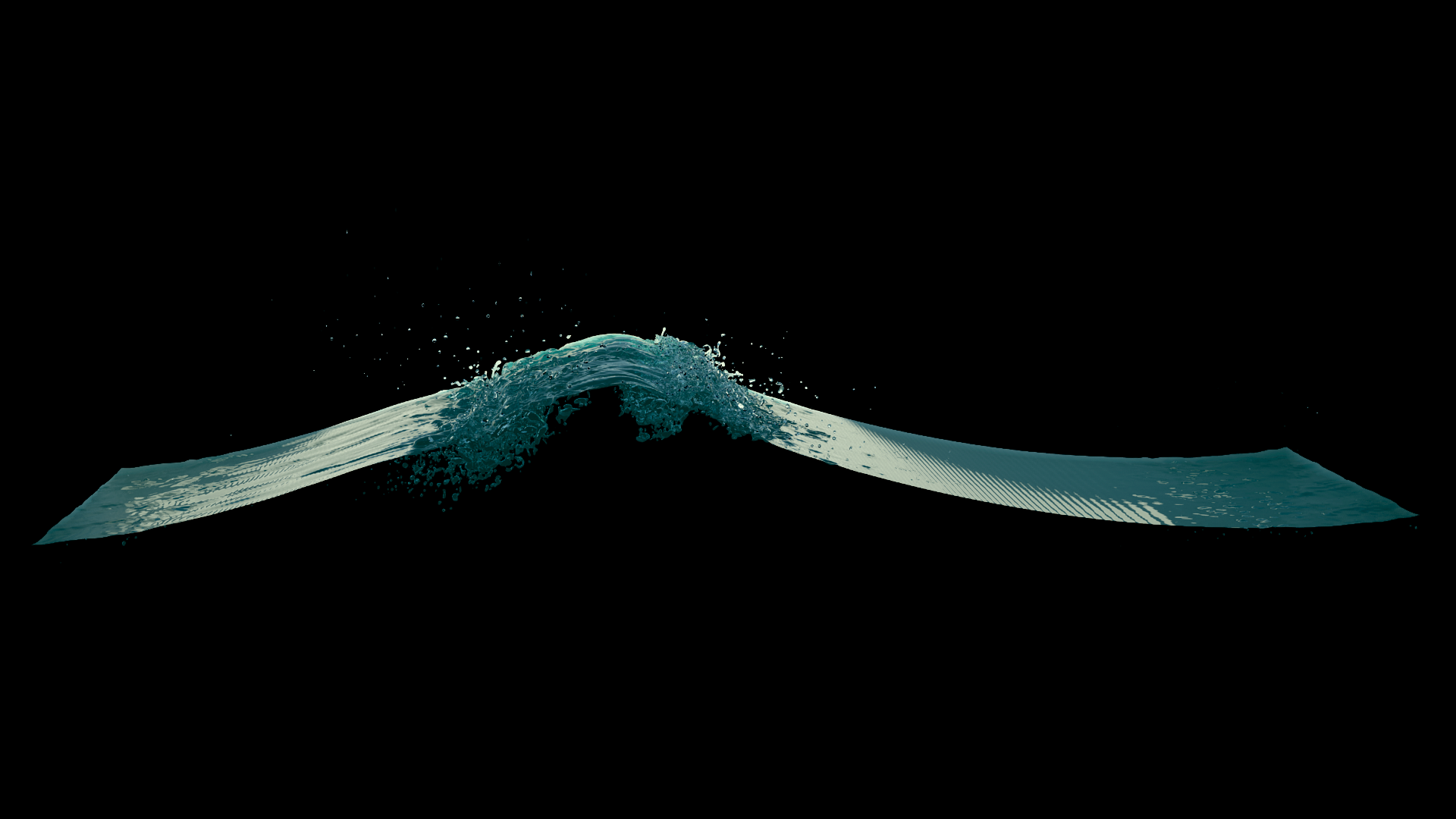}}
\pdfdest name{viz83} FitR height 2.3cm 
\hyperlink{viz84}{\includegraphics[trim = 0mm .00cm 0mm
.00cm,clip=true,angle=0,scale=0.06]{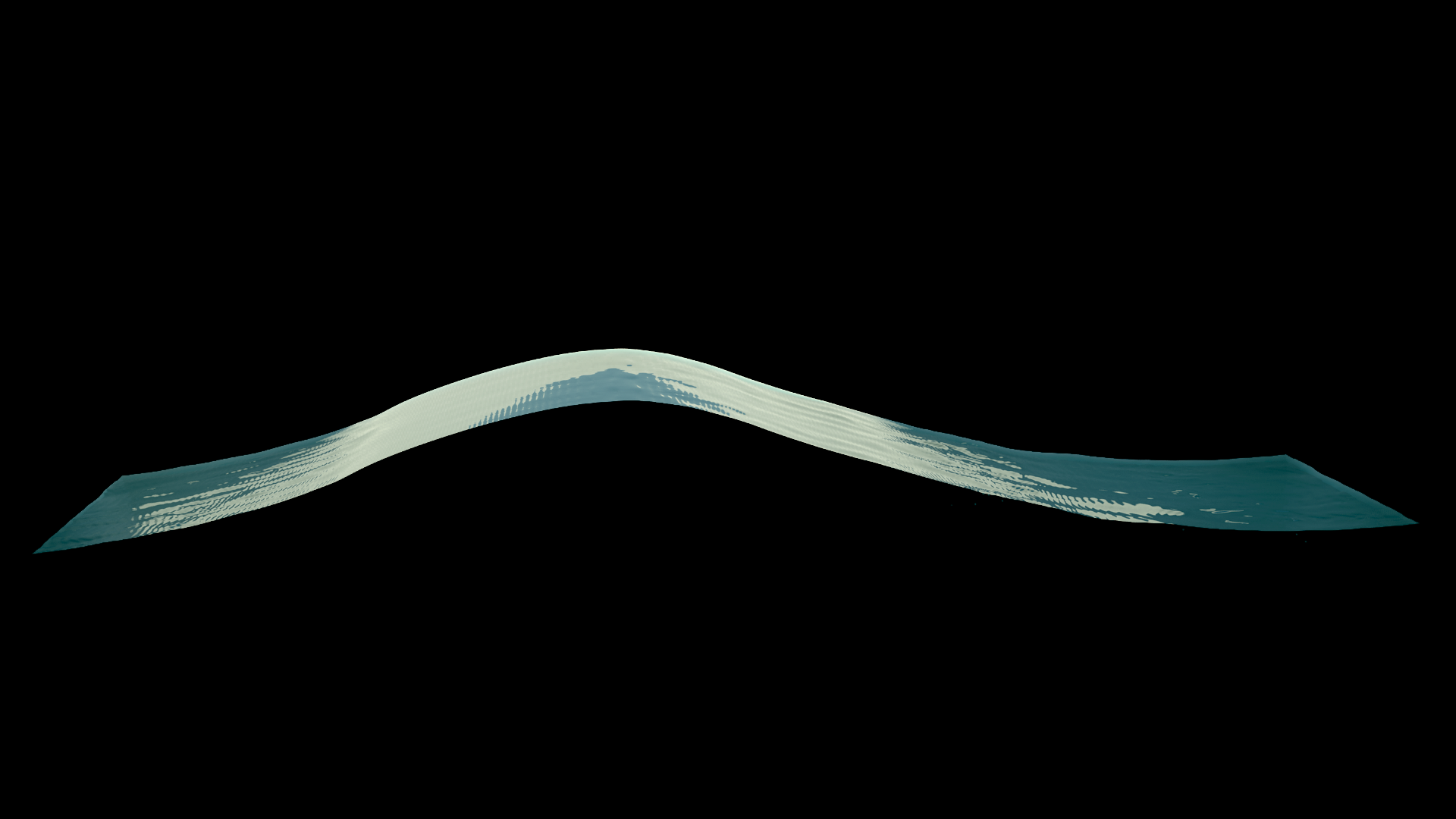}}\\
\pdfdest name{viz84} FitR height 2.3cm 
\caption{\label{fig:3d} Three-Dimensional renderings of the 50\% iso-contour of the
volume fraction. Cases $VSP^F$, $SP^F$, $P^F$, and $WP^F$ are shown in
columns A, B, C and D respectively.  The frames here correspond to those in Figures~\ref{fig:vsp_vmt}-\ref{fig:vsp_bpv}, \ref{fig:sp_vmt}-\ref{fig:sp_bpv}, \ref{fig:p_vmt}-\ref{fig:p_bpv}, and \ref{fig:wp_vmt}-\ref{fig:wp_bpv}, respectively. Note: the electronic version of this document contains high resolution frames that can be zoomed in on by clicking
the desired frame.}
\end{figure*}

Figure~\ref{fig:3d} shows the free surface visualized with the 50\%
iso contour of the volume fraction at eight times for each
of the plunging breaking cases. Columns A-D show results
from cases $VSP^F$, $SP^F$, $P^F$ , and $WP^F$ , respectively.
The electronic version of the document has zoomable
images which enables the reader to enlarge a panel
of Figure~\ref{fig:3d}  to the full window size by clicking it. Each
of the frames has been selected to correspond to the following
stages in the breaking cycle: \\
{\bf 1. Vertical Front Face:} The wave becomes asymmetric
with the front face becoming vertical, and the the tip begins
to form. \\
{\bf 2. Pinch Off:} The tip of the plunging breaker touches
down on the surface below, causing an ovular cavity of
air to be pinched off. \\
{\bf 3. Splash-up:} The touch down of the jet formed in the
pinch-off stage forces up a secondary jet. Since, the jet in
the pinch-off stage impacts the surface almost vertically,
the splash up is nearly vertical. \\
{\bf 4. Interaction:} The nearly vertical splash-up, which
may reach a height greater than the wave crest, interacts
with the advancing wave crest and causes the entrained
air cavity to tilt. This three-way interaction allows the
cavity to reconnect with the free-surface. This reconnection
provides an escape path for the air. \\
{\bf 5. Tip Break Off:} The advancing wave crest is now past
the splash-up region and carries the splash-up region beneath
the crest with the wave. The portion of the splash up
that reached a height greater than the wave crest is
either ÔwhippedÕ forward or broken off. \\
{\bf 6. Tip Impact:} The portion of the splash up above the
wave crest that remains attached is slammed down onto
the front face, and the portion of splash up that broke free
crashes into the front face. \\
{\bf 7. Dual Breaking:} Forward and backward plunging
events are evident, giving the wave a mushroom top
look. It is the time at which the large bubble cloud is
formed and entrainment occurs at both the forward and
rear plunging locations. \\
{\bf 8. Spilling:} A sequence of weaker and weaker plunging
and spilling events occur on the front face until a steady
progressive wave is formed.

\begin{figure}
\centering
\includegraphics[width=\linewidth]{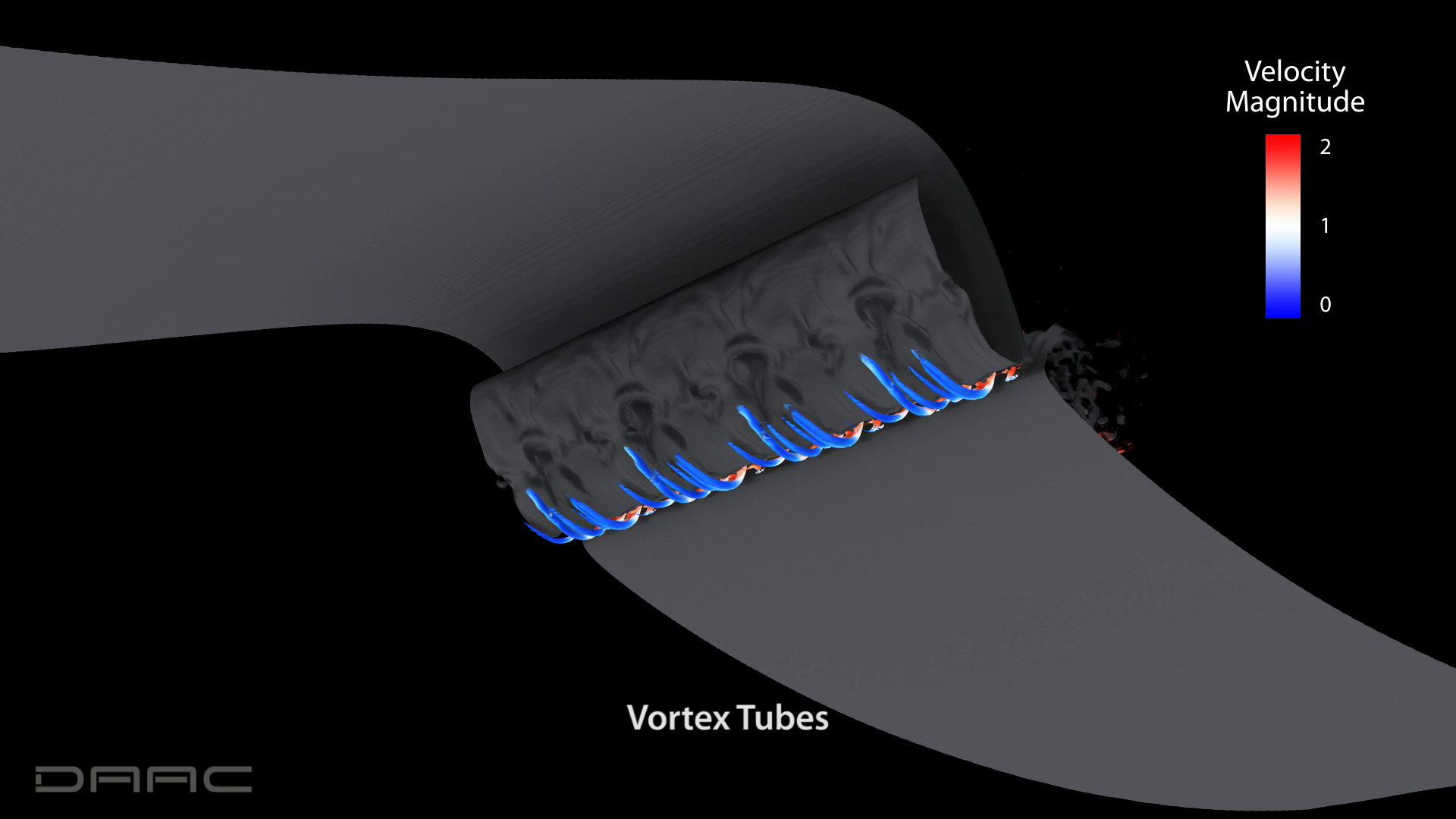}
\caption{\label{fig:tubes} Vortex tubes apparent on back of ovular cavity of air for case $VSP^F$}
\end{figure}

In addition to the contours of the volume fraction,
contours of the vorticity and velocity magnitude
show interesting three-dimensional structures. Figure~\ref{fig:tubes}
shows the magnitude of the velocity overlaid on the free surface
for case $VSP^F$ at $t = 14.7$. Evidence of vortex
tubes, and hairpin like structure are evident on the back
of the initial cavity of air entrained by the primary plunging
event. Similar structures have been observed in high definition
video of the large waves breaking on beaches
\citep{walker09}.

\section{Volume Integrated Energetics and Air Entrainment}
\begin{figure}
\centering
\includegraphics[width=\linewidth]{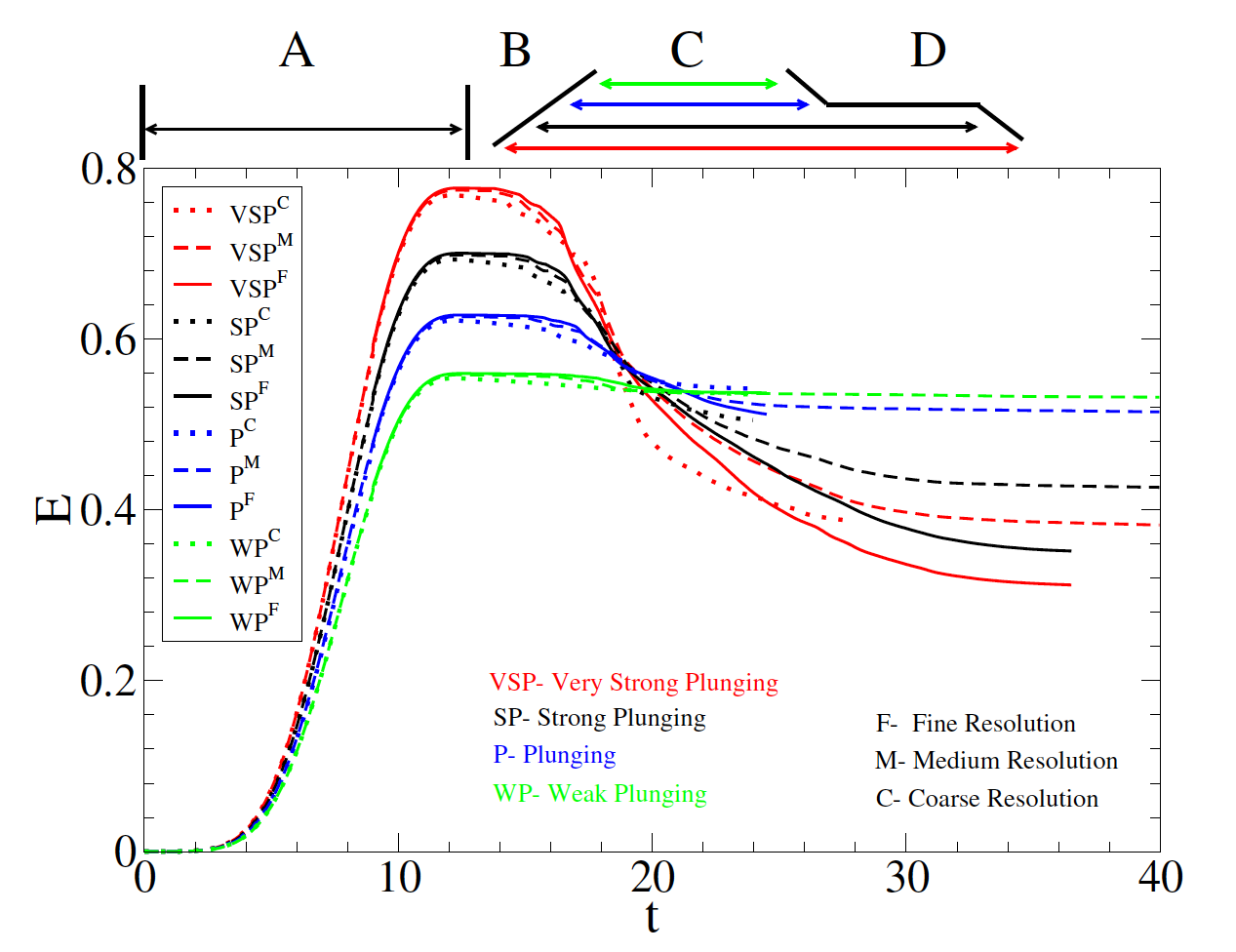}
\caption{\label{fig:evolution} Volume integrated energy: potential plus kinetic (see Equation~\ref{eq:intE}).}
\end{figure}

\begin{figure}
(a) \includegraphics[width=0.95\linewidth]{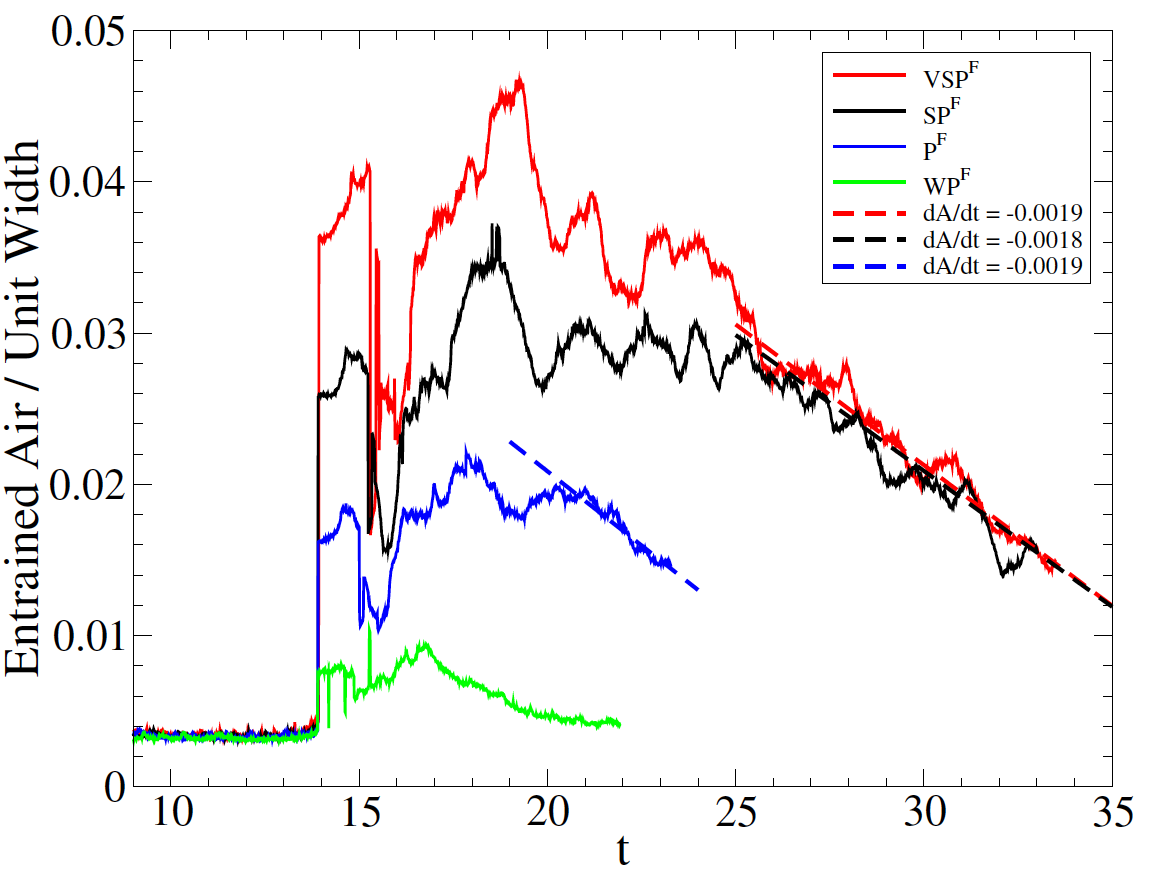} \\
(b) \includegraphics[width=0.95\linewidth]{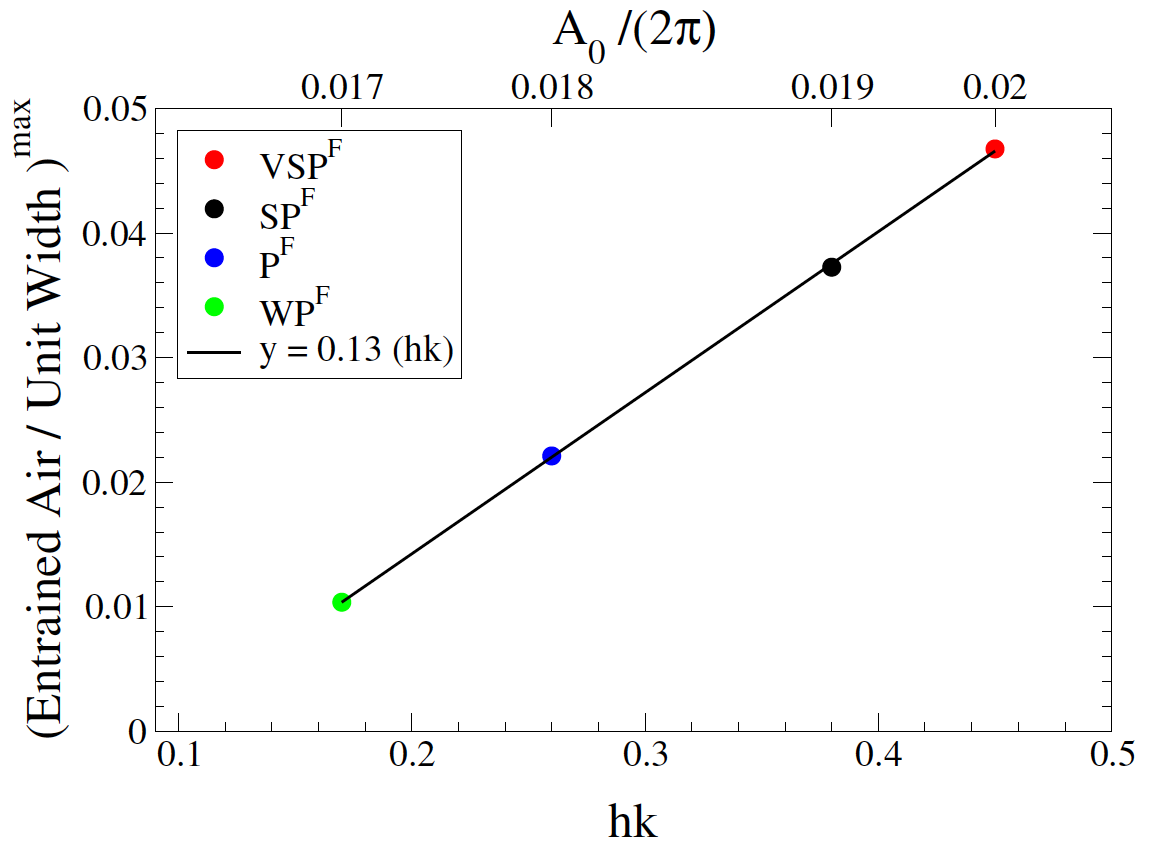}
\caption{\label{fig:air} Air beneath free-surface, defined by the 50\% iso contour
of the volume fraction, for cases $VSP^F$, $SP^F$, $P^F$,
$WP^F$. (a) Air entrained as a function of time. (b) Maximum air entrained as
a function of normalized pinch-off height $kh$.} 
\end{figure}

Figure~(\ref{fig:evolution}) is a plot the the total energy,  
\begin{equation} 
E(t)=\int_V dV  \left(\frac{\rho(t)}{2}u_i(t)u_i(t) +
\frac{\left[\rho(t)-\rho(0)\right] z}{F_r^2} \right) \; , 
\label{eq:intE}
\end{equation}
over time. Four distinct stages are evident in Figure~\ref{fig:evolution},including: (A) Atmospheric forcing; (B) Potential flow
before breaking; (C) Breaking which consists of plunging, spilling and splash-up events; and (D) Potential flow
after breaking. Note the unequal length of stages B, C,and D, shown at the top of Figure~\ref{fig:evolution}. Figure~\ref{fig:evolution} also
illustrates the grid resolution necessary to correctly resolve strong plunging events. The minimum resolution is
on the order of 1000 grid points per wave length. Stage C is the most interesting as it characterized by turbulent
motions, air entrainment (bubbles), and droplet formation, which leads to complex interactions between the
kinetic and potential energies. The VSP and SP cases lose over half their total energies due to the effects of
breaking. Figure~\ref{fig:air} shows the total amount of air that is entrained per unit width for finest resolution breaking-wave studies. By our definition, the amount of air that is entrained is based on how much air is surrounded by the outermost 50\% iso contour of the volume fraction. Based on this definition, there will always be small amounts of
entrained air as is evident for $t < 14$. The first entraining events are due to the tip of the plunging breaking wave
closing off an air cavity. For VSP, SP, P, and WP-type breaking, this occurs at $t \approx 14$, $t \approx 14.5$, $t \approx 15$, and
$t \approx 16.5$, respectively. The first entraining event is followed by sudden dip when the 50\% iso contour opens up
to the air due to the violent impact of the jet. Slightly later entraining events are due to splash ups, especially
for the VSP ($16 < t < 25$) and SP ($16 < t < 25$) cases. During these splash-up events, the amount of entrained air is nearly constant, which indicates that entraining balances degassing through the free surface. After the main splash-ups, air is entrained at the toe of spilling region. During this stage, the air that had been entrained
by plunging and splash ups degasses more quickly than air can be entrained by spilling. This occurs for
$t > 25$ for the VSP and SP cases, for $t > 22$ for the P case,and for
$t > 20$ for the WP case. When viewing Figure~\ref{fig:air}, it is helpful to refer to Figure~\ref{fig:3d}, which shows
three-dimensional views of the wave breaking, and Figures~\ref{fig:vsp_vmt}-\ref{fig:wp_bpv}, which show various energy terms. Similar air
entrainment calculations had been performed by \cite{nfa5} who have good agreement with laboratory
measurements of the void fraction behind a wetted transom.

The maximum air entrained beneath the free surface does not occur at pinch-off. Inspection of Figure~\ref{fig:air}(a)
shows that it occurs well after the initial cavity of air is entrained. Figure~\ref{fig:air}(b) shows that \underline{maximum} value of air
entrained as a function of the pinch off height, $h$. That is the difference in elevation between the crest and the point
at which the toe impacts the front face \citep{drazen08}. The maximum air entrained per unit width is a linear function of the pinch-off height.

\begin{figure}
(a) \includegraphics[width=0.85\linewidth]{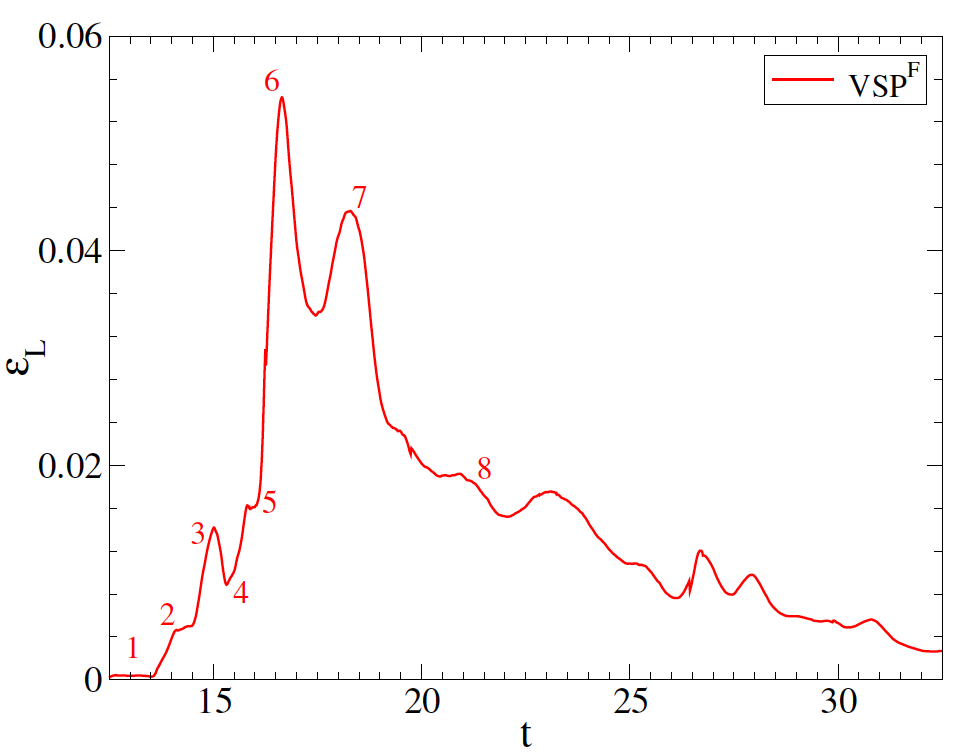} \\
(b) \includegraphics[width=0.85\linewidth]{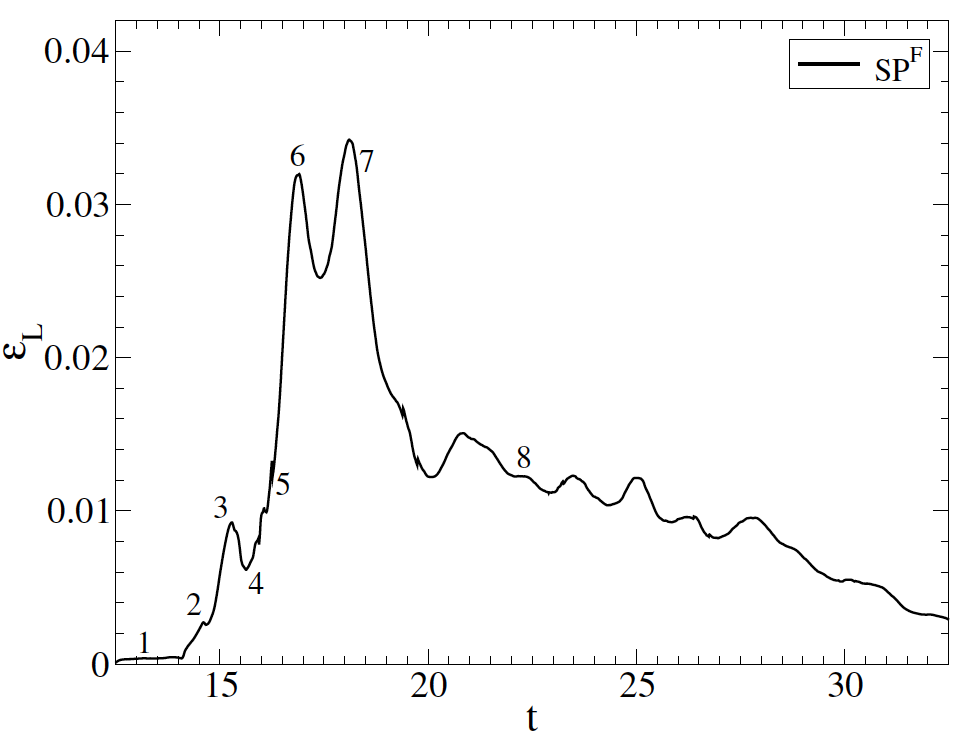} \\
(c) \includegraphics[width=0.85\linewidth]{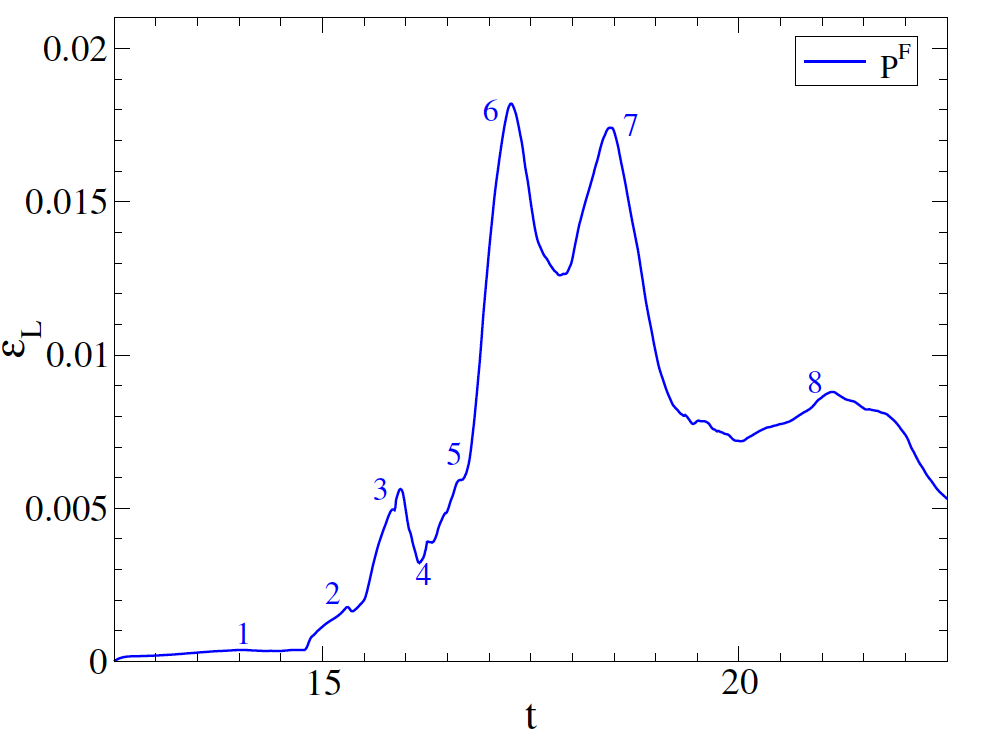} \\
(d) \includegraphics[width=0.85\linewidth]{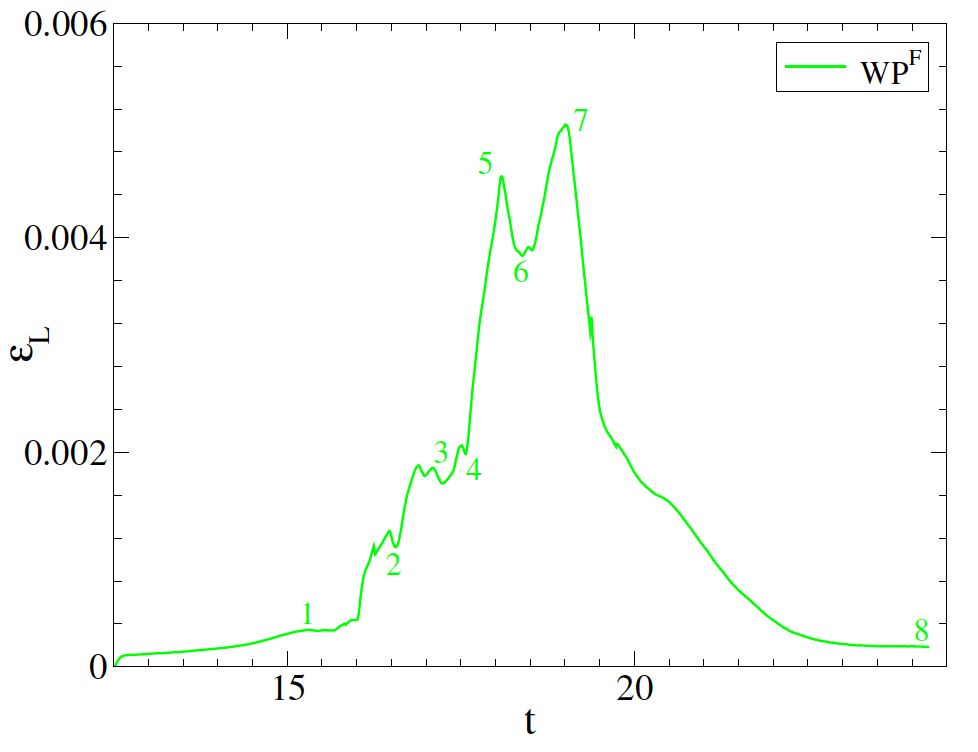}
\caption{\label{fig:energyrate} Rate of change of Total Energy per unit width
shown. (a) $VSP^F$, (b) $SP^F$, (c) $P^F$, (d) $WP^F$. Points 1-
7 correspond to the rows of Figures~\ref{fig:vsp_vmt}-\ref{fig:vsp_bpv}, \ref{fig:sp_vmt}-\ref{fig:sp_bpv}, \ref{fig:p_vmt}-\ref{fig:p_bpv}, and \ref{fig:wp_vmt}-\ref{fig:wp_bpv}, respectively.} 
\end{figure}

Figure~\ref{fig:energyrate}(a)-(d) shows the rate of change of the total energy per unit width for cases
$VSP^ F$, $SP^F$, $P^F$, and $WP^F$, respectively. Each figure has 8 points labeled which correspond to distinct changes in the rate
of change of the total energy. Inspection of Figure~\ref{fig:air}(b) shows that the volume of air beneath the free surface
changes dramatically at the same times that the large changes in the dissipation rate are observed. The next
section develops a statistical framework in which to analyze immiscible multi-phase flows, which is then applied
to the breaking waves simulated here to further analyze the mechanisms which are responsible for air entrainment and changes in the dissipation rate.

\section{Statistical Energy Analysis of Multiphase Flows}
\label{sec:RD} 
In the subsequent discussion, angle brackets  and single primes are used to denote mean
and fluctuating volume-weighted (Reynolds) values, viz.,
\begin{equation} 
\label{eq:RD} 
u_i = \left< u_i \right>  +u'_i   \, .
\end{equation}
While a tilde and double primes denote mean and fluctuating mass-weighted
(Favre) values, viz.,
\begin{equation} 
\label{eq:FV} 
u_i = \tilde{u}_i +u''_i  
\end{equation}
where $\tilde{u}_i = \left<\rho u_i \right> / \left< \rho \right>$. Here it is
noted that the mean mass-weighted velocity is the sum of the Reynolds mean plus
the normalized mass flux,  $\tilde{u}_i=\left<u_i\right> + \left<\rho' u_i'
\right> / \left< \rho \right>$.
%
%
\subsection{Mean and Turbulent Kinetic Energy} 
\label{sec:mke_tke}
The total kinetic energy at location $x_i$ and time $t$ is: 
\begin{equation} 
KE \equiv \frac{1}{2}
\rho(x_i,t) u_i(x_i,t) u_i(x_i,t) \,.  
\label{eq:KE} 
\end{equation}

If the kinetic energy is volume averaged then the mean and fluctuating
components are: 
\begin{equation} 
m.k.e. \equiv \frac{1}{2}\left<\rho\right>\left<u_{i}\right>\left<u_i\right>\;,
\label{eq:mkeV} 
\end{equation}
and
\begin{equation} 
t.k.e.  \equiv \frac{1}{2}\left<\rho\right>\left<u'_{i} u'_i\right>
+\left<u_i\right>\left<\rho' u'_i\right> + \frac{1}{2}\left<\rho' u'_{i}
u'_i\right> \, .  
\label{eq:tkeV} 
\end{equation}

If the kinetic energy is mass averaged, then the mean and fluctuating components
are: 
\begin{multline} 
m.k.e. \equiv \frac{1}{2}\left<\rho\right> \tilde{u}_i \tilde{u}_i =
\frac{1}{2}\left<\rho\right> \left<u_i\right>\left<u_i\right>   \\ +
\left<u_i\right> \left<\rho'u_i'\right> + \frac{1}{2} \frac{\left<\rho'
u_i' \right>^2}{ \left< \rho \right>} \; ,
\label{eq:mkeF} 
\end{multline}
and
\begin{equation} 
t.k.e. \equiv \frac{1}{2}\left<\rho\right>\left<u'_{i}u'_i\right> +
\frac{1}{2} \left<\rho'u'_i u'_i\right> -\frac{1}{2}\frac{\left<\rho'
u'_i\right>^2}{\left<\rho\right>} \, . 
\label{eq:tkeF} 
\end{equation}

\begin{figure}
(a) \includegraphics[width=\linewidth]{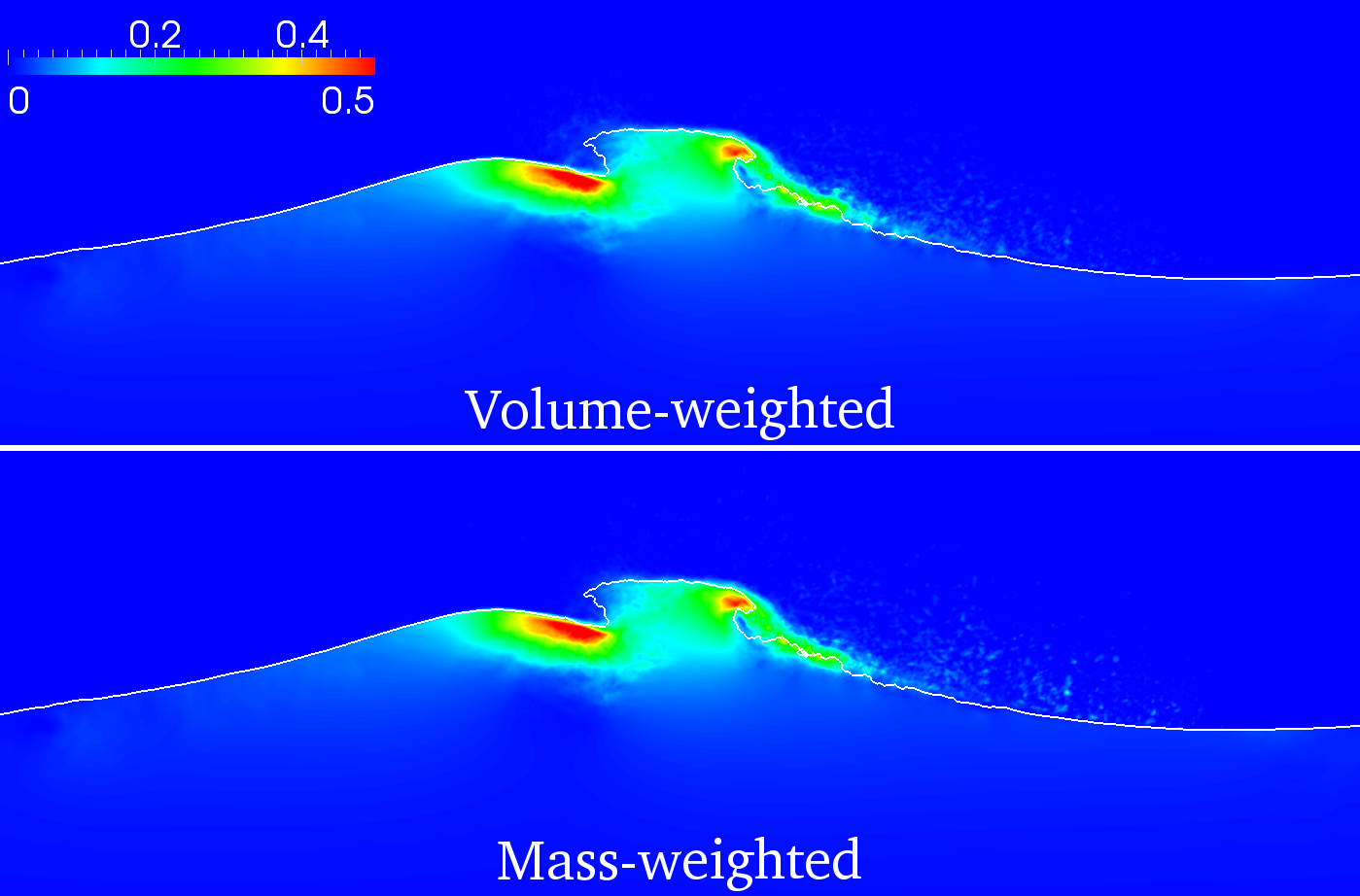} \\
(b) \includegraphics[width=\linewidth]{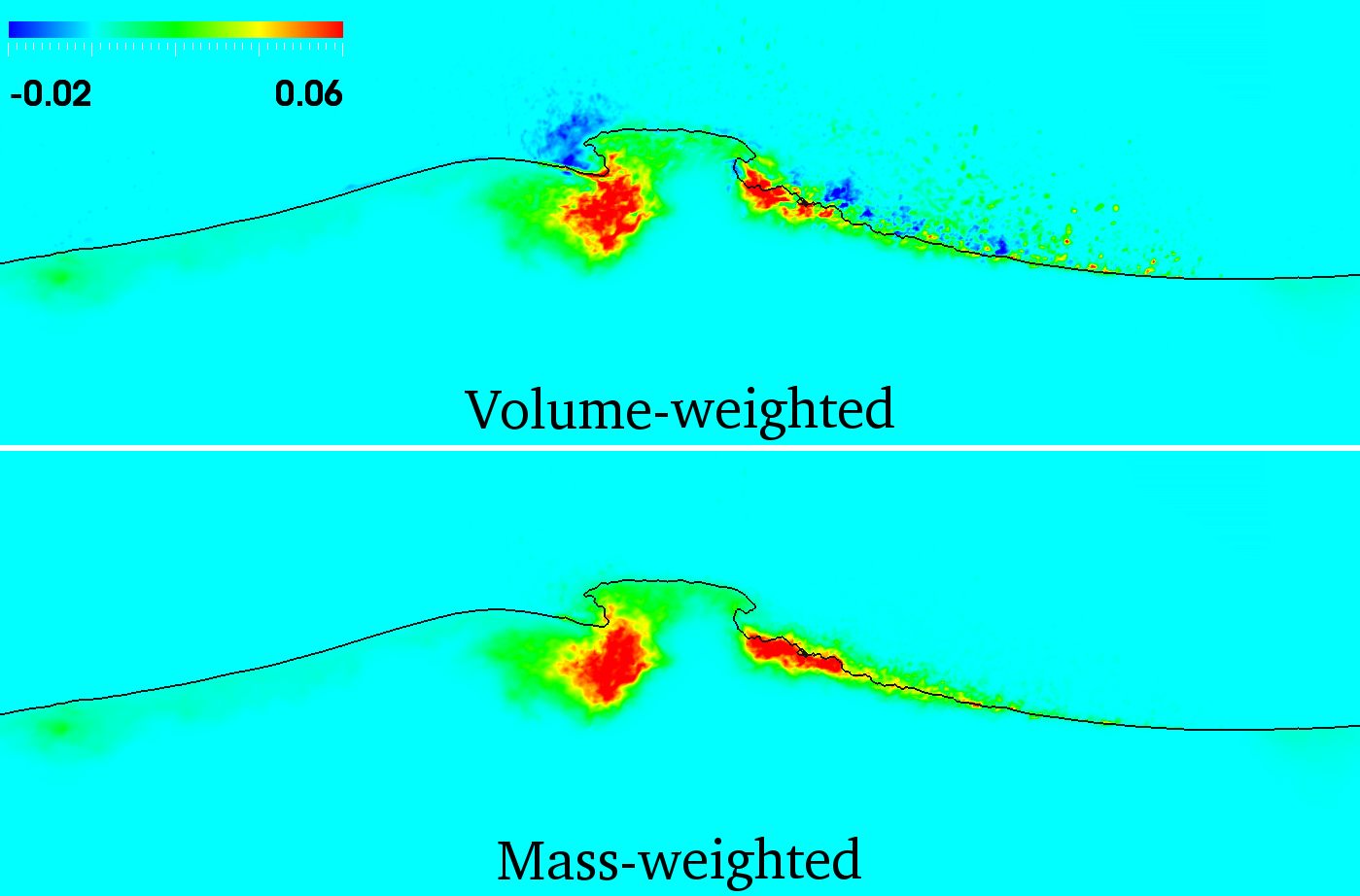}
\caption{\label{fig:energy} Difference in (a) m.k.e. and (b) t.k.e. using volume-weighted averages
Eqs.~\ref{eq:mkeV}-\ref{eq:tkeV} and mass-weighted averages Eqs.~\ref{eq:mkeF}-\ref{eq:tkeF}, for case
$VSP^F$ at t=18.50.} 
\end{figure}

Figure~\ref{fig:energy}(a)-(b) show the m.k.e.\ and t.k.e.\ using volume-weighted (Reynolds) and mass-weighted (Favre)
averages for case $VSP^F$ at t = 18.5. The differences in the m.k.e.\ are barely discernible, but the t.k.e.\ shows significant differences. In particular, the
t.k.e.\ based on Reynolds averaging is negative in several regions in the air, whereas the t.k.e.\ based on Favre averaging is positive everywhere. In addition, the magnitude of the
t.k.e.\ based on Favre averaging is much greater in the water than the air, but the same cannot be said for the
Reynolds-averaged t.k.e.. In the subsequent analysis mass-weighted (Favre) averaging is employed.

\subsection{Mean Kinetic Energy Balance}

An evolution equation for the mean kinetic energy can be written as \citep{chassaing02}:
\begin{multline} 
\left[ \frac{\partial}{\partial t}+
\frac{\partial}{\partial x_j} \tilde{u}_j \right]\left(m.k.e.\right)=\\
P+B+V-\frac{\partial T_{ik}}{\partial x_k}-\varepsilon \;\;, 
\label{eq:mke_evolution} 
\end{multline}
where $\varepsilon$ is the viscous dissipation rate, which is not
modeled here.  $P$ is the shear production of turbulence defined as:
\begin{equation} 
\label{eq:production}
P  \equiv  A_{ik} \frac{\partial \left<
\tilde{u}_i \right>}{\partial x_k} \; , 
\end{equation}
where
\begin{equation} 
\label{eq:tensor}
 A_{ik} = \left<\rho\right>
\left<u'_iu'_k\right> +\left<\rho'u'_iu'_k\right> -
\frac{\left<\rho'u'_i\right>\left<\rho'u'_k\right>}{\left< \rho \right>} .
\end{equation}
$B$ is the mean buoyant exchange between kinetic and potential energy, defined
as:
\begin{equation} 
\label{eq:buoyancy} B \equiv -
\frac{\left<\rho\right>\tilde{u}_3}{Fr^2}.  
\end{equation}
$V$ is the volume expansion work, defined as:
\begin{equation} 
\label{eq:expansion} 
V \equiv \left< p \right>
\frac{\partial}{\partial x_k} \left( \frac{\left< \rho' u'_k \right>}{\left<
\rho \right>}   \right) \, ,
\end{equation}
and $\frac{\partial T_{ik}}{\partial x_k}$ is the transport, defined as:
\begin{equation} 
\label{eq:transport} 
T_{ik} \equiv A_{ik} \tilde{u}_i + \left<
p \right> \tilde{u}_k \; .  
\end{equation}
Here, attention is returned to Figure~\ref{fig:energyrate}(a)-(d). Recall that the points labeled 1-8 are the points at which large
changes in the dissipation rate are observed. An analysis using the statistical framework developed above is
used to garner further insight into the physical processes responsible for the observed changes.

\begin{figure*}
\centering
{\bf A \hspace{135pt} B  \hspace{135pt} C }\\
\vspace{5pt}
\begin{minipage}[c]{0.01\textwidth}
\hspace{6pt}\\
\end{minipage}
\begin{minipage}[c]{0.98\textwidth}
\centering
\includegraphics[trim = 0mm 0cm 0mm
0cm,clip=true,angle=0,scale=0.159]{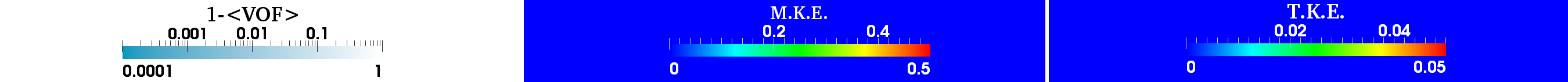}\\
\end{minipage}
\begin{minipage}[c]{0.01\textwidth}
{\bf 1}\\
\end{minipage}
\begin{minipage}[c]{0.98\textwidth}
\centering
\includegraphics[trim = 0mm .6cm 0mm
1.25cm,clip=true,angle=0,scale=0.159]{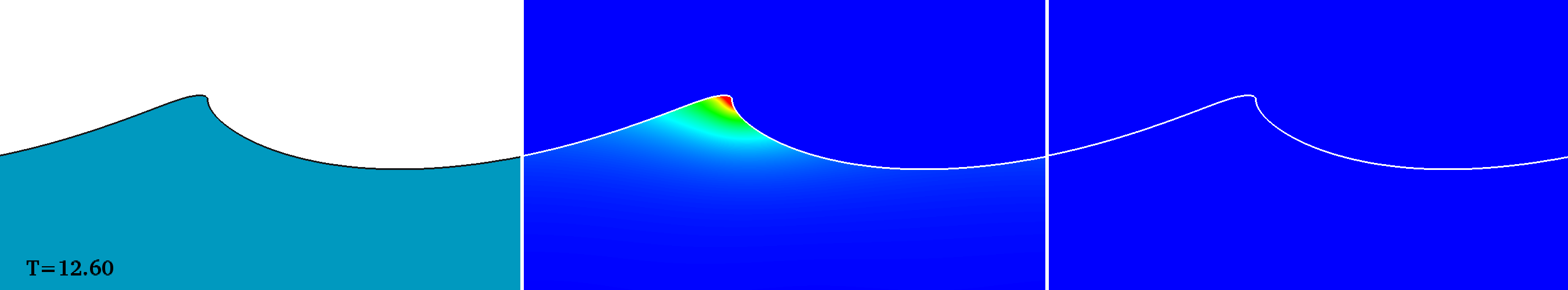}\\
\end{minipage}
\begin{minipage}[c]{0.01\textwidth}
{\bf 2}\\
\end{minipage}
\begin{minipage}[c]{0.98\textwidth}
\centering
\includegraphics[trim = 0mm .6cm 0mm
1.25cm,clip=true,angle=0,scale=0.159]{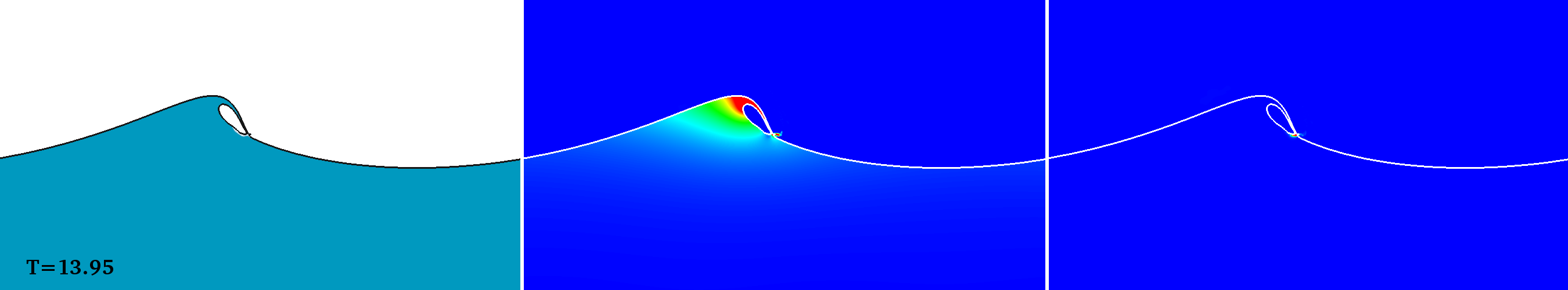}\\
\end{minipage}
\begin{minipage}[c]{0.01\textwidth}
{\bf 3}\\
\end{minipage}
\begin{minipage}[c]{0.98\textwidth}
\centering
\includegraphics[trim = 0mm .6cm 0mm
1.25cm,clip=true,angle=0,scale=0.159]{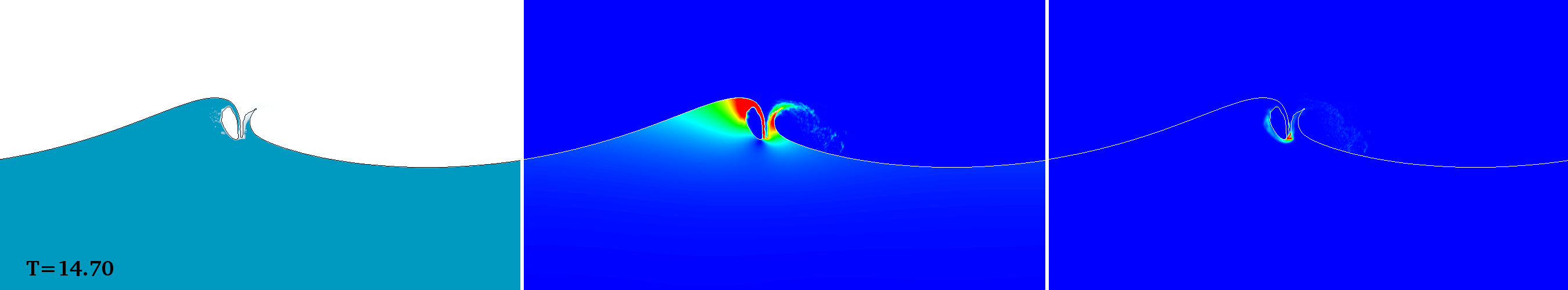}\\
\end{minipage}
\begin{minipage}[c]{0.01\textwidth}
{\bf 4}\\
\end{minipage}
\begin{minipage}[c]{0.98\textwidth}
\centering
\includegraphics[trim = 0mm .6cm 0mm
1.25cm,clip=true,angle=0,scale=0.159]{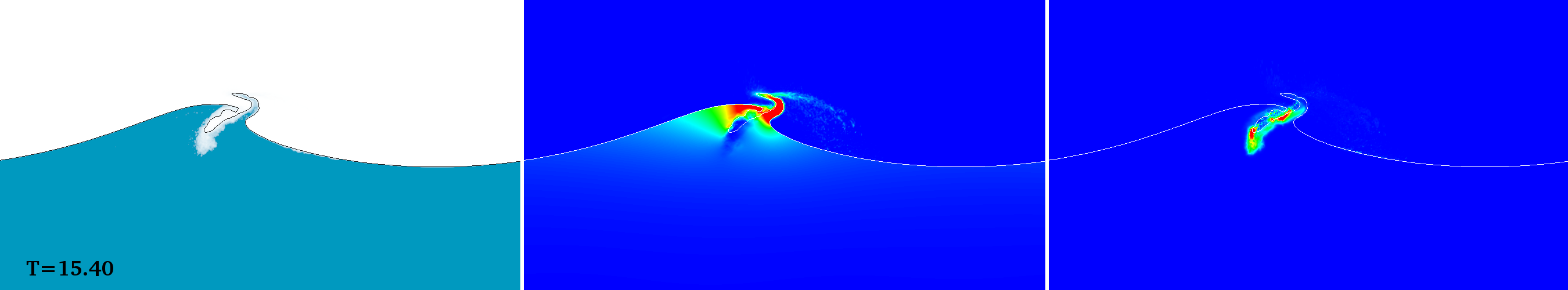}\\
\end{minipage}
\begin{minipage}[c]{0.01\textwidth}
{\bf 5}\\
\end{minipage}
\begin{minipage}[c]{0.98\textwidth}
\centering
\includegraphics[trim = 0mm .6cm 0mm
1.25cm,clip=true,angle=0,scale=0.159]{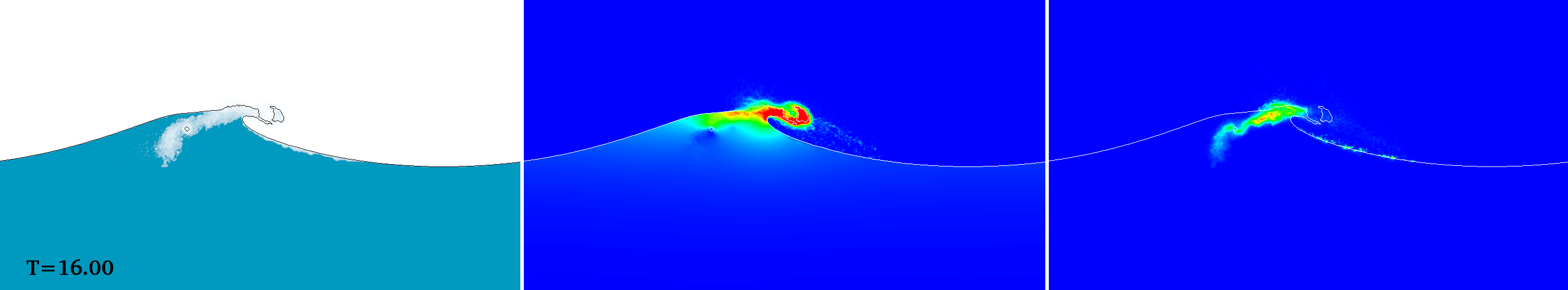}\\
\end{minipage}
\begin{minipage}[c]{0.01\textwidth}
{\bf 6}\\
\end{minipage}
\begin{minipage}[c]{0.98\textwidth}
\centering
\includegraphics[trim = 0mm .6cm 0mm
1.25cm,clip=true,angle=0,scale=0.159]{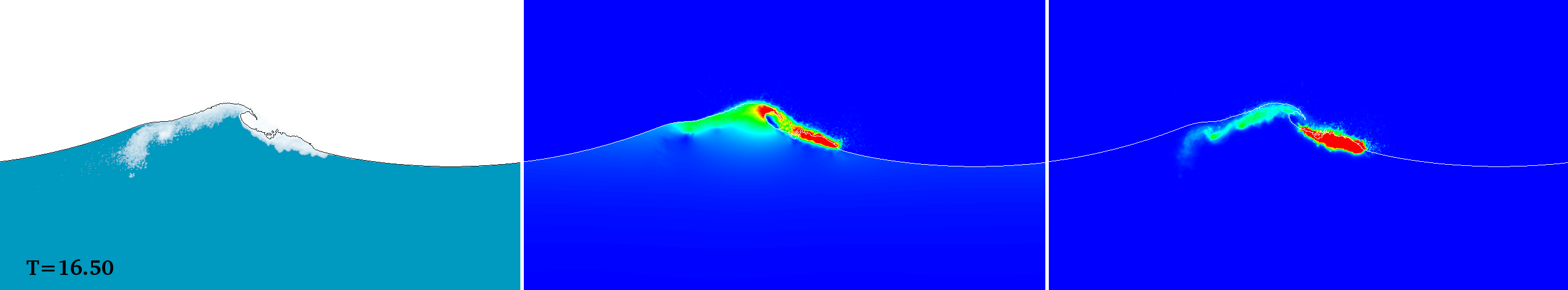}\\
\end{minipage}
\begin{minipage}[c]{0.01\textwidth}
{\bf 7}\\
\end{minipage}
\begin{minipage}[c]{0.98\textwidth} 
\centering
\includegraphics[trim = 0mm .6cm 0mm
1.25cm,clip=true,angle=0,scale=0.159]{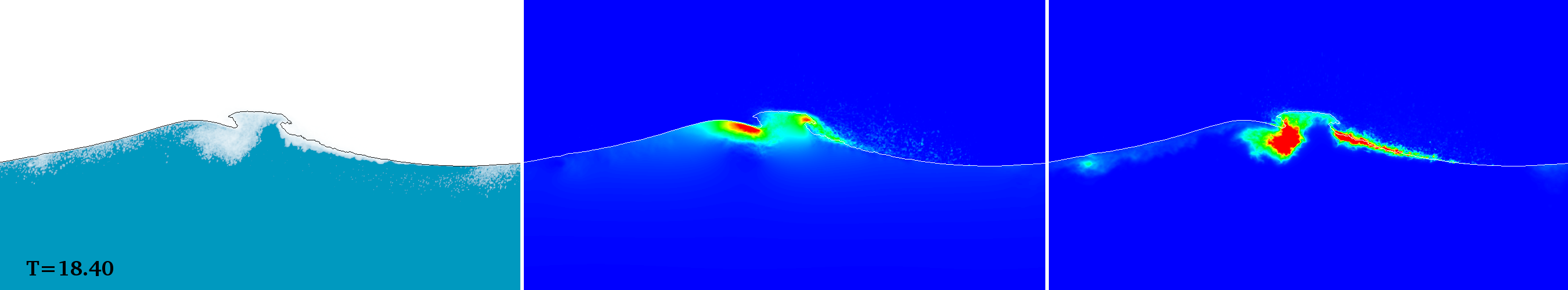}\\
\end{minipage}
\begin{minipage}[c]{0.01\textwidth}
{\bf 8}\\
\end{minipage}
\begin{minipage}[c]{0.98\textwidth} 
\centering
\includegraphics[trim = 0mm .6cm 0mm
1.25cm,clip=true,angle=0,scale=0.159]{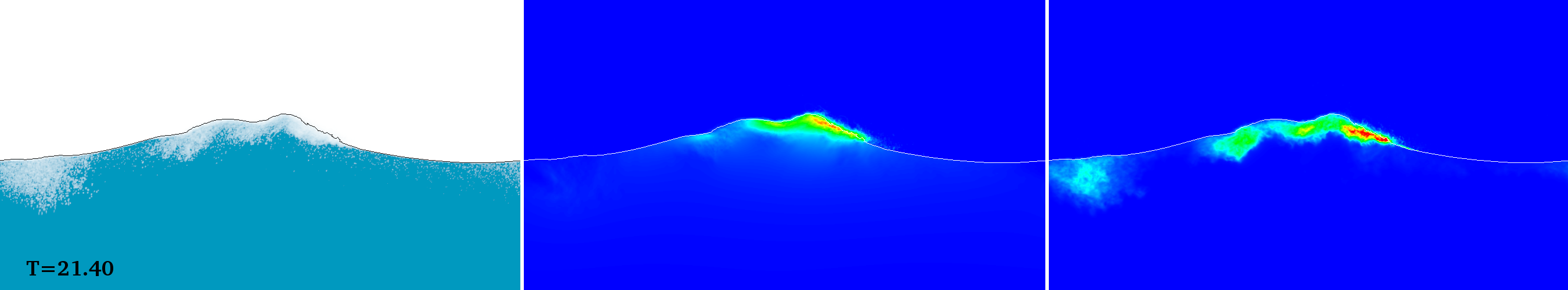}\\
\end{minipage}
\caption{\label{fig:vsp_vmt} Volume of Air, Mean Kinetic Energy (Eq. \ref{eq:mkeF}), and
Turbulent Kinetic Energy (Eq. \ref{eq:tkeF}) for case $VSP^F$. The non-dimensional time is given in
the lower left corner of column A and the contour levels are given at
the top of each column.}
\end{figure*}

\begin{figure*}
\centering
{\bf A \hspace{135pt} B  \hspace{135pt} C }\\
\vspace{5pt}
\begin{minipage}[c]{0.01\textwidth}
\hspace{6pt}\\
\end{minipage}
\begin{minipage}[c]{0.98\textwidth}
\centering
\includegraphics[trim = 0mm 0cm 0mm
0cm,clip=true,angle=0,scale=0.159]{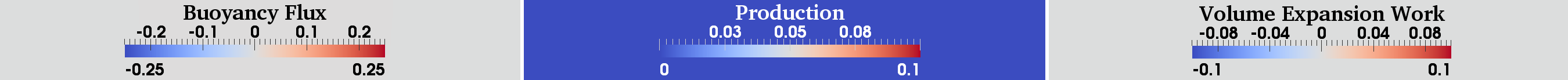}\\
\end{minipage}
\begin{minipage}[c]{0.01\textwidth}
{\bf 1}\\
\end{minipage}
\begin{minipage}[c]{0.98\textwidth}
\centering
\includegraphics[trim = 0mm .6cm 0mm
1.25cm,clip=true,angle=0,scale=0.159]{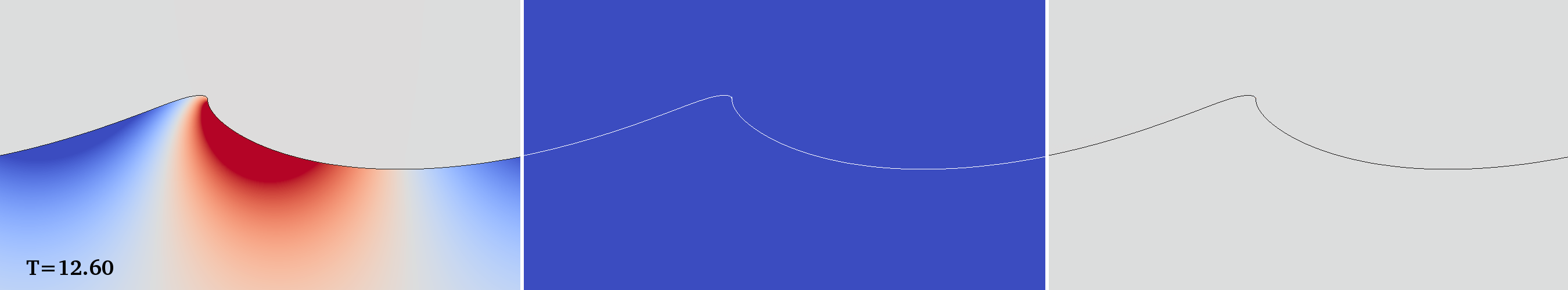}\\
\end{minipage}
\begin{minipage}[c]{0.01\textwidth}
{\bf 2}\\
\end{minipage}
\begin{minipage}[c]{0.98\textwidth}
\centering
\includegraphics[trim = 0mm .6cm 0mm
1.25cm,clip=true,angle=0,scale=0.159]{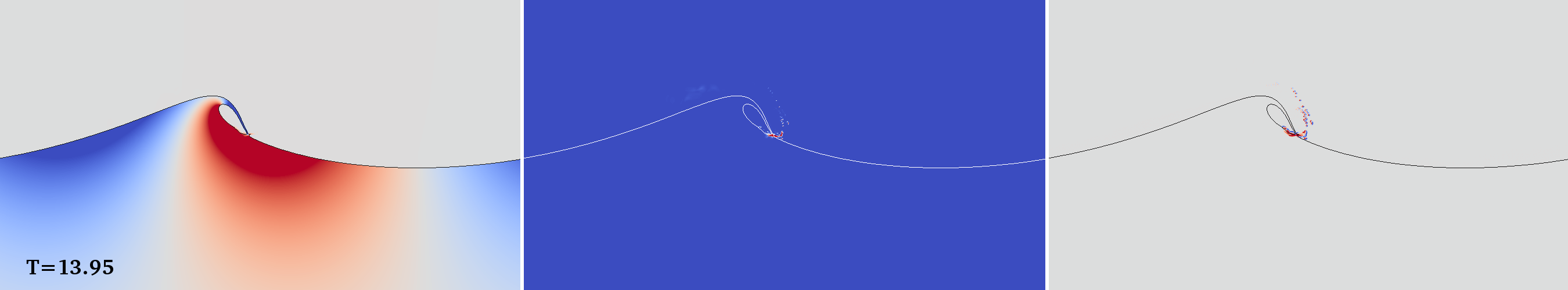}\\
\end{minipage}
\begin{minipage}[c]{0.01\textwidth}
{\bf 3}\\
\end{minipage}
\begin{minipage}[c]{0.98\textwidth}
\centering
\includegraphics[trim = 0mm .6cm 0mm
1.25cm,clip=true,angle=0,scale=0.159]{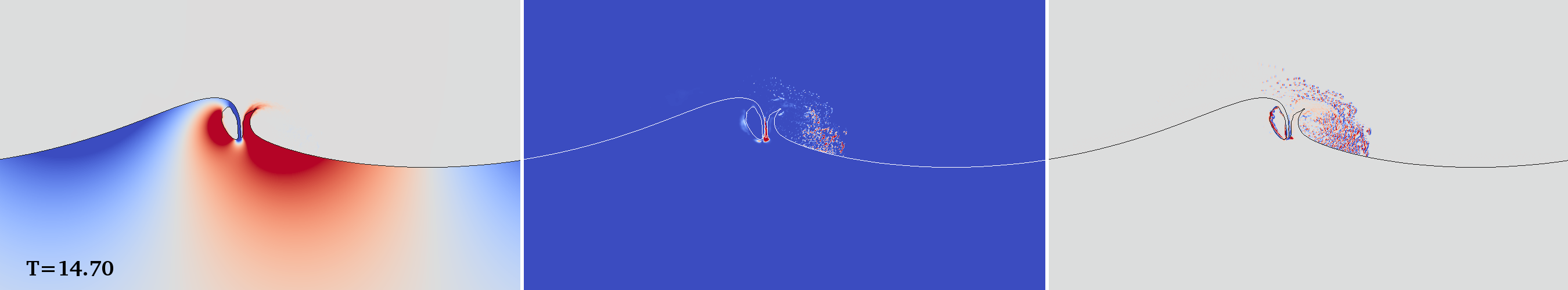}\\
\end{minipage}
\begin{minipage}[c]{0.01\textwidth}
{\bf 4}\\
\end{minipage}
\begin{minipage}[c]{0.98\textwidth}
\centering
\includegraphics[trim = 0mm .6cm 0mm
1.25cm,clip=true,angle=0,scale=0.159]{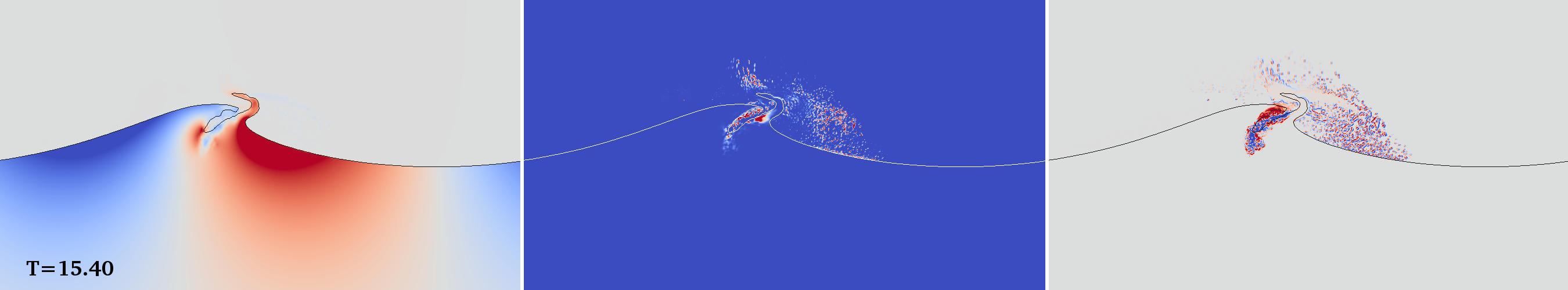}\\
\end{minipage}
\begin{minipage}[c]{0.01\textwidth}
{\bf 5}\\
\end{minipage}
\begin{minipage}[c]{0.98\textwidth}
\centering
\includegraphics[trim = 0mm .6cm 0mm
1.25cm,clip=true,angle=0,scale=0.159]{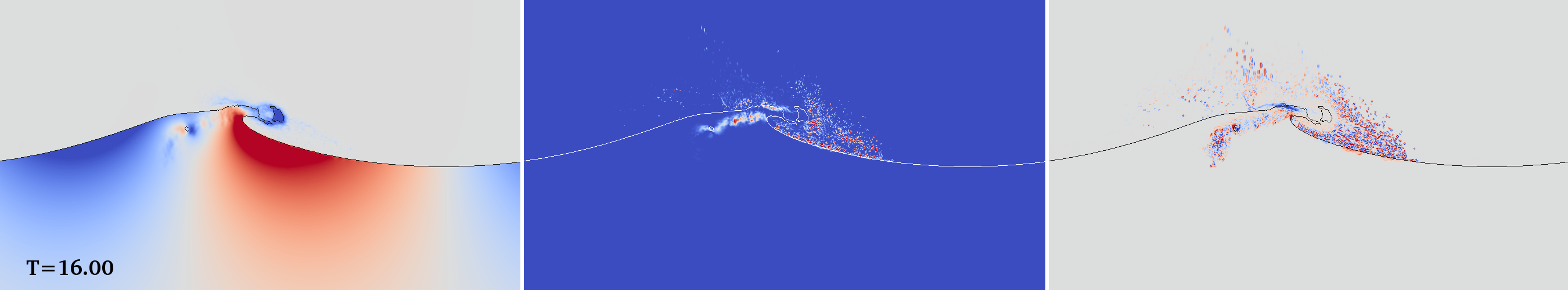}\\
\end{minipage}
\begin{minipage}[c]{0.01\textwidth}
{\bf 6}\\
\end{minipage}
\begin{minipage}[c]{0.98\textwidth}
\centering
\includegraphics[trim = 0mm .6cm 0mm
1.25cm,clip=true,angle=0,scale=0.159]{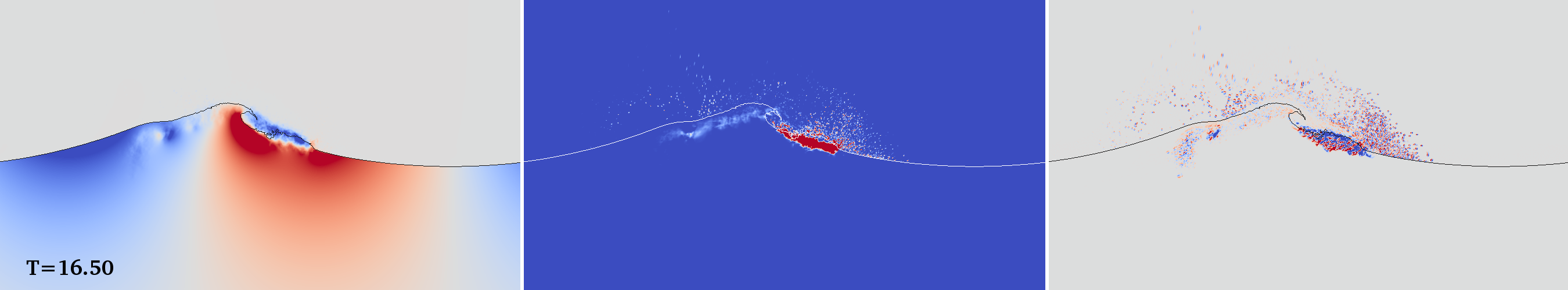}\\
\end{minipage}
\begin{minipage}[c]{0.01\textwidth}
{\bf 7}\\
\end{minipage}
\begin{minipage}[c]{0.98\textwidth} 
\centering
\includegraphics[trim = 0mm .6cm 0mm
1.25cm,clip=true,angle=0,scale=0.159]{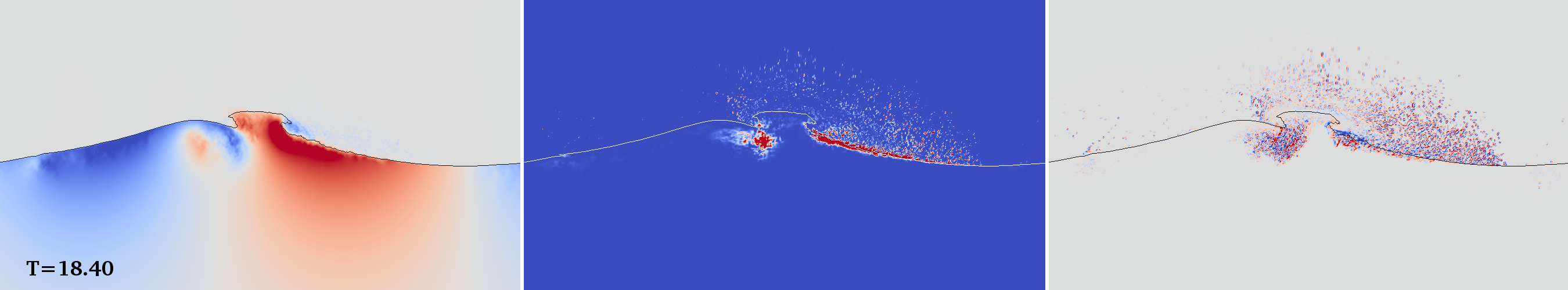}\\
\end{minipage}
\begin{minipage}[c]{0.01\textwidth}
{\bf 8}\\
\end{minipage}
\begin{minipage}[c]{0.98\textwidth} 
\centering
\includegraphics[trim = 0mm .6cm 0mm
1.25cm,clip=true,angle=0,scale=0.159]{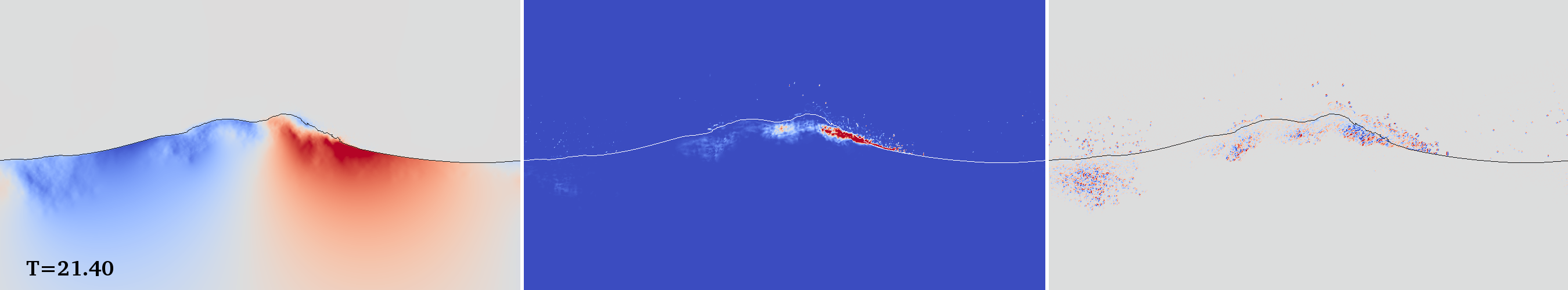}\\
\end{minipage}
\caption{\label{fig:vsp_bpv}
Buoyant Flux, Shear Production of T.K.E., Volume Expansion Work
for case $VSP^F$.   The non-dimensional time is given in
the lower left corner of column A and the contour levels are given at
the top of each column.}
\end{figure*}

\begin{figure*}
\centering
{\bf A \hspace{135pt} B  \hspace{135pt} C }\\
\vspace{5pt}
\begin{minipage}[c]{0.01\textwidth}
\hspace{6pt}\\
\end{minipage}
\begin{minipage}[c]{0.98\textwidth}
\centering
\includegraphics[trim = 0mm 0cm 0mm
0cm,clip=true,angle=0,scale=0.159]{figures/contours/VSP/vsp_sp_vmt_legend.png}\\
\end{minipage}
\begin{minipage}[c]{0.01\textwidth}
{\bf 1}\\
\end{minipage}
\begin{minipage}[c]{0.98\textwidth}
\centering
\includegraphics[trim = 0mm .6cm 0mm
1.25cm,clip=true,angle=0,scale=0.159]{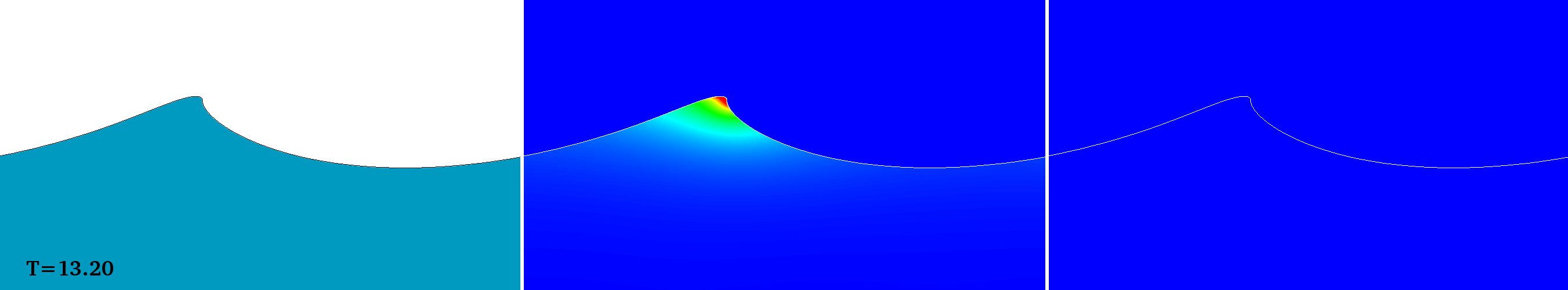}\\
\end{minipage}
\begin{minipage}[c]{0.01\textwidth}
{\bf 2}\\
\end{minipage}
\begin{minipage}[c]{0.98\textwidth}
\centering
\includegraphics[trim = 0mm .6cm 0mm
1.25cm,clip=true,angle=0,scale=0.159]{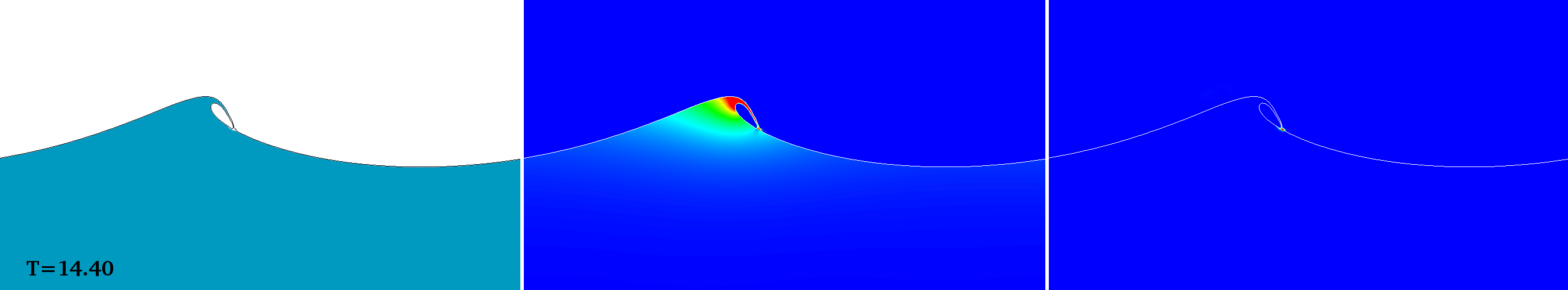}\\
\end{minipage}
\begin{minipage}[c]{0.01\textwidth}
{\bf 3}\\
\end{minipage}
\begin{minipage}[c]{0.98\textwidth}
\centering
\includegraphics[trim = 0mm .6cm 0mm
1.25cm,clip=true,angle=0,scale=0.159]{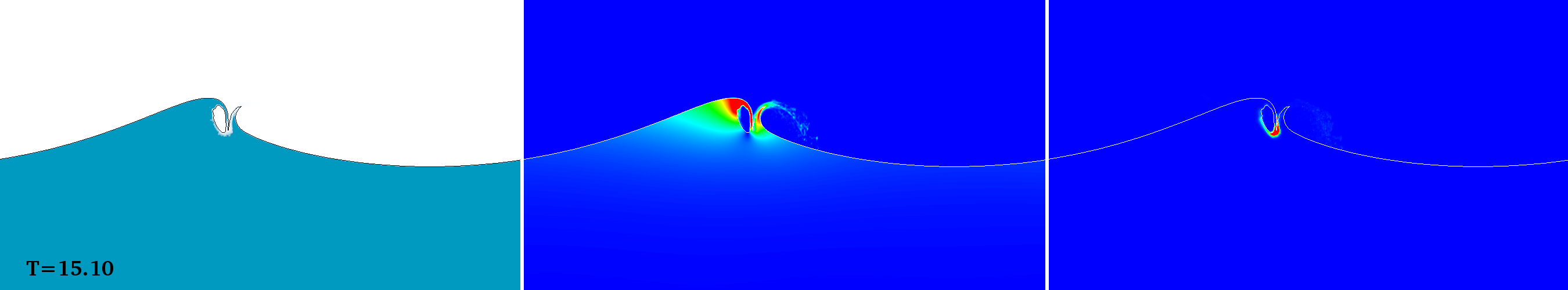}\\
\end{minipage}
\begin{minipage}[c]{0.01\textwidth}
{\bf 4}\\
\end{minipage}
\begin{minipage}[c]{0.98\textwidth}
\centering
\includegraphics[trim = 0mm .6cm 0mm
1.25cm,clip=true,angle=0,scale=0.159]{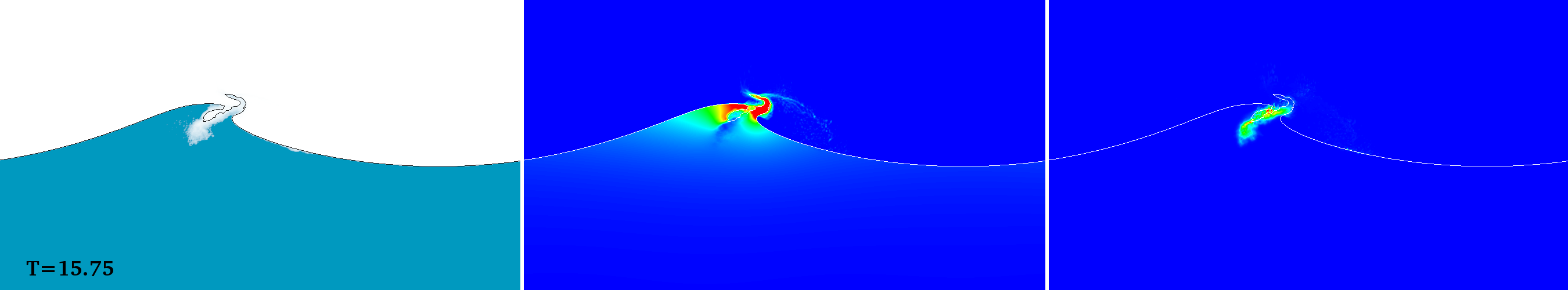}\\
\end{minipage}
\begin{minipage}[c]{0.01\textwidth}
{\bf 5}\\
\end{minipage}
\begin{minipage}[c]{0.98\textwidth}
\centering
\includegraphics[trim = 0mm .6cm 0mm
1.25cm,clip=true,angle=0,scale=0.159]{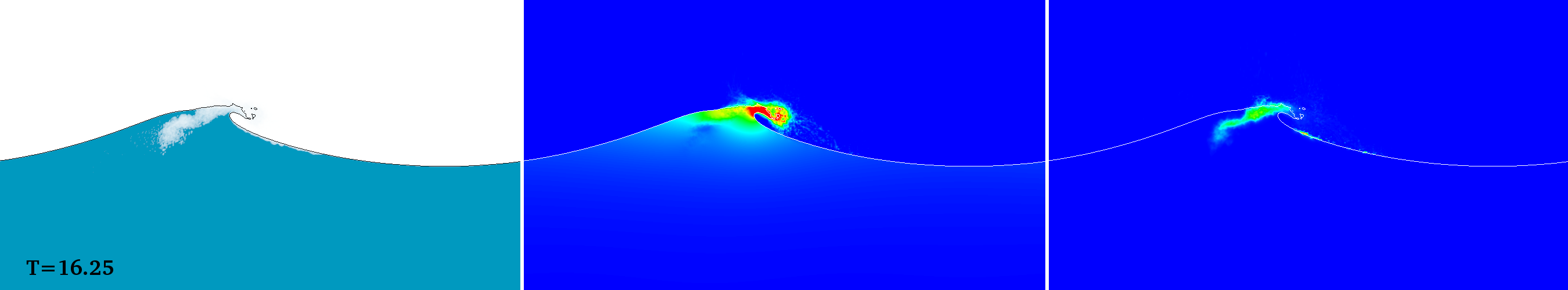}\\
\end{minipage}
\begin{minipage}[c]{0.01\textwidth}
{\bf 6}\\
\end{minipage}
\begin{minipage}[c]{0.98\textwidth}
\centering
\includegraphics[trim = 0mm .6cm 0mm
1.25cm,clip=true,angle=0,scale=0.159]{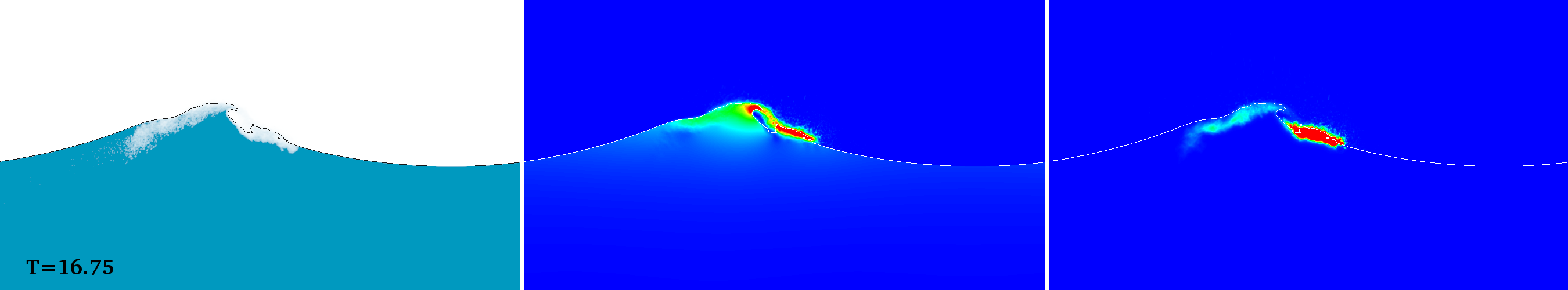}\\
\end{minipage}
\begin{minipage}[c]{0.01\textwidth}
{\bf 7}\\
\end{minipage}
\begin{minipage}[c]{0.98\textwidth} 
\centering
\includegraphics[trim = 0mm .6cm 0mm
1.25cm,clip=true,angle=0,scale=0.159]{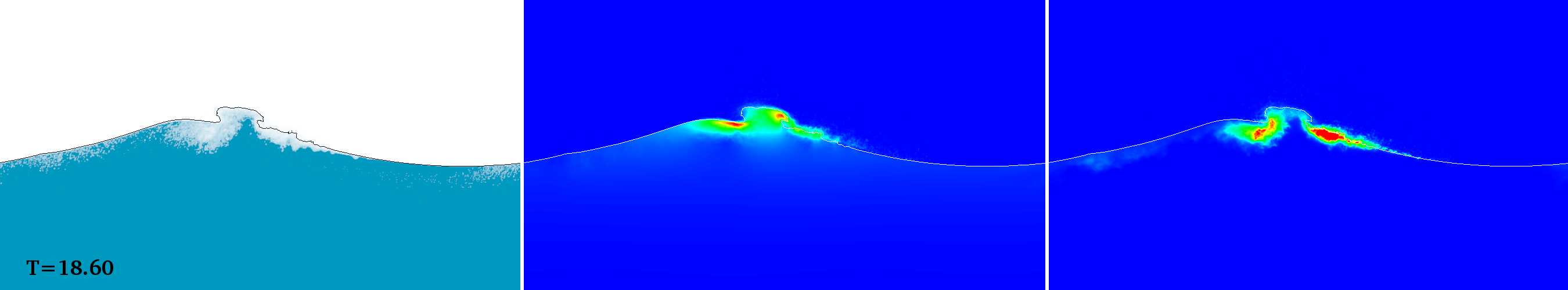}\\
\end{minipage}
\begin{minipage}[c]{0.01\textwidth}
{\bf 8}\\
\end{minipage}
\begin{minipage}[c]{0.98\textwidth} 
\centering
 \includegraphics[trim = 0mm .6cm 0mm
1.25cm,clip=true,angle=0,scale=0.159]{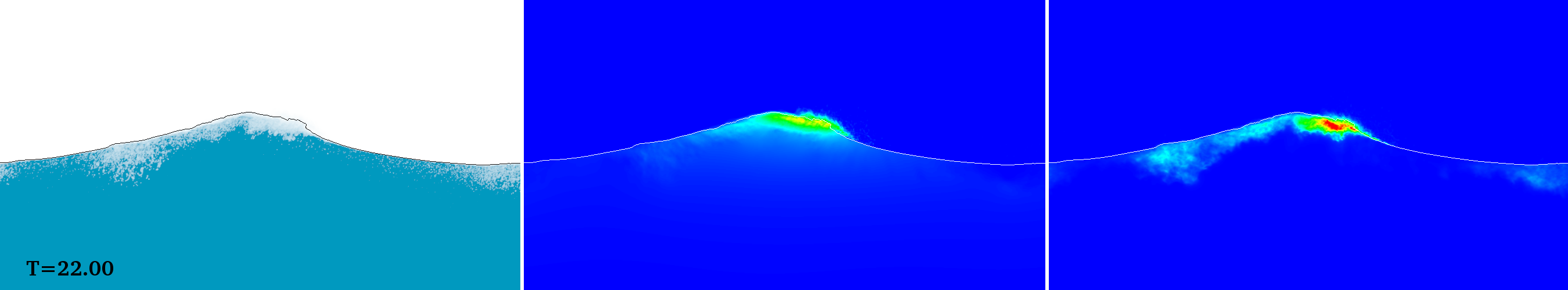}\\
\end{minipage}
\caption{\label{fig:sp_vmt}
Volume of Air, Mean Kinetic Energy (Eq. \ref{eq:mkeF}), and
Turbulent Kinetic Energy (Eq. \ref{eq:tkeF}) for case $SP^F$. The non-dimensional time is given in
the lower left corner of column A and the contour levels are given at
the top of each column.}
\end{figure*}

\begin{figure*}
\centering
{\bf A \hspace{135pt} B  \hspace{135pt} C }\\
\vspace{5pt}
\begin{minipage}[c]{0.01\textwidth}
\hspace{6pt}\\
\end{minipage}
\begin{minipage}[c]{0.98\textwidth}
\centering
\includegraphics[trim = 0mm 0cm 0mm
0cm,clip=true,angle=0,scale=0.159]{figures/contours/VSP/vsp_sp_bpv_legend.png}\\
\end{minipage}
\begin{minipage}[c]{0.01\textwidth}
{\bf 1}\\
\end{minipage}
\begin{minipage}[c]{0.98\textwidth}
\centering
\includegraphics[trim = 0mm .6cm 0mm
1.25cm,clip=true,angle=0,scale=0.159]{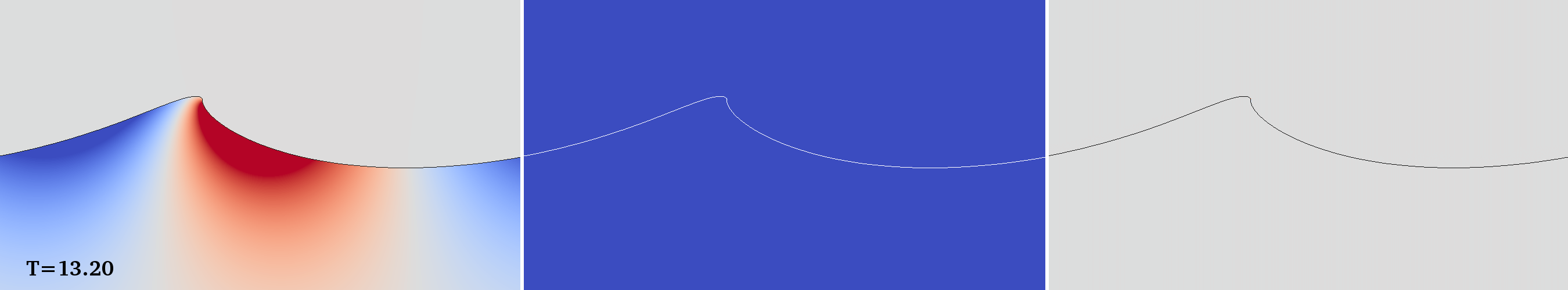}\\
\end{minipage}
\begin{minipage}[c]{0.01\textwidth}
{\bf 2}\\
\end{minipage}
\begin{minipage}[c]{0.98\textwidth}
\centering
\includegraphics[trim = 0mm .6cm 0mm
1.25cm,clip=true,angle=0,scale=0.159]{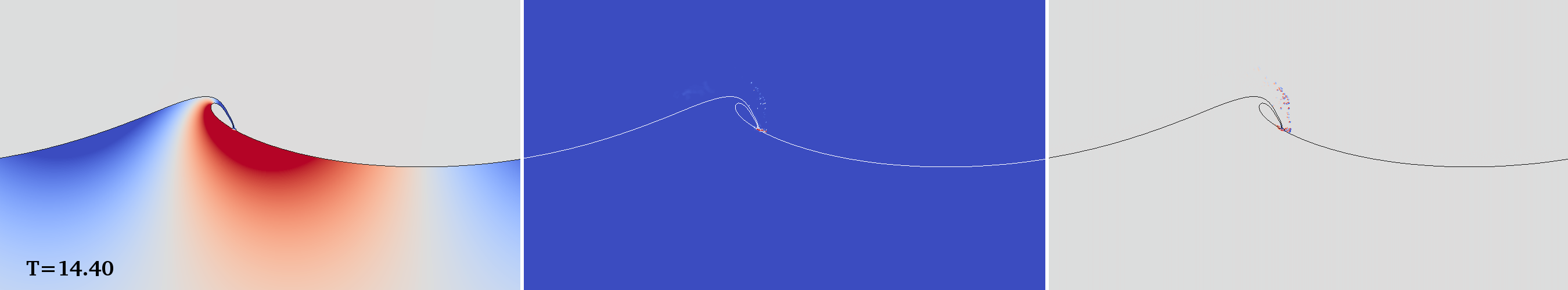}\\
\end{minipage}
\begin{minipage}[c]{0.01\textwidth}
{\bf 3}\\
\end{minipage}
\begin{minipage}[c]{0.98\textwidth}
\centering
\includegraphics[trim = 0mm .6cm 0mm
1.25cm,clip=true,angle=0,scale=0.159]{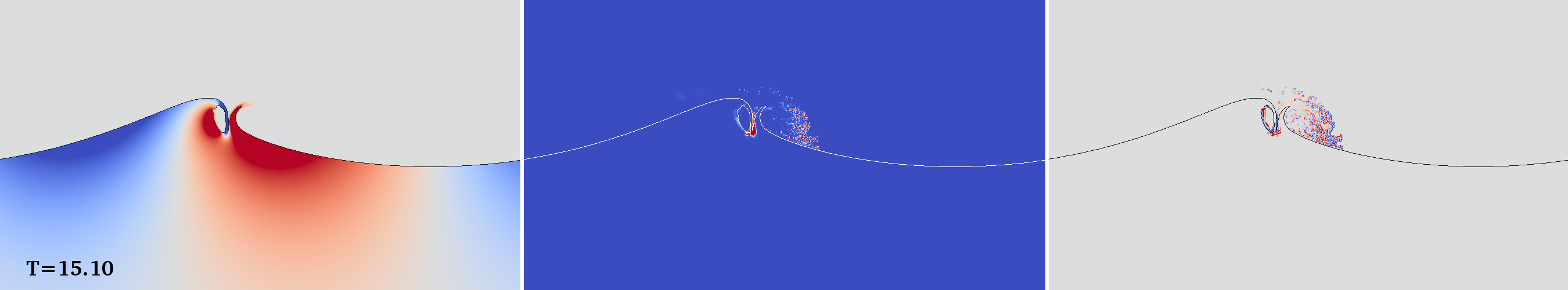}\\
\end{minipage}
\begin{minipage}[c]{0.01\textwidth}
{\bf 4}\\
\end{minipage}
\begin{minipage}[c]{0.98\textwidth}
\centering
\includegraphics[trim = 0mm .6cm 0mm
1.25cm,clip=true,angle=0,scale=0.159]{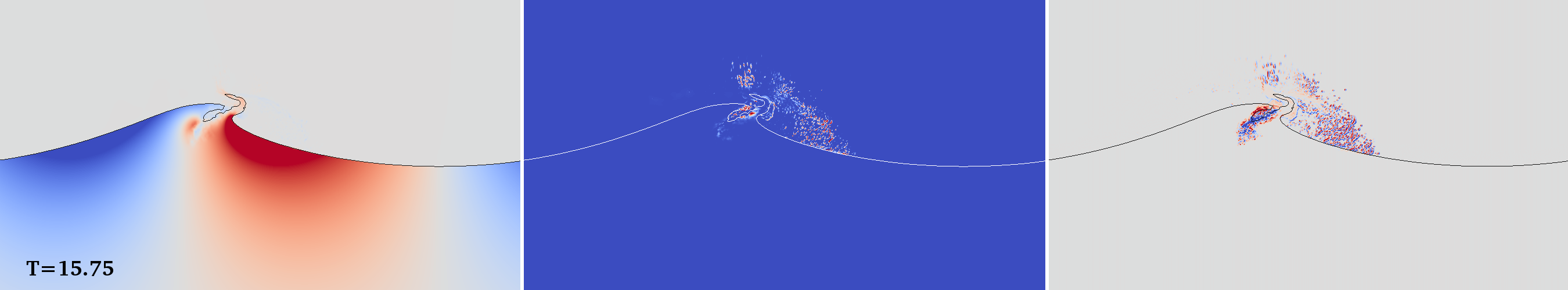}\\
\end{minipage}
\begin{minipage}[c]{0.01\textwidth}
{\bf 5}\\
\end{minipage}
\begin{minipage}[c]{0.98\textwidth}
\centering
\includegraphics[trim = 0mm .6cm 0mm
1.25cm,clip=true,angle=0,scale=0.159]{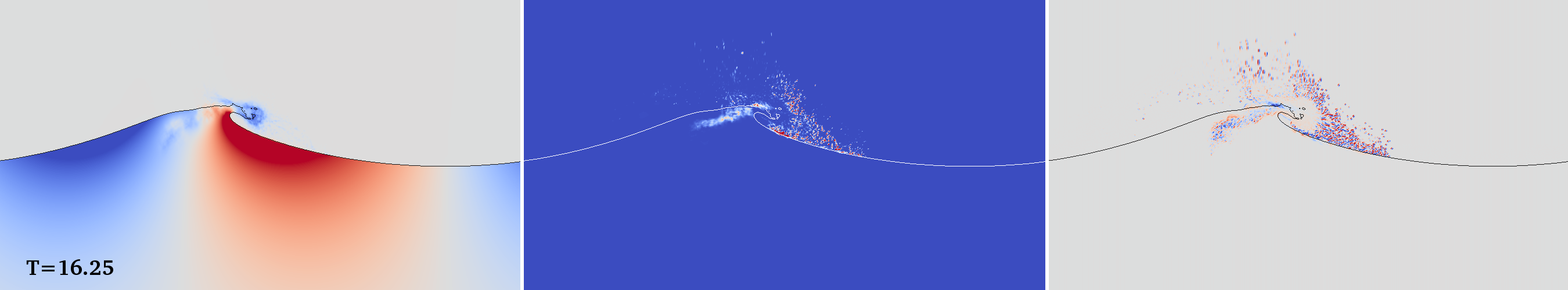}\\
\end{minipage}
\begin{minipage}[c]{0.01\textwidth}
{\bf 6}\\
\end{minipage}
\begin{minipage}[c]{0.98\textwidth}
\centering
\includegraphics[trim = 0mm .6cm 0mm
1.25cm,clip=true,angle=0,scale=0.159]{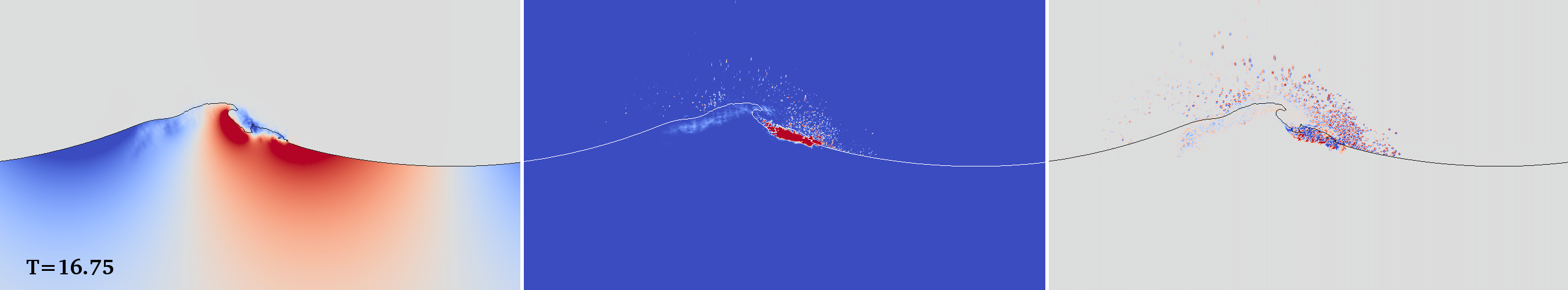}\\
\end{minipage}
\begin{minipage}[c]{0.01\textwidth}
{\bf 7}\\
\end{minipage}
\begin{minipage}[c]{0.98\textwidth} 
\centering
\includegraphics[trim = 0mm .6cm 0mm
1.25cm,clip=true,angle=0,scale=0.159]{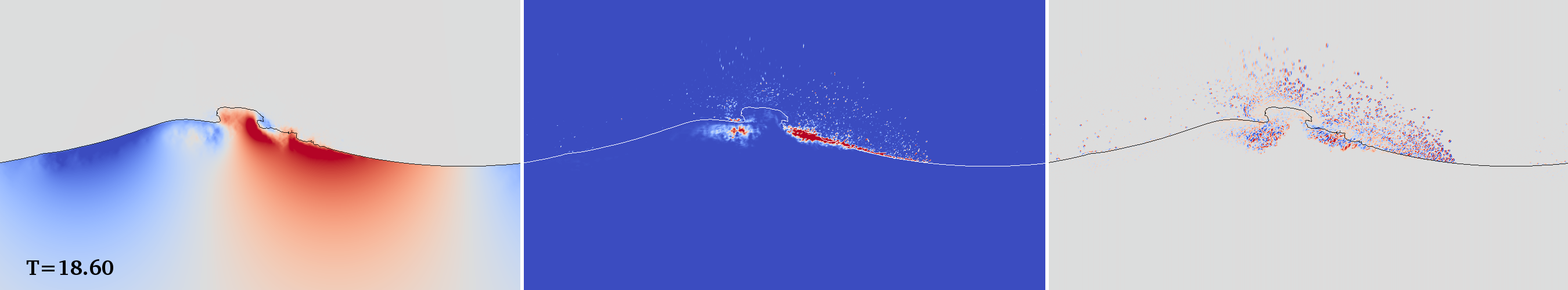}\\
\end{minipage}
\begin{minipage}[c]{0.01\textwidth}
{\bf 8}\\
\end{minipage}
\begin{minipage}[c]{0.98\textwidth} 
\centering
 \includegraphics[trim = 0mm .6cm 0mm
1.25cm,clip=true,angle=0,scale=0.159]{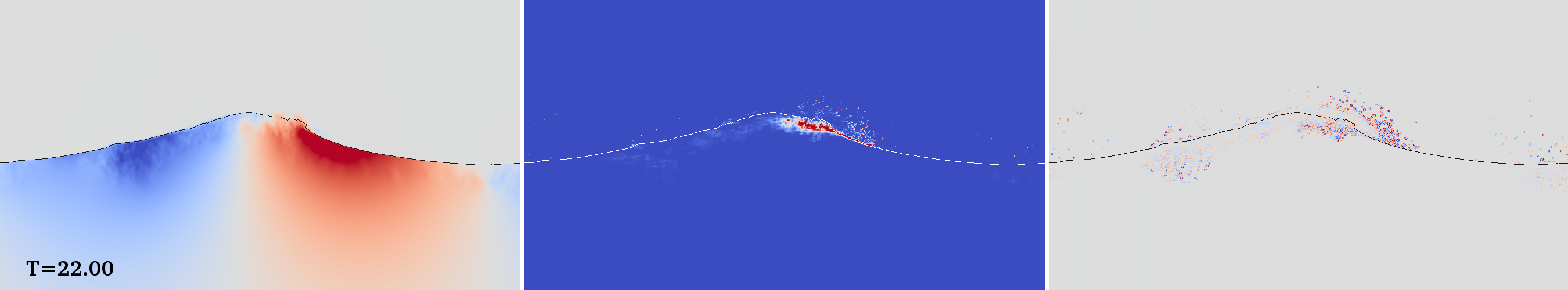}\\
\end{minipage}
\caption{\label{fig:sp_bpv}
Buoyant Flux, Shear Production of T.K.E., Volume Expansion Work
for case $SP^F$.   The non-dimensional time is given in
the lower left corner of column A and the contour levels are given at
the top of each column.}
\end{figure*}

\begin{figure*}
\centering
{\bf A \hspace{135pt} B  \hspace{135pt} C }\\
\vspace{5pt}
\begin{minipage}[c]{0.01\textwidth}
\hspace{6pt}\\
\end{minipage}
\begin{minipage}[c]{0.98\textwidth}
\centering
\includegraphics[trim = 0mm 0cm 0mm
0cm,clip=true,angle=0,scale=0.159]{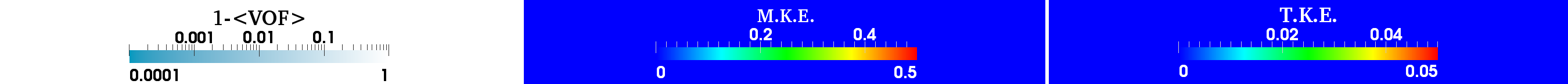}\\
\end{minipage}
\begin{minipage}[c]{0.01\textwidth}
{\bf 1}\\
\end{minipage}
\begin{minipage}[c]{0.98\textwidth}
\centering
\includegraphics[trim = 0mm .6cm 0mm
1.25cm,clip=true,angle=0,scale=0.159]{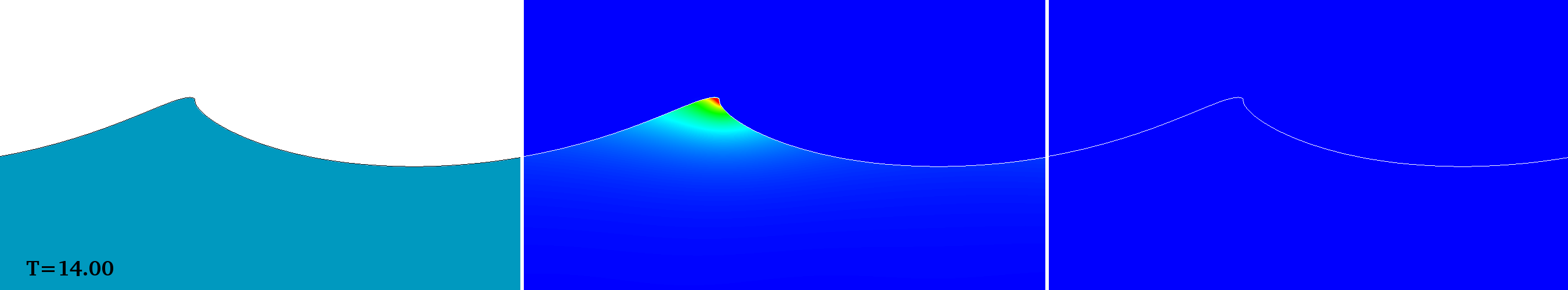}\\
\end{minipage}
\begin{minipage}[c]{0.01\textwidth}
{\bf 2}\\
\end{minipage}
\begin{minipage}[c]{0.98\textwidth}
\centering
\includegraphics[trim = 0mm .6cm 0mm
1.25cm,clip=true,angle=0,scale=0.159]{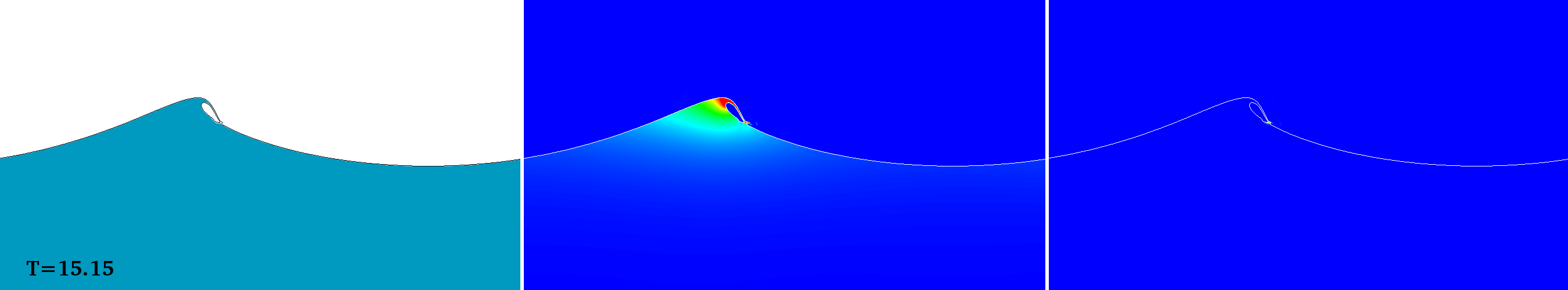}\\
\end{minipage}
\begin{minipage}[c]{0.01\textwidth}
{\bf 3}\\
\end{minipage}
\begin{minipage}[c]{0.98\textwidth}
\centering
\includegraphics[trim = 0mm .6cm 0mm
1.25cm,clip=true,angle=0,scale=0.159]{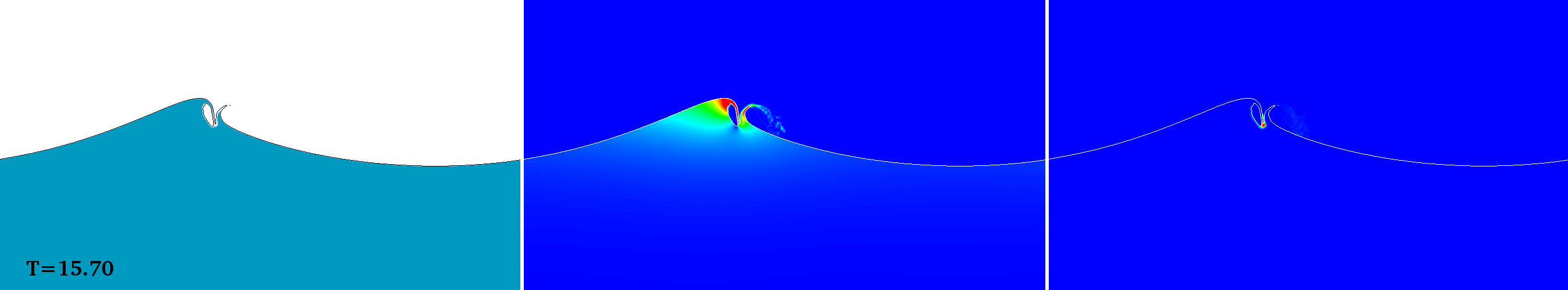}\\
\end{minipage}
\begin{minipage}[c]{0.01\textwidth}
{\bf 4}\\
\end{minipage}
\begin{minipage}[c]{0.98\textwidth}
\centering
\includegraphics[trim = 0mm .6cm 0mm
1.25cm,clip=true,angle=0,scale=0.159]{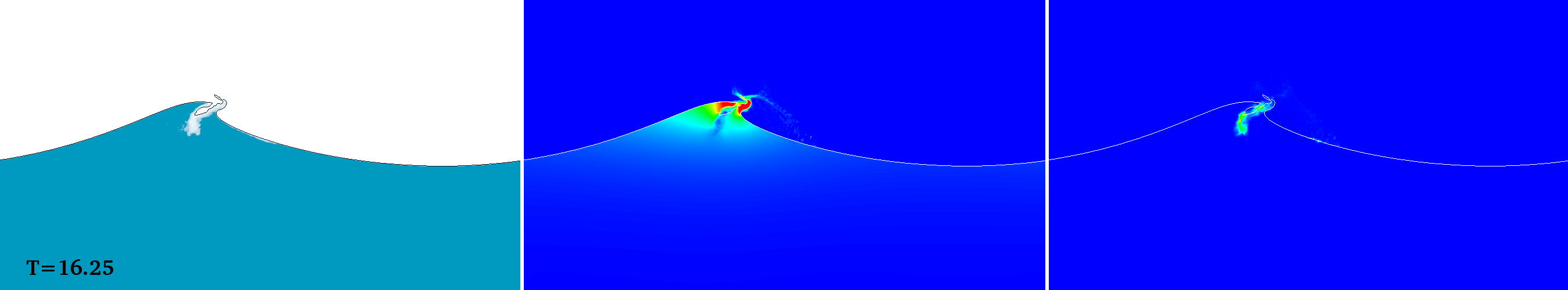}\\
\end{minipage}
\begin{minipage}[c]{0.01\textwidth}
{\bf 5}\\
\end{minipage}
\begin{minipage}[c]{0.98\textwidth}
\centering
\includegraphics[trim = 0mm .6cm 0mm
1.25cm,clip=true,angle=0,scale=0.159]{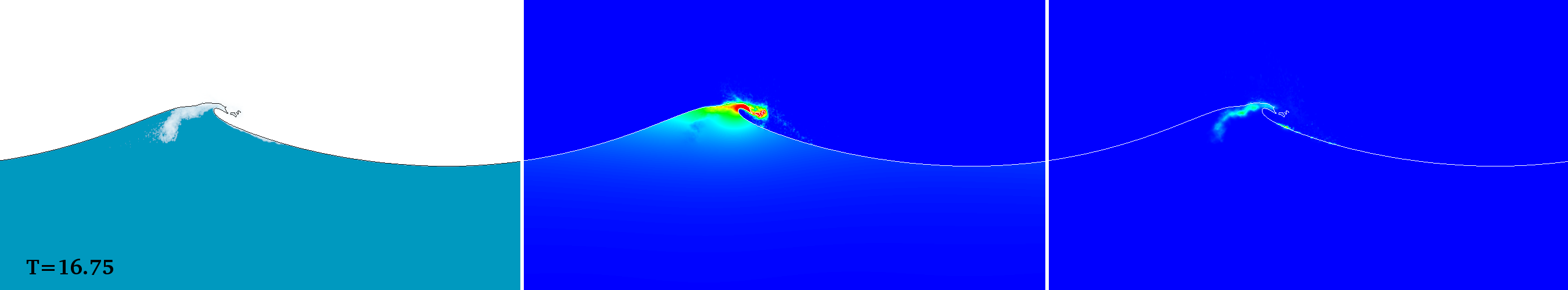}\\
\end{minipage}
\begin{minipage}[c]{0.01\textwidth}
{\bf 6}\\
\end{minipage}
\begin{minipage}[c]{0.98\textwidth}
\centering
\includegraphics[trim = 0mm .6cm 0mm
1.25cm,clip=true,angle=0,scale=0.159]{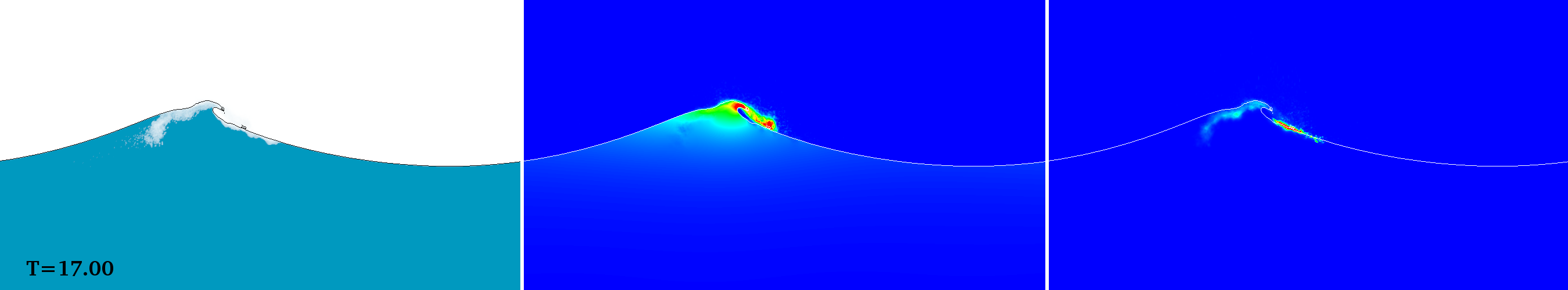}\\
\end{minipage}
\begin{minipage}[c]{0.01\textwidth}
{\bf 7}\\
\end{minipage}
\begin{minipage}[c]{0.98\textwidth} 
\centering
\includegraphics[trim = 0mm .6cm 0mm
1.25cm,clip=true,angle=0,scale=0.159]{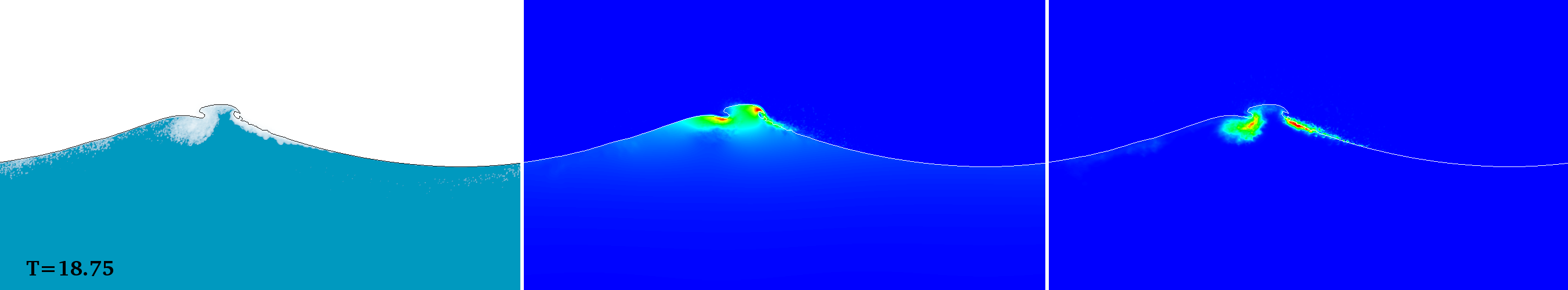}\\
\end{minipage}
\begin{minipage}[c]{0.01\textwidth}
{\bf 8}\\
\end{minipage}
\begin{minipage}[c]{0.98\textwidth} 
\centering
 \includegraphics[trim = 0mm .6cm 0mm
1.25cm,clip=true,angle=0,scale=0.159]{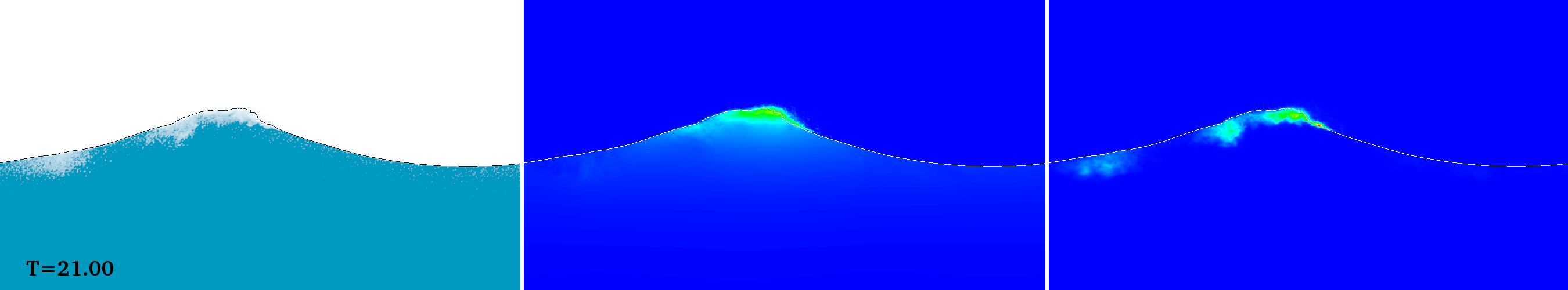}\\
\end{minipage}
\caption{\label{fig:p_vmt}
Volume of Air, Mean Kinetic Energy (Eq. \ref{eq:mkeF}), and
Turbulent Kinetic Energy (Eq. \ref{eq:tkeF}) for case $P^F$. The non-dimensional time is given in
the lower left corner of column A and the contour levels are given at
the top of each column.}
\end{figure*}

\begin{figure*}
\centering
{\bf A \hspace{135pt} B  \hspace{135pt} C }\\
\vspace{5pt}
\begin{minipage}[c]{0.01\textwidth}
\hspace{6pt}\\
\end{minipage}
\begin{minipage}[c]{0.98\textwidth}
\centering
\includegraphics[trim = 0mm 0cm 0mm
0cm,clip=true,angle=0,scale=0.159]{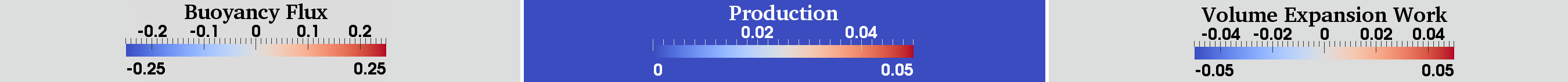}\\
\end{minipage}
\begin{minipage}[c]{0.01\textwidth}
{\bf 1}\\
\end{minipage}
\begin{minipage}[c]{0.98\textwidth}
\centering
\includegraphics[trim = 0mm .6cm 0mm
1.25cm,clip=true,angle=0,scale=0.159]{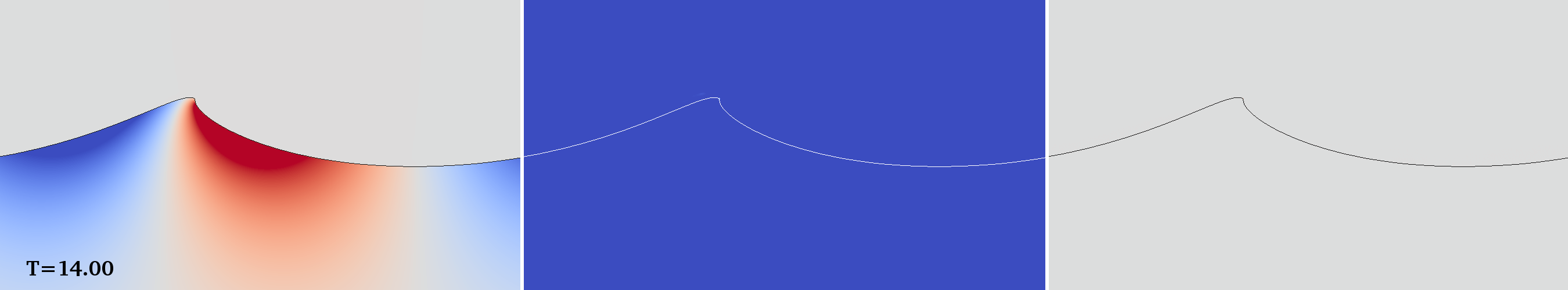}\\
\end{minipage}
\begin{minipage}[c]{0.01\textwidth}
{\bf 2}\\
\end{minipage}
\begin{minipage}[c]{0.98\textwidth}
\centering
\includegraphics[trim = 0mm .6cm 0mm
1.25cm,clip=true,angle=0,scale=0.159]{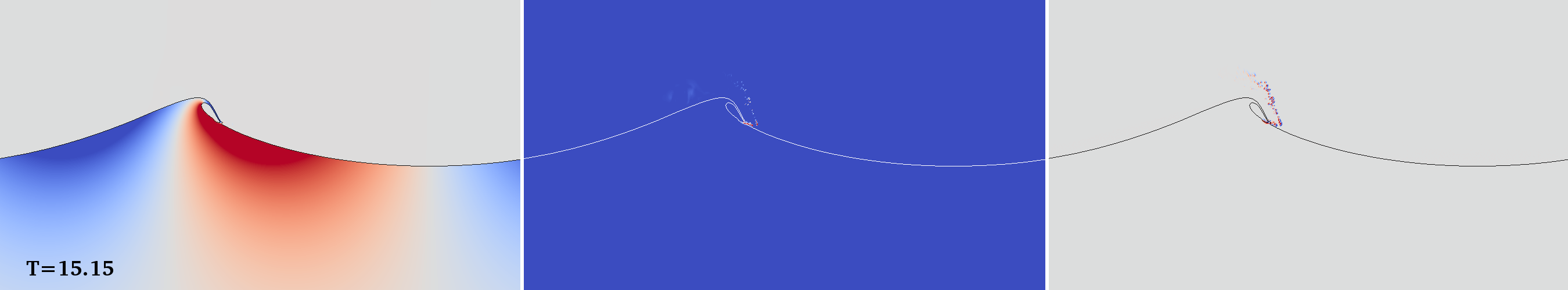}\\
\end{minipage}
\begin{minipage}[c]{0.01\textwidth}
{\bf 3}\\
\end{minipage}
\begin{minipage}[c]{0.98\textwidth}
\centering
\includegraphics[trim = 0mm .6cm 0mm
1.25cm,clip=true,angle=0,scale=0.159]{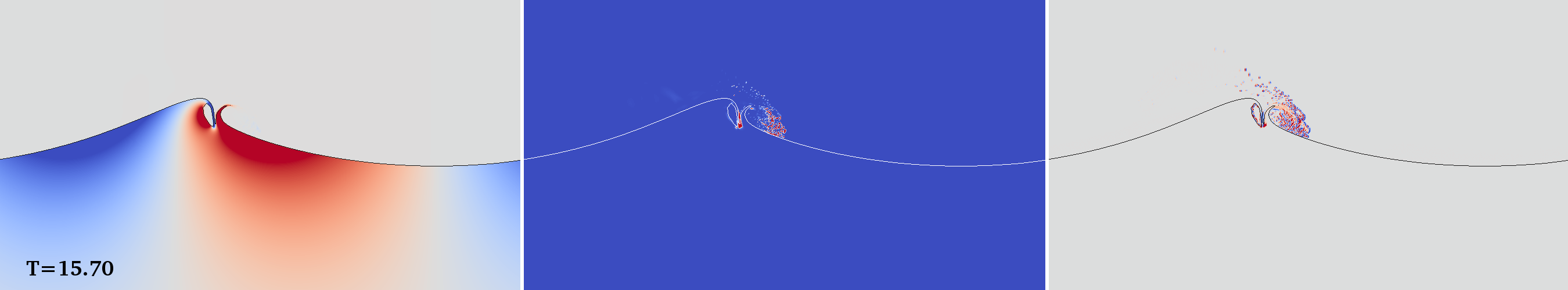}\\
\end{minipage}
\begin{minipage}[c]{0.01\textwidth}
{\bf 4}\\
\end{minipage}
\begin{minipage}[c]{0.98\textwidth}
\centering
\includegraphics[trim = 0mm .6cm 0mm
1.25cm,clip=true,angle=0,scale=0.159]{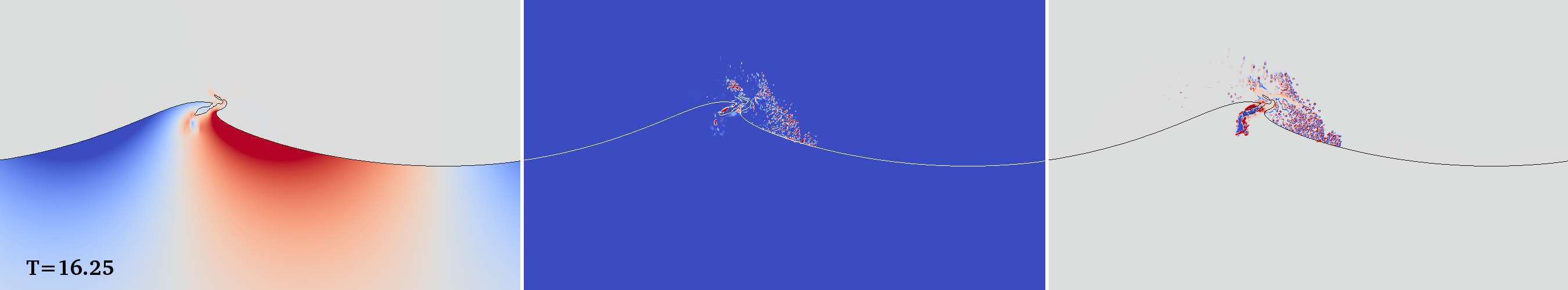}\\
\end{minipage}
\begin{minipage}[c]{0.01\textwidth}
{\bf 5}\\
\end{minipage}
\begin{minipage}[c]{0.98\textwidth}
\centering
\includegraphics[trim = 0mm .6cm 0mm
1.25cm,clip=true,angle=0,scale=0.159]{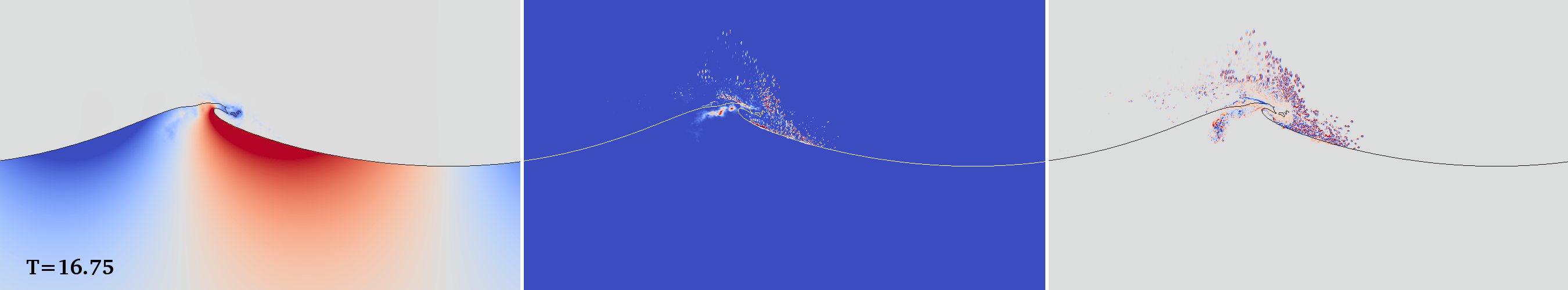}\\
\end{minipage}
\begin{minipage}[c]{0.01\textwidth}
{\bf 6}\\
\end{minipage}
\begin{minipage}[c]{0.98\textwidth}
\centering
\includegraphics[trim = 0mm .6cm 0mm
1.25cm,clip=true,angle=0,scale=0.159]{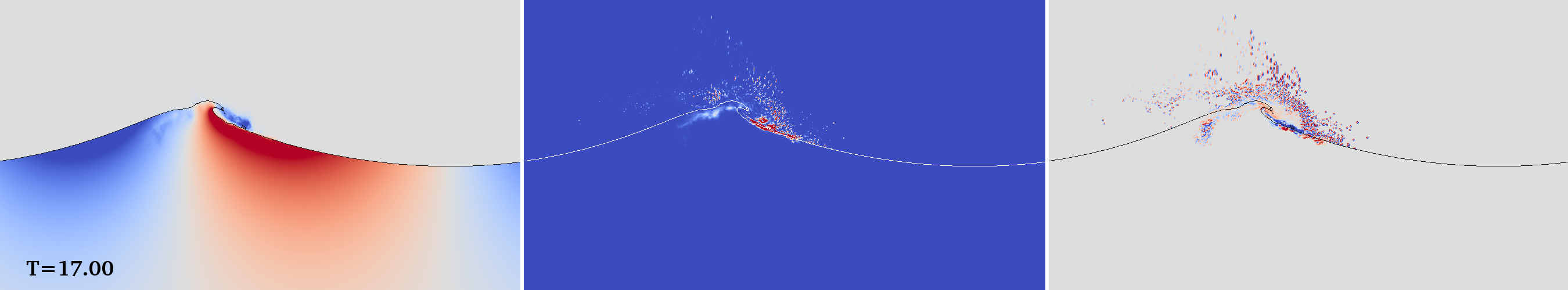}\\
\end{minipage}
\begin{minipage}[c]{0.01\textwidth}
{\bf 7}\\
\end{minipage}
\begin{minipage}[c]{0.98\textwidth} 
\centering
\includegraphics[trim = 0mm .6cm 0mm
1.25cm,clip=true,angle=0,scale=0.159]{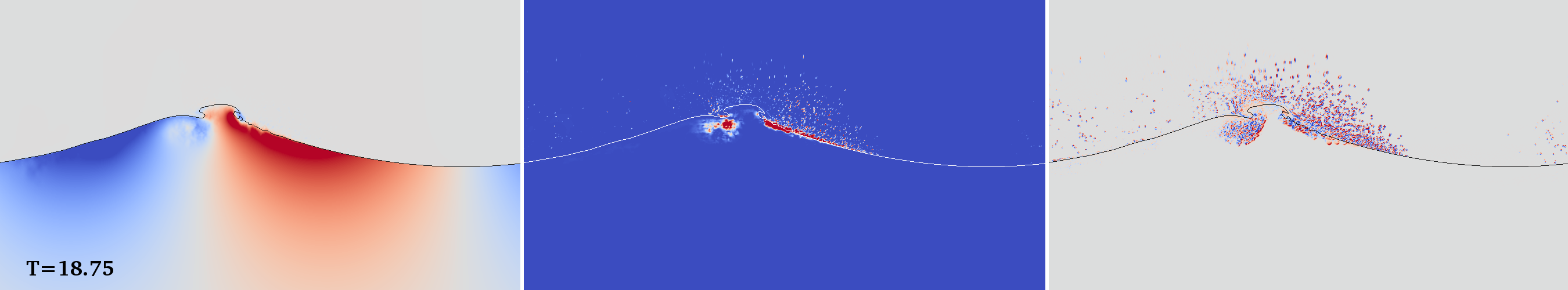}\\
\end{minipage}
\begin{minipage}[c]{0.01\textwidth}
{\bf 8}\\
\end{minipage}
\begin{minipage}[c]{0.98\textwidth} 
\centering
 \includegraphics[trim = 0mm .6cm 0mm
1.25cm,clip=true,angle=0,scale=0.159]{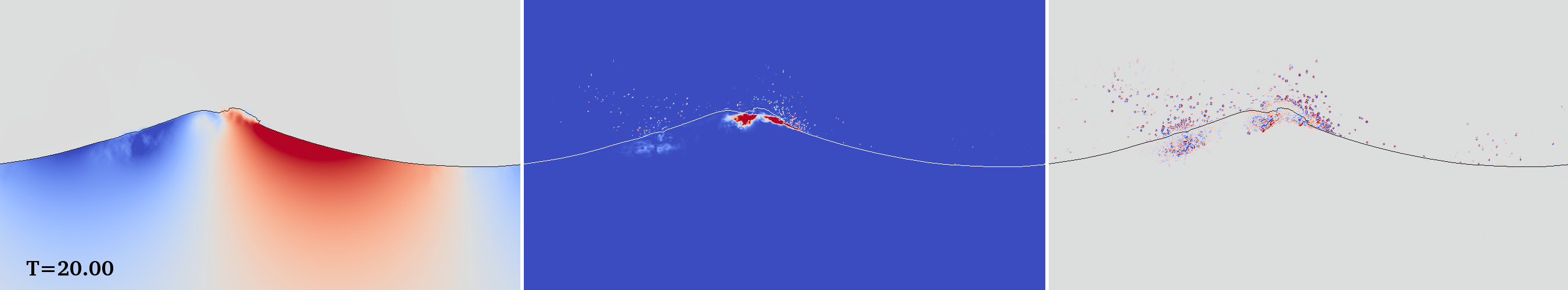}\\
\end{minipage}
\caption{\label{fig:p_bpv}
Buoyant Flux, Shear Production of T.K.E., Volume Expansion Work
for case $P^F$.   The non-dimensional time is given in
the lower left corner of column A and the contour levels are given at
the top of each column.}
\end{figure*}

\begin{figure*}
\centering
{\bf A \hspace{135pt} B  \hspace{135pt} C }\\
\vspace{5pt}
\begin{minipage}[c]{0.01\textwidth}
\hspace{6pt}\\
\end{minipage}
\begin{minipage}[c]{0.98\textwidth}
\centering
\includegraphics[trim = 0mm 0cm 0mm
0cm,clip=true,angle=0,scale=0.159]{figures/contours/P/p_wp_vmt_legend.png}\\
\end{minipage}
\begin{minipage}[c]{0.01\textwidth}
{\bf 1}\\
\end{minipage}
\begin{minipage}[c]{0.98\textwidth}
\centering
\includegraphics[trim = 0mm .6cm 0mm
1.25cm,clip=true,angle=0,scale=0.159]{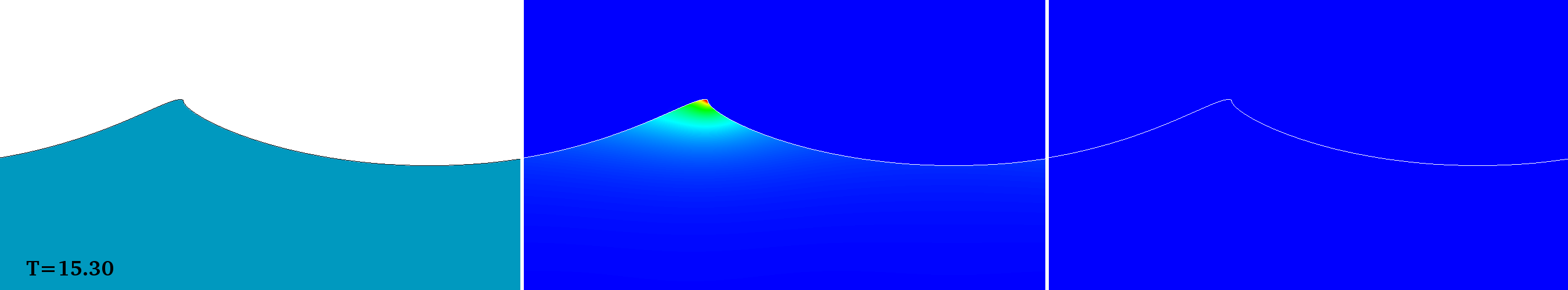}\\
\end{minipage}
\begin{minipage}[c]{0.01\textwidth}
{\bf 2}\\
\end{minipage}
\begin{minipage}[c]{0.98\textwidth}
\centering
\includegraphics[trim = 0mm .6cm 0mm
1.25cm,clip=true,angle=0,scale=0.159]{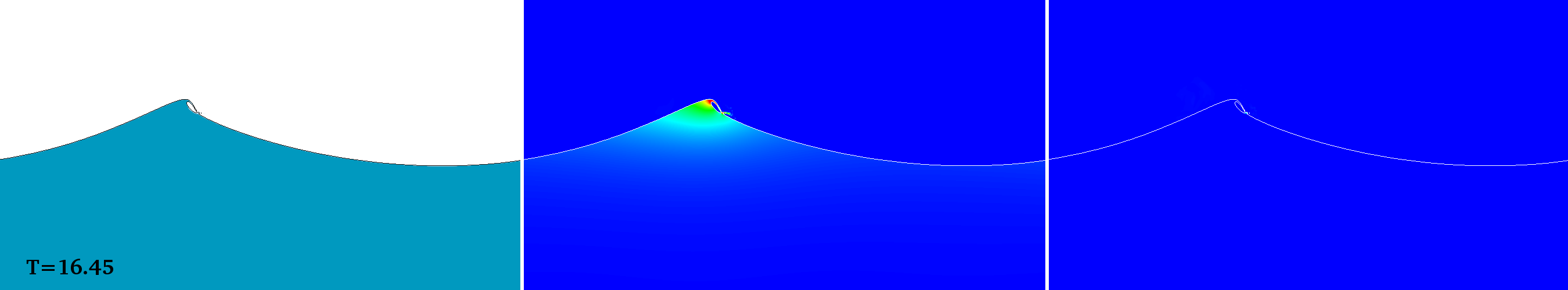}\\
\end{minipage}
\begin{minipage}[c]{0.01\textwidth}
{\bf 3}\\
\end{minipage}
\begin{minipage}[c]{0.98\textwidth}
\centering
\includegraphics[trim = 0mm .6cm 0mm
1.25cm,clip=true,angle=0,scale=0.159]{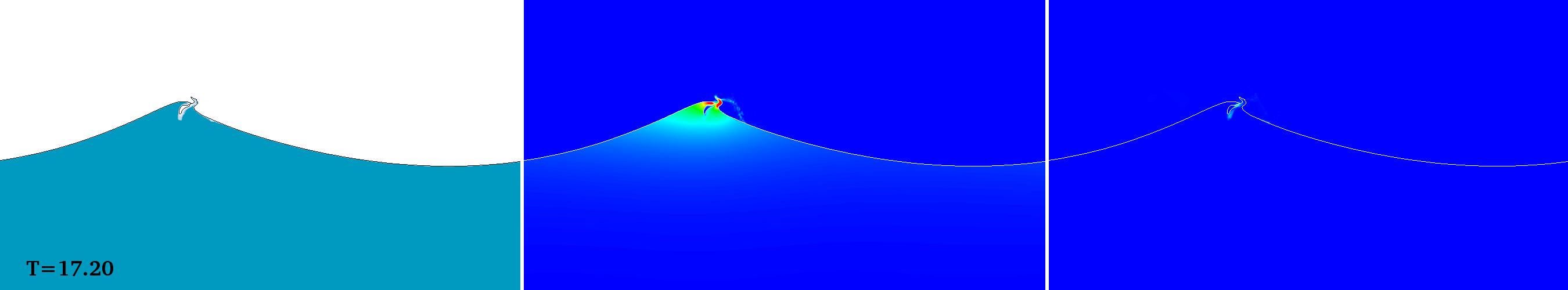}\\
\end{minipage}
\begin{minipage}[c]{0.01\textwidth}
{\bf 4}\\
\end{minipage}
\begin{minipage}[c]{0.98\textwidth}
\centering
\includegraphics[trim = 0mm .6cm 0mm
1.25cm,clip=true,angle=0,scale=0.159]{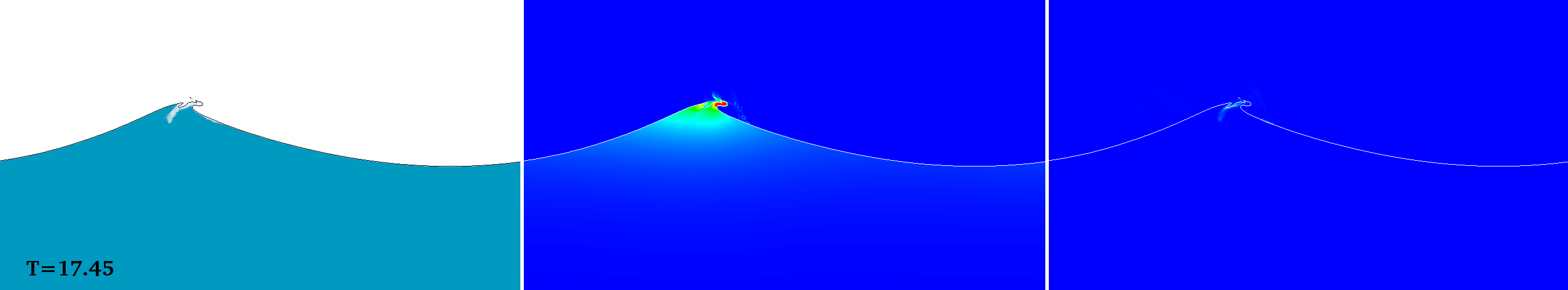}\\
\end{minipage}
\begin{minipage}[c]{0.01\textwidth}
{\bf 5}\\
\end{minipage}
\begin{minipage}[c]{0.98\textwidth}
\centering
\includegraphics[trim = 0mm .6cm 0mm
1.25cm,clip=true,angle=0,scale=0.159]{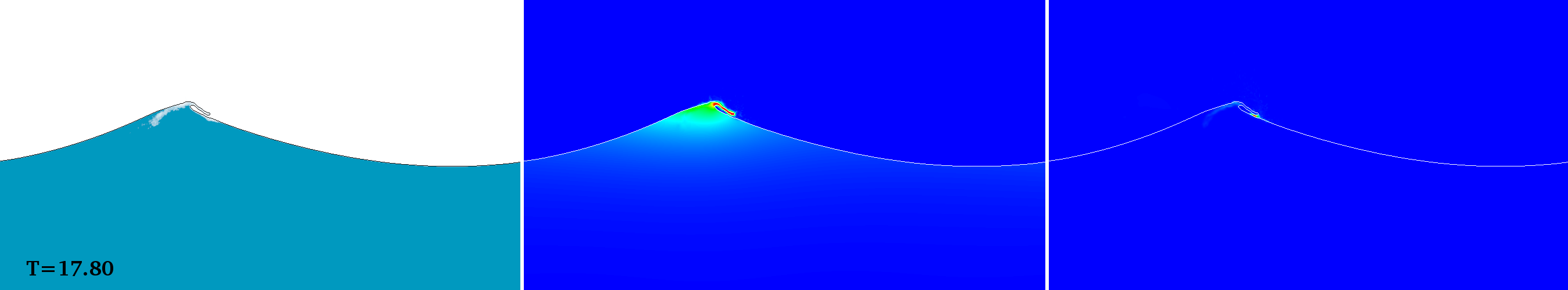}\\
\end{minipage}
\begin{minipage}[c]{0.01\textwidth}
{\bf 6}\\
\end{minipage}
\begin{minipage}[c]{0.98\textwidth}
\centering
\includegraphics[trim = 0mm .6cm 0mm
1.25cm,clip=true,angle=0,scale=0.159]{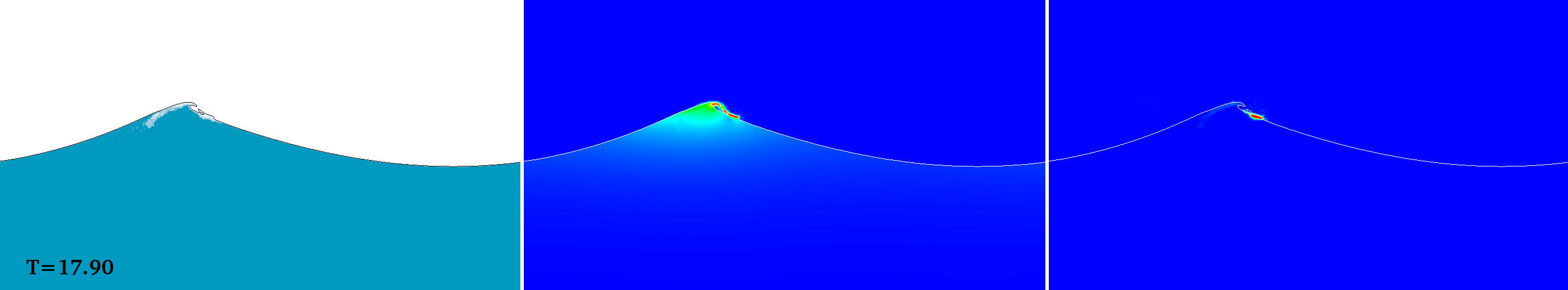}\\
\end{minipage}
\begin{minipage}[c]{0.01\textwidth}
{\bf 7}\\
\end{minipage}
\begin{minipage}[c]{0.98\textwidth} 
\centering
\includegraphics[trim = 0mm .6cm 0mm
1.25cm,clip=true,angle=0,scale=0.159]{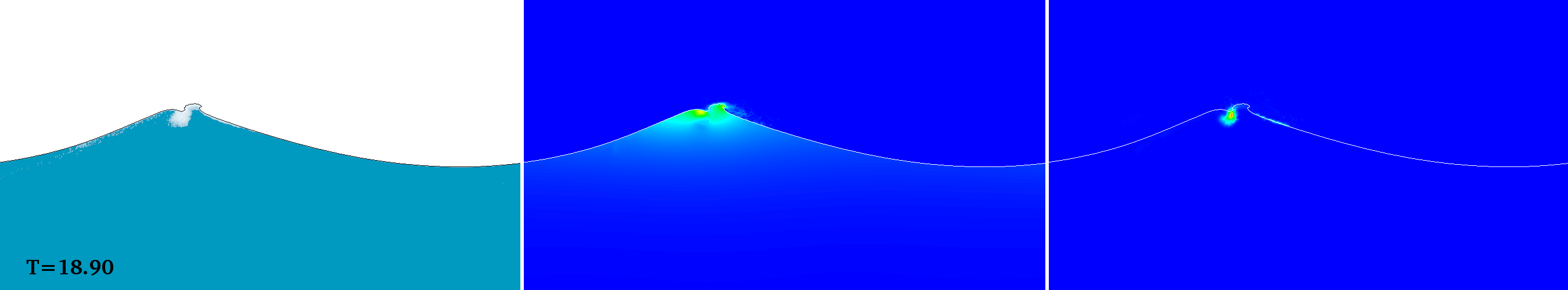}\\
\end{minipage}
\begin{minipage}[c]{0.01\textwidth}
{\bf 8}\\
\end{minipage}
\begin{minipage}[c]{0.98\textwidth} 
\centering
 \includegraphics[trim = 0mm .6cm 0mm
1.25cm,clip=true,angle=0,scale=0.159]{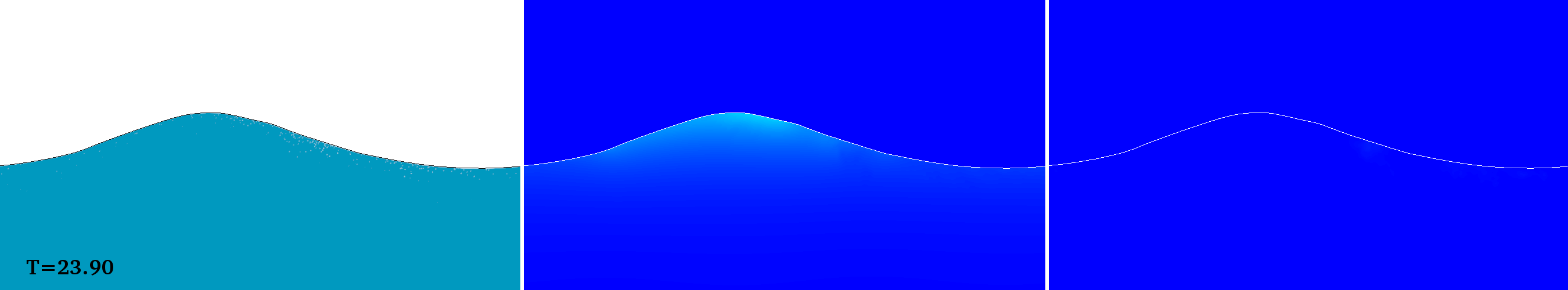}\\
\end{minipage}
\caption{\label{fig:wp_vmt}
Volume of Air, Mean Kinetic Energy (Eq. \ref{eq:mkeF}), and
Turbulent Kinetic Energy (Eq. \ref{eq:tkeF}) for case $WP^F$. The non-dimensional time is given in
the lower left corner of column A and the contour levels are given at
the top of each column.}
\end{figure*}

\begin{figure*}
\centering
{\bf A \hspace{135pt} B  \hspace{135pt} C }\\
\vspace{5pt}
\begin{minipage}[c]{0.01\textwidth}
\hspace{6pt}\\
\end{minipage}
\begin{minipage}[c]{0.98\textwidth}
\centering
\includegraphics[trim = 0mm 0cm 0mm
0cm,clip=true,angle=0,scale=0.159]{figures/contours/P/p_wp_bpv_legend.png}\\
\end{minipage}
\begin{minipage}[c]{0.01\textwidth}
{\bf 1}\\
\end{minipage}
\begin{minipage}[c]{0.98\textwidth}
\centering
\includegraphics[trim = 0mm .6cm 0mm
1.25cm,clip=true,angle=0,scale=0.159]{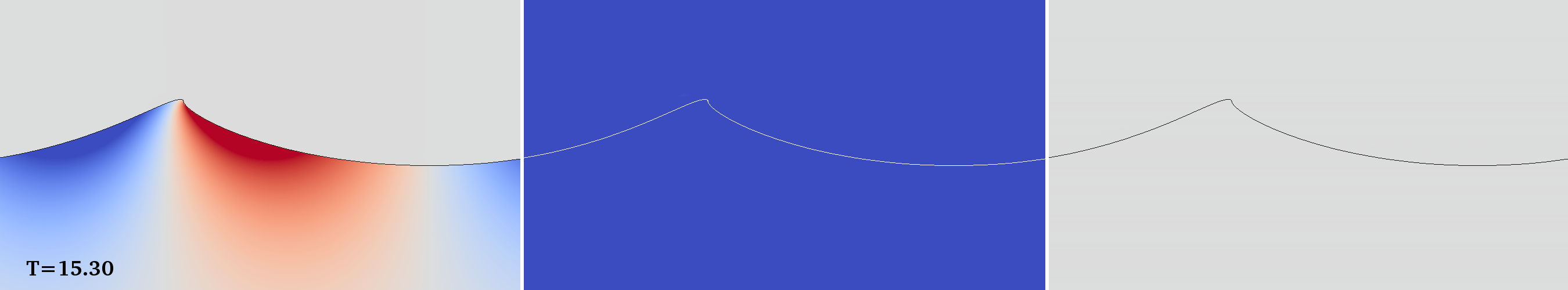}\\
\end{minipage}
\begin{minipage}[c]{0.01\textwidth}
{\bf 2}\\
\end{minipage}
\begin{minipage}[c]{0.98\textwidth}
\centering
\includegraphics[trim = 0mm .6cm 0mm
1.25cm,clip=true,angle=0,scale=0.159]{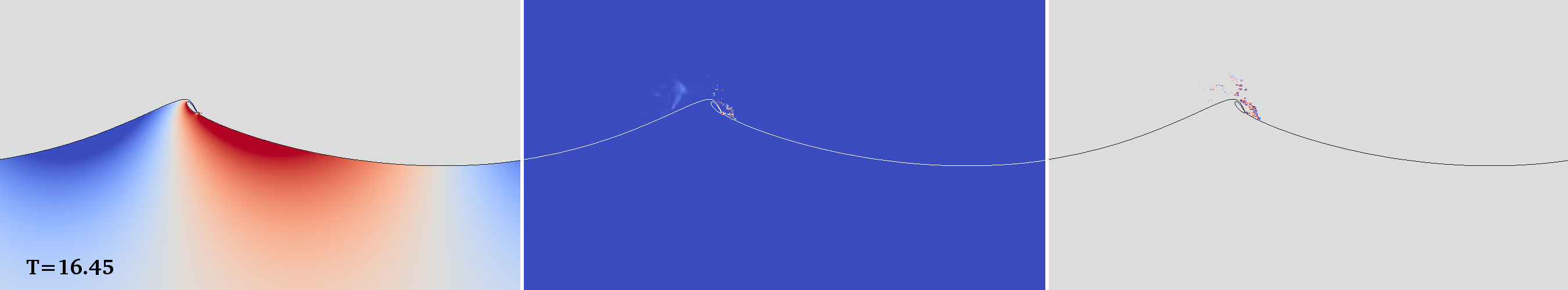}\\
\end{minipage}
\begin{minipage}[c]{0.01\textwidth}
{\bf 3}\\
\end{minipage}
\begin{minipage}[c]{0.98\textwidth}
\centering
\includegraphics[trim = 0mm .6cm 0mm
1.25cm,clip=true,angle=0,scale=0.159]{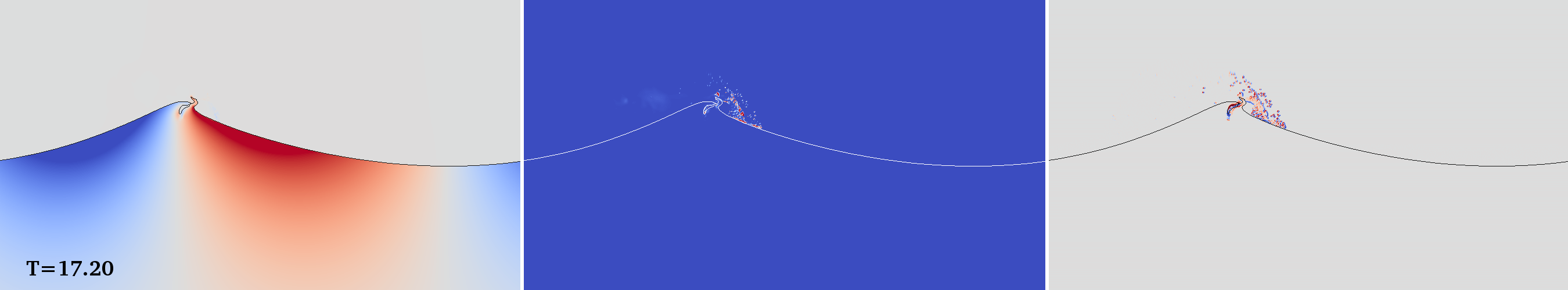}\\
\end{minipage}
\begin{minipage}[c]{0.01\textwidth}
{\bf 4}\\
\end{minipage}
\begin{minipage}[c]{0.98\textwidth}
\centering
\includegraphics[trim = 0mm .6cm 0mm
1.25cm,clip=true,angle=0,scale=0.159]{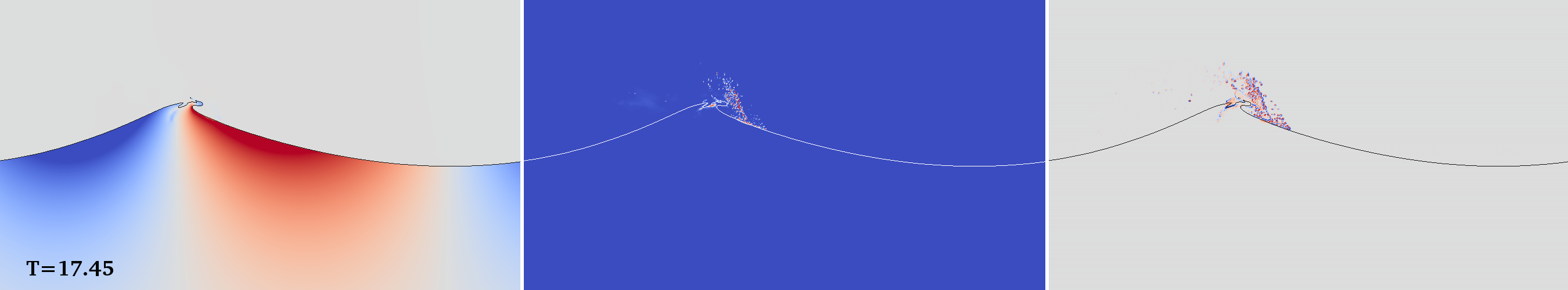}\\
\end{minipage}
\begin{minipage}[c]{0.01\textwidth}
{\bf 5}\\
\end{minipage}
\begin{minipage}[c]{0.98\textwidth}
\centering
\includegraphics[trim = 0mm .6cm 0mm
1.25cm,clip=true,angle=0,scale=0.159]{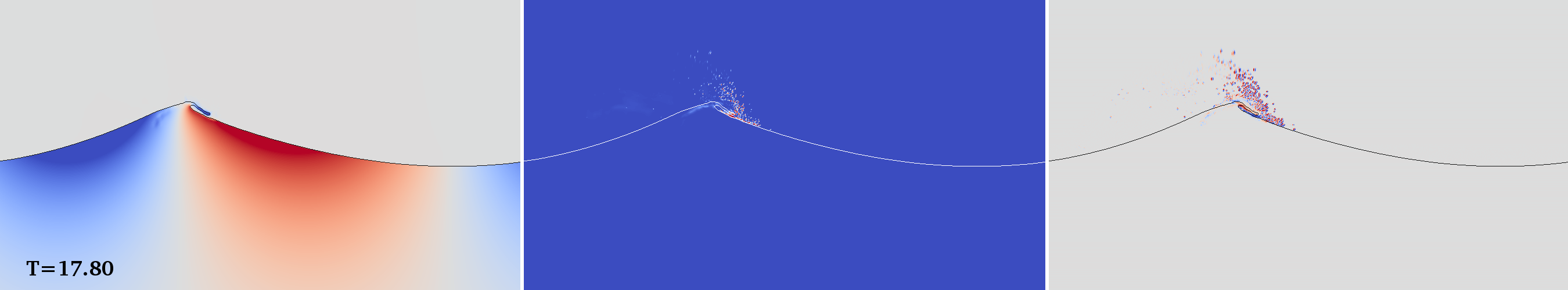}\\
\end{minipage}
\begin{minipage}[c]{0.01\textwidth}
{\bf 6}\\
\end{minipage}
\begin{minipage}[c]{0.98\textwidth}
\centering
\includegraphics[trim = 0mm .6cm 0mm
1.25cm,clip=true,angle=0,scale=0.159]{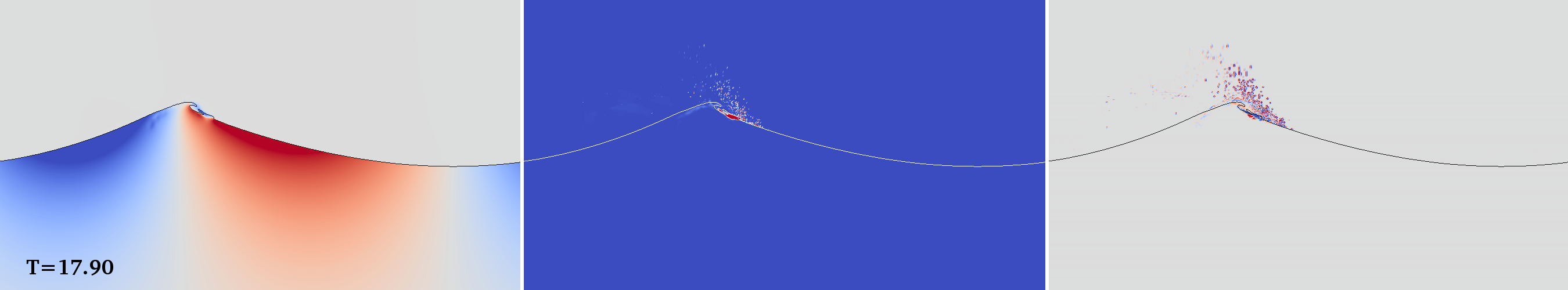}\\
\end{minipage}
\begin{minipage}[c]{0.01\textwidth}
{\bf 7}\\
\end{minipage}
\begin{minipage}[c]{0.98\textwidth} 
\centering
\includegraphics[trim = 0mm .6cm 0mm
1.25cm,clip=true,angle=0,scale=0.159]{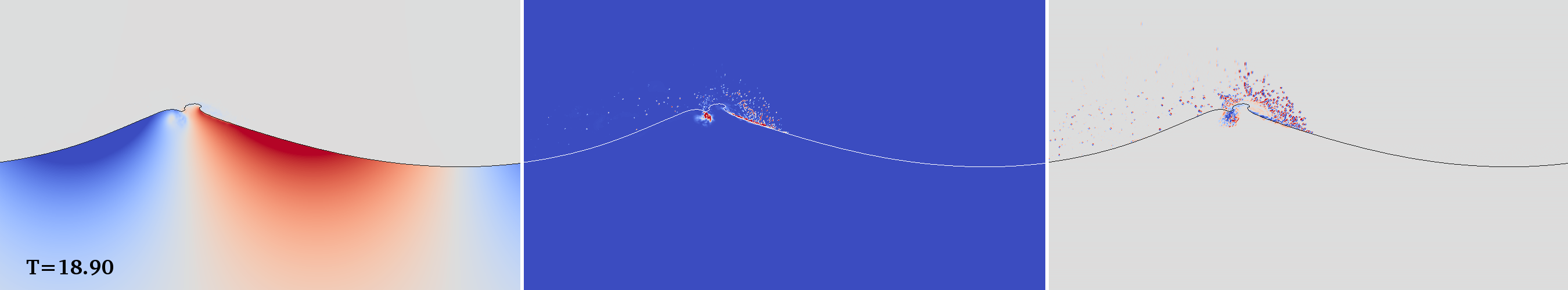}\\
\end{minipage}
\begin{minipage}[c]{0.01\textwidth}
{\bf 8}\\
\end{minipage}
\begin{minipage}[c]{0.98\textwidth} 
\centering
 \includegraphics[trim = 0mm .6cm 0mm
1.25cm,clip=true,angle=0,scale=0.159]{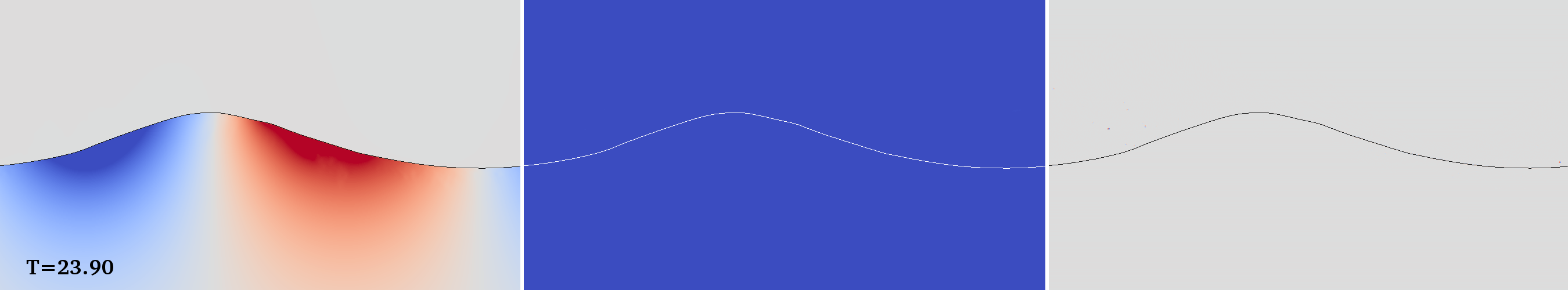}\\
\end{minipage}
\caption{\label{fig:wp_bpv}
Buoyant Flux, Shear Production of T.K.E., Volume Expansion Work
for case $WP^F$.   The non-dimensional time is given in
the lower left corner of column A and the contour levels are given at
the top of each column.}
\end{figure*}

Figures~\ref{fig:vsp_vmt}, \ref{fig:sp_vmt}, \ref{fig:p_vmt}, and \ref{fig:wp_vmt} show the volume fraction, mean kinetic energy, and turbulent kinetic energy for cases $VSP^F$, $SP^F$, $P^F$, and $WP^F$, respectively. Figures~\ref{fig:vsp_bpv}, \ref{fig:sp_bpv}, \ref{fig:p_bpv}, and \ref{fig:wp_bpv} show buoyant flux, turbulent kinetic energy production due to shear, and volume expansion work for the same cases. All quantities are based on span-wise Favre averaging. Each row is labeled and corresponds to the points labeled in
Figures~\ref{fig:energyrate}(a)-(d).

Recall in the discussion of Figure~\ref{fig:3d} eight stages were discussed and in each of the subsequent figures the rows
correspond respectively to these stages: 1. Vertical Front Face; 2. Pinch Off; 3. Splash-up; 4. Interaction; 5. Tip
Break Off; 6. Tip Impact; 7. Dual Breaking; 8. Spilling.

In column C of Figures~\ref{fig:vsp_vmt}, \ref{fig:sp_vmt}, \ref{fig:p_vmt}, and \ref{fig:wp_vmt}  turbulent kinetic energy is greatest on the front face of the
wave in the toe region of splashing events (row 6). Column B of Figures~\ref{fig:vsp_bpv}, \ref{fig:sp_bpv}, \ref{fig:p_bpv}, and \ref{fig:wp_bpv} show that
production of turbulent kinetic energy is also greatest in this region. Based on column A of Figures~\ref{fig:vsp_vmt}, \ref{fig:sp_vmt}, \ref{fig:p_vmt}, and \ref{fig:wp_vmt}, 
there are three types of air entraining events: (1) Breakup of the cavity formed by the first plunging
event; (2) Breakup of the cavity formed by the splash-up event; and (3) Entrainment at the toe of the spilling
breaking. Plumes of air are periodically entrained and left behind. The greatest amount of air is entrained by
the VSP and SP plunging breakers. The depth of the of air entrainment is proportional to the wave amplitude.
Based on column A of Figures~\ref{fig:vsp_vmt}, \ref{fig:sp_vmt}, \ref{fig:p_vmt}, and \ref{fig:wp_vmt}, the mean kinetic energy is highest in the crest of the first
plunging and splashing events. At later time, turbulent kinetic energy is transformed into mean kinetic energy
beneath the crest of the spilling breaking region. Based on column A of Figures~\ref{fig:vsp_bpv}, \ref{fig:sp_bpv}, \ref{fig:p_bpv}, and \ref{fig:wp_bpv}, the
buoyant flux is positive on the front face of waves and negative on the back faces, as expected. The structure
of the buoyant flux indicates that energy is being transferred from the primary gravity wave to shorter waves on
the front face of the primary gravity wave. The volume expansion work in column C of Figures~\ref{fig:vsp_bpv}, \ref{fig:sp_bpv}, \ref{fig:p_bpv}, and \ref{fig:wp_bpv} is less organized than the production of turbulent kinetic energy, but it too is greatest in the splashing
region.

Stage 7 is the stage in which forward and backward plunging events are evident. Significant
t.k.e.\ is generated at both the forward and rear plunging locations. Note that the m.k.e., t.k.e., air entrainment, production,and volume expansion are all larger at the rear breaking
region. It is also the region in which the large bubble cloud is formed. The buoyancy sets up a rear recirculation zone, which pumps air down. In Figure~\ref{fig:sp_vmt}(A7) notice the rear breaking region has a positive B region beneath the negative region which causes the water to push up on the free-surface and provide a path way for air to be driven into.

It is after this stage that the distinct deep and shallow air volumes are evident in Figures~\ref{fig:vsp_vmt}(A8), \ref{fig:sp_vmt}(A8)
and \ref{fig:p_vmt}(A8). The presence of the deep and shallow bubble clouds has been observed in the experiments of \citet{rapp90} and are discussed in the review of \citet{melville96}.

\section{Computational Improvements}
\label{sec:cmp_imp}

The Numerical Flow Analysis (NFA) code was originally designed to provide turnkey capabilities to model
the flow around breaking waves around ship hulls, including plunging and spilling breaking waves, the formation of spray, and the entrainment of air \citep[see][]{dommermuth07,dommermuth08}, In the most recent development cycle the types of problems for which it is well suited have been significantly extended. To
provide some perspective to the scope of problems NFA can accurately simulate, consider that there are five other
symposium papers based, all or in part, on NFA simulations, including: (1) Numerical predictions of seaway by \citet{nfa4}; (2) Comparisons of model-scale experimental measurements and computational predictions for the transom wave of a large-scale transom model by \citet{nfa5}; (3) Parameterization of the internal-wave field generated by a submarine and its turbulent wake in a uniformly stratified fluid by \citet{nfa3}; (4) A comparison of measured
and predicted wave impact pressures from breaking and non-breaking waves by \citet{nfa2}; and (5) A
comparison of experimental measurements and computational predictions of a deep-V planing hull by \citet{nfa6}.

In addition to the advances in the numerical methods which expanded the capabilities, one significant
modification to the inter-processor data communication routines and sub-domain layout has been implemented.
This modification leverages the small number of blocks (or sub-domains) in any one direction that result from
the three-dimensional domain decomposition and allows 1-dimensional sub-communicators to replace the global
communicator in MPI collective data operations. The modifications to the data-communication improve the
execution time in terms of wall-clock hours and CPU hours required per unknown by a factor of 2 or more.
More importantly the elimination of the collective data operations allows NFA to scale to thousands of processors.

\begin{figure}
\centering
\includegraphics[width=\linewidth]{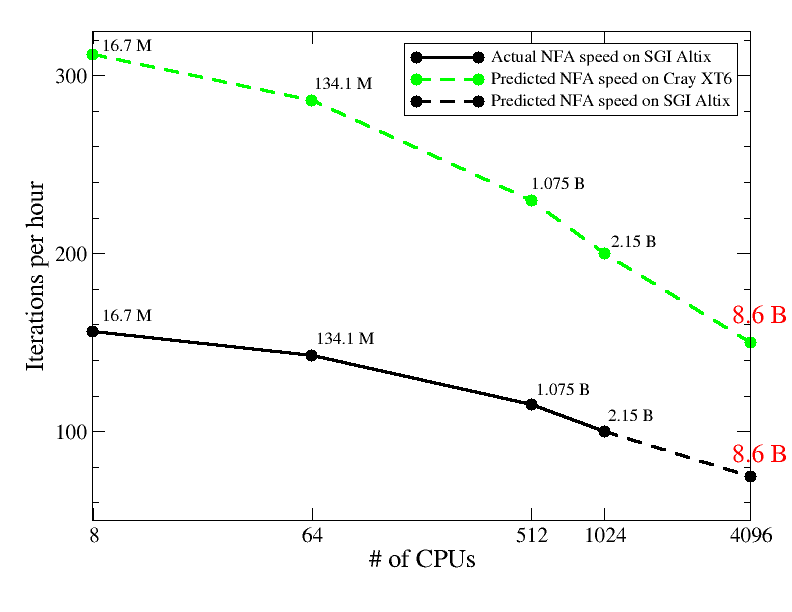}
\caption{\label{fig:scaling}
The number of iterations (time steps) per hour achieved while the
number of grid points (unknowns) per core is fixed at $128 \times 128 \times 128$ and the number
of CPU's (or cores) is increased.}
\end{figure}

To demonstrate the ability of NFA to scale to thou-sands of processors, a weak scaling analysis is performed. Weak scaling, in the context of high performance computing, is a measure of how an algorithm performs when the problem size per node is held constant and the number of nodes is increased. In this type of
analysis as the number of cores increases, the number of unknowns increases in direct proportion. Here the problem size is $128^3$ unknowns per node and the iterations per hour achieved on 8, 64, 512, and 1024 nodes is timed. To add perspective as to total number of unknowns, the case with 1024 cores has 2.15 billion unknowns and the case
with 8 cores has 16.7 million unknowns. The analysis has been carried out on the the SGI Altix ICE cluster (Diamond located at the ERDC). The results are shown in Figure~\ref{fig:scaling}, along with the predicted performance on the new Cray XT6 platforms. In summary, Figure~\ref{fig:scaling} shows that in its current form NFA can efficiently simulate
jobs with 2+ billion unknowns in a reasonable number of wall-clock hours.

The scalings in Figure~\ref{fig:scaling} are based on normal working conditions with many users running jobs. On a dedicated machine, the node layout can be optimized leading to an additional factor of two increase in performance. Recent simulations on the SGI ALTIX ICE platform, discussed in \citet{nfa5}, achieved nearly 340 iteration per hour with 1.05 billion unknowns. Assuming that the speed of the new Cray XT6 is twice that of the SGI ALTIX and that dedicated access time is available,a factor of four speed increase can be expected. It is reasonable to assume that in this environment 10 billion cell jobs can achieve 100 iterations/hour.

\section{Conclusions}

The formulation for a canonical deep-water breaking wave is developed. In conjunction, an atmospheric pressure forcing technique which is able to generate fully non-linear progressive waves is developed and validated. Together they enabled a systematic study of deep-water plunging breaking waves.

\section{Acknowledgments} The Office of Naval Research supports this research.
Dr.~Patrick Purtell is the program manager.  This work is supported in part by a
grant of computer time from the \href{http://www.hpcmo.hpc.mil/}{DOD High Performance Computing Modernization
Program}.  The numerical simulations have been
performed on the Cray XT4 at the U.S. Army Engineering Research and Development
Center (ERDC).  Animations of the breaking wave simulations are available at
\href{http://www.youtube.com/waveanimations}{http://www.youtube.com/waveanimations}.

\bibliography{28onr} 
\bibliographystyle{28onr}

\end{document}